AGH University of Science and Technology
Faculty of Materials Science and Ceramics
Department of Physical Chemistry and Modelling

Ph.D. Dissertation

# DIFFUSION AND PHASE EVOLUTION IN MULTI-COMPONENT MULTI-PHASE ALLOYS

Ph.D. candidate:
Marek Zajusz

Supervisor:
Prof. dr hab. inż. Marek Danielewski

Acknowledges:

This dissertation would be impossible without support of many persons, in this place I would like to express my gratitude to them. First and foremost to my supervisor, prof. dr hab. inż. Marek Danielewski for his guidance, patience, time, and friendship during my postgraduate studies. His mentorship was priceless in providing knowledge and experience required in my research. I would also like to thank prof. Andriy Gusak for his encouragement and variable discussions. I also thank other members of prof. Danielewski research group, especially dr Bogusław Bożek and dr hab. inż. Katarzyna Tkacz-Śmiech prof. AGH for their valuable help and assistance in completing the dissertation.

My special thanks go to the ACMiN, especially dr inż. Katarzyna Berent, for help in SEM imagining and EDS measurements. I also thank dr inż. Dariusz Zientara from the Department of Ceramics and Refractories, WIMiC AGH from his help in implementing HIP procedure and to Mr Tadeusz Krzywda from Institute of Advanced Manufacturing Technology in Kraków for EDM machining.

Finally, I would like to present my hearty thanks to my family, friends from KLIKA and colleagues from my department for their support.

This work is supported by a National Science Center (Poland) decision DEC-011/02/A/ST8/00280.

Marek Zajusz

## Table of contents

## Abstract

The diffusion in ternary, multiphase systems was studied theoretically and experimentally in Ni-Cr-Al system at 1200°C. The samples were prepared by the multiple method. It has been shown that the method allows obtaining good quality, planar, reproducible and oxide-free diffusion couples by applying HIP procedure. For the first time the concentration profiles were measured by the wide-line EDS analysis.

Two models were applied: 1) multi-phase multi-component model based on the bi-velocity method in R1 and 2) dual-scale two-phase model, that includes interdiffusion in matrix (global scale in R1) and the diffusion between matrix and precipitates (local scale in R3). The models were used in numerical experiments in which the diffusion paths, the local entropy production rate and the volume fractions of the phases present in the system were calculated. For the first time the Kirkendall plane shift in two-phase system was computed and compared with experimental findings. It has been shown that the dual-scale two-phase model of interdiffusion in two-phase zone is consistent with generalized multi-phase multi-component model.

The multi-phase, multi-component model was applied to simulate diffusion and the results were compared with experiment. Experimental and numerical results are in agreement. It has been found that: 1) the type 0 interphase boundary (IB) moves during diffusion, contrary to the Morral theorem, 2) in the case when the terminal compositions of the two-phase alloys are close to the phase boundary then the single β-phase zone can grow between two-phase zones and diffusion path enters the single-phase region, 3) the diffusion path strongly depends on the shape of the phase boundary line and the accuracy of thermodynamical data is a key factor in modelling and 4) the recently reported occurrence of the local extrema in the distribution of entropy production at the IBs is a result of numerical errors and is related to the discontinuity of concentration at IB. Such errors can be eliminated as is shown in dual-scale two-phase model.

The new challenges for theory end experiment that follow from this work are presented in summary.

## Lexicon

The symbols and nomenclature used in publications dedicated to diffusion in ternary systems is not unified and may confuse even the most careful reader. The main reason for this is the fact that phenomenon “occurs” in two different spaces: real space assigned to the physical sample and the space embedded in concentration triangle.

To avoid ambiguity, the uniform nomenclature based on [1-3] are used here.

**Single-phase zone**: The one –phase area in the real space, typically observed at cross-sections.

**Two-phase zone**: The area in real space containing two phases.

**Interphase boundary** (IB): The interface between two areas showing different phase compositions in real space.

**(Single-, two-, three-) phase region:** The part of the phase diagram related to the stability of single, two or three phases.

**Phase boundary**: The line on phase diagram which is a margin of a two-phase region.

**Conode** or **tie-line**: A straight line passing through the two-phase region that connects points at the phase boundaries being in equilibrium with each other.

**Concentration profile**: A graphical representation of concentration/molar ratio, experimental or simulated, versus position in the sample.

**Diffusion path**: Graphical representation of the concentration in form: $c_2 = c_2(c_1)$ mapped on to concentration triangle. In a case of semi-infinitive diffusion couple it connects terminal compositions of the diffusion couple on the phase diagram [4, 5].

## List of key symbols

$i = 1,2,3$ - components: $\mathrm{r} = 3$;

$j = \alpha, \beta, \gamma$ - phases;

$N_i = N_i(t,x)$ - overall molar fraction of the *i*-th component;

$N_i^j = N_i^j(t,x)$ - molar fraction of the *i*-th component in the *j*-th phase;

$c_i = c_i(t,x)$ - overall concentration of the *i*-th component;

$c_i^j = c_i^j(t,x)$ - concentration of the *i*-th component in the *j*-th phase;

$c^j = c^j(t,x) = \sum_i c_i^j$ - overall concentration in the *j*-th phase;

$\varphi^j = \varphi^j(t,x)$ - volume fraction of the j-phase;

$\varphi = \varphi^\alpha(t,x)$ - volume fraction of the $\alpha$-phase;

$\varphi^* = \varphi^{*\alpha}(t,x)$ - molar fraction of the $\alpha$-phase;

$\Omega_i^j = const$ - partial molar volume of the *i*-th component in the *j*-th phase;

$\Omega^j = \Omega^j(t,x)$ - molar volume of the *j*-th phase;

$\Omega = \Omega(t,x)$ - overall molar volume;

$B_i^j = const$ - mobility of the atoms of *i*-th component in the *j*-th phase;

$D_i^j = const$ - intrinsic diffusion coefficient of *i*-th component in the *j*-th phase;

$J_i^j = J_i^j(t,x)$ - diffusion flux of the *i*-th component in the *j*-th phase;

$\tilde{J}_i^j = \tilde{J}_i^j(t,x)$ - interdiffusion flux of the *i*-th component in the *j*-th phase;

$\tilde{J}_i = \tilde{J}_i(t,x)$ - interdiffusion flux of the *i*-th component in the two-phase zone;

$\upsilon^j = \upsilon^j(t,x)$ - drift velocity in the *j*-th phase;

$\upsilon = \upsilon(t,x)$ - drift velocity due to reaction in the two-phase zone;

$\dot{\sigma} = \dot{\sigma}(t,x)$ - local entropy production rate.

# Chapter 1. Introduction

The interdiffusion in multi-component systems controls many processes, like, nitriding, aluminizing, degradation of protective layers and therefore, an understanding of this process is very important both from theoretical and practical (technological) points of view. The diffusion in such systems can be followed by formation of multi-phase zone of complex morphology. In ternary systems the Gibbs phase rule allows a formation of two-phase zone. The diffusion in such zone is significantly different from the interdiffusion in the single phase. The reduction of the degrees of freedom from 2 to 1 may cause appearance of new effects, like zigzag diffusion, horns, jump of concentration without typical interphase boundary.

The interest in the ternary Ni-Cr-Al system is due to the applications of nickel-based superalloys in gas turbines and jet engines working at high temperatures. These alloys have superior mechanical properties and high heat-resistance at high temperatures [6, 7]. The alloys from this system are a basis for the development of creep resistance alloys. Therefore there is a big interest in the diffusion in these alloys.

The interest in the Ni-Cr-Al system is reflected by numerous experimental and theoretical studies of its phase diagram [8-23]. Most of them are confined to Ni-rich corner at high temperatures, i.e. above 1000°C. For this system three solid phases are of interest: austenitic $\gamma$(Ni) phase, austenitic $\gamma'$($Ni_3Al$) phase and tetragonal β(NiAl) phase.

Unfortunately due to high level of complication, the bibliography dedicated to diffusion in Ni-Cr-Al alloys is limited and there are only few experimental results.

This dissertation is to fill this gap. It is divided into six sections. In the first section review of the state of the art regarding interdiffusion in multi-component multi-phase systems is presented. Main objectives and motivation of this work are presented in section two. Generalized model of interdiffusion in two-phase systems and the multiscale two-phase model which links the processes occurring at macro and micro scale are presented in section three.

Experiments and modelling of interdiffusion in ternary Ni-Cr-Al systems are presented in sections four and five.

## 1.1 Theory of diffusion in multiphase systems

### 1.1.1 Multi-phase diffusion theorems

The rules for diffusion in ternary systems were summarized in the theorems formulated by Kirkaldy and Brown, (theorems 1-17) [24], and subsequently supplemented by Morral (theorems A1-A11) [3]. Below the theorems that relates to the diffusion in two-phase zone are quoted in original form (1-4, 8-17, A3-A5) using authors original numeration. The theorems for diffusion in single-phase region are omitted (5-7, A1-A2). The theorems related to the interphase boundaries (A6-A11) will be presented in the next section.

The Kirkaldy and Morrals theorems [3, 24]:

1 Diffusion penetration curves can be mapped onto the ternary isotherm (as diffusion path) for all times as stationary lines.

2 Calculated paths on ternary isotherms remain invariant as the four diffusion constants, (*interdiffusion coefficients*), are varied in direct proportion, (valid for semi-infinite couple).

3 Diffusion paths cannot be mapped back into the $C_1$–$C_2$–$\lambda$ ($\lambda = x/t^{0.5}$) space without the reintroduction of diffusion data.

4 Diffusion path for an infinite couple on the ternary isotherm must cross the straight line joining the terminal compositions at least once.

8 Diffusion path on the ternary isotherm is defined uniquely only by its terminal compositions.

9 There is no theoretical restriction that prevents *different* paths radiating from one terminal composition from crossing.

12 To the extent that lateral diffusion and non-uniformity of layer interfaces can be ignored or averaged out, the diffusion paths involving two-phase regions may be approximated by a stationary path connecting a continuous series of local equilibria.

13 Diffusion path that passes through a two-phase region coincident with a tie line contains a planar interface whose local equilibrium specification is given by that tie-line.

14 Diffusion path that passes into a two-phase region from a single-phase one at an angle to the tie lines and returned immediately to that same single-phase describes a region of isolated precipitations.

15 Diffusion path that passes into a two-phase region from a single-phase one at an angle to the tie lines but exits into another phase represents a columnar or a columnar-plus, isolated precipitate two-phase zone.

16 Diffusion path in a two-phase region may not reverse its order of crossing of the tie lines.

17 Paths that pass through three-phase triangles must do so along a straight line representing at its extremes the local equilibrium existing at a planar multiphase interface.

A3 Diffusion paths can be divided into segments corresponding to diffusion couple regions. In most cases, the segments are connected by a jump (i.e., a discontinuity) in composition.

A4 The end regions of a diffusion couple can contain one, two, or three phases, but intermediate regions can contain only one or two phases.

A5 Boundaries between microstructural regions can be typed by the number of phases that change on crossing the boundary. For ternaries the possible types are 0, 1, 2, and 3.

### 1.1.2 Interphase boundary in multi-phase systems

The interphase boundaries, (IB), separate the phase zones in multi-phase diffusion couple. Typically, the IB is planar, however the boundaries of multiphase regions are often irregular. Therefore the systematization of various types of boundaries is very useful for interpretation of results. For ternary diffusion couples the types of boundaries were classified by Morral. Initially Morral defined three types of the IB type 0, 1, and 2 [25]. Later the type 3 boundary was added [3].

The number (between 0 and 3), indicates how many phases change at the interface. Possible diffusion paths for types 0 to 3 are shown in Fig. 1. and possess the following properties [3]:

a) **Type 0 boundary**: According to theorems A6-A8, this type of boundary is observed at the interface (also called *pseudo-interface*) between two-phase zones having the same phase composition. It is always connected with concentration jump along conode, and is stationary, its velocity should be 0. Type 0 boundaries occur only at the initial diffusion couple interface. The type 0 interphase boundary imply zigzag diffusion path, (Fig. 1a), and zigzag diffusion paths with horns (Fig. 1b).

b) **Type 1 boundary**: According to theorem A9, in ternary systems such interphase boundary occurs between single-phase and two-phase zone, eg. α|α+β. In a typical case, when the single-phase zone grows, then a jump of composition is observed at the interphase (Fig. 1c). However, there is an exception when a microstructural boundary separates a two-phase zone that grows into a one-phase zone. In such case [3] predicts that the diffusion path contains no jump of the concentration (Fig. 1d) and the single-phase segment of the diffusion path is tangent to the phase boundary. To distinguish those kinds of type 1 IBs, the notation 1a and 1b will be used in this thesis respectively for the cases with and without concentration jumps.

c) **Type 2 boundaries** according to theorem A10, occurs when a single-phase zone transform into another single-phase, eg. α|β, with a jump of the diffusion path along the tie-line (Fig. 1e), or when one two-phase zone transform into another two-phase zone eg. α+ γ|β+γ, with a diffusion path jump through tie-triangle on the phase diagram, from one side to the other (Fig. 1f).

d) **Type 3 boundary**, α|β+γ according to theorem A11, in ternary systems occurs only when the diffusion path jumps through the tie-triangle, from its corner to the opposite two-phase field (Fig. 1g).

All above theoretical rules were confirmed theoretically by [26].

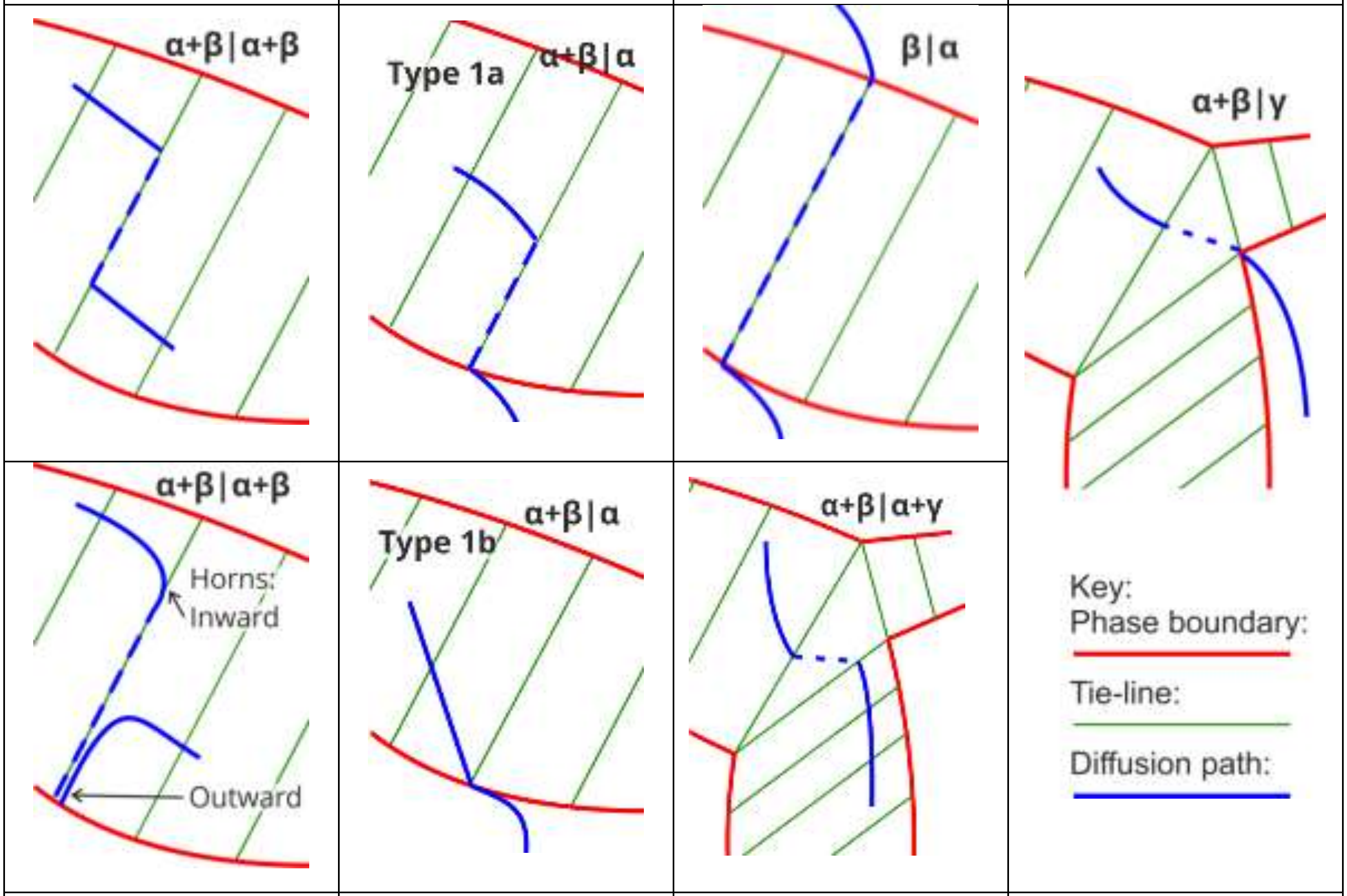

| Type 0 IB | Type 1 IB | Type 2 IB | Type 3 IB |
|---|---|---|---|
| a) zigzag path: for equal D;<br>b) zigzag with horns; | c) single-phase region is growing;<br>d) two-phase region is growing; | e) single-phase transition into another phase;<br>f) two-phase transition trough the tie-triangle; | g) diffusion path jump from the tie-triangle, from the single-phase corner to the two-phase region. |

**Fig. 1. Diffusion paths known for three component system, after [3].**

## 1.2 Diffusion modeling in two-phase ternary systems

Below various models of diffusion in ternary systems are presented in the historical order.

### 1.2.1 Model by Roper and Whittle (1981)

The historically first mathematical model of interdiffusion between single-phase and two-phase alloys base on the following assumptions [27]:

a) diffusion couple is semi-infinite i.e. diffusion zone is much smaller than dimension of the diffusion couple;
b) component 1 is the majority element, or solvent, and components 2 and 3 are the minority elements, or solutes;
c) one alloy is a single-phase α, and the second is a two-phase α+β alloy, in which α-phase is dominant;
d) diffusion through the minority β-phase is neglected;
e) equilibrium between α and β in the two-phase region/zone is maintained by dissolution or growth of β phase;
f) the concentration of the third component in α-phase in two-phase zone is constant (Fig. 2);
g) cross-diffusion coefficients, $\tilde{D}_{32}^{\alpha}$ and $\tilde{D}_{23}^{\alpha}$, in α-phase are negligible;
h) overall concentration in α-phase is constant $\sum c_i^{\alpha} = \sum c_i = c$ .

Evolution of the concentration and phase composition is determined by the mass conservation law in the Onsager formalism:

$$\frac{\partial N_i}{\partial t} = \sum_{j=2}^{3} \frac{\partial}{\partial x}\left( \tilde{D}_{ij} \frac{\partial N_j}{\partial x} \right), \quad i = 2,3 . \tag{1}$$

where $N_i$ is an overall molar ratio of the *i*-th component and $\tilde{D}_{ij}$ is an element of 2x2 interdiffusivity matrix in α-phase, $\tilde{D} = \left( \tilde{D}_{ij} \right)_{i,j=2,3}$ .

To solve Eq. (1) the diffusion in each part of diffusion couple was treated separately. Diffusion in the single-phase zone (phase α) was described by analytical solution [28, 29]. In order to solve Eq. (1) in two-phase zone more simplifications were postulated.

1) Because diffusion in β-phase is negligible, the concentration gradients in Eq. (1) are the concentration gradients in α-phase.

2) In the two-phase zone Eqs. (1) become:

$$\frac{\partial}{\partial t}\left(\varphi N_2^\alpha + (1-\varphi) N_2^\beta\right) = \frac{\partial}{\partial x}\left(\tilde{D}_{22}\frac{\partial N_2^\alpha}{\partial x}\right) + \frac{\partial}{\partial x}\left(\tilde{D}_{23}\frac{\partial N_3^\alpha}{\partial x}\right), \quad (2)$$

$$\frac{\partial}{\partial t}\left(\varphi N_3^\alpha + (1-\varphi) N_3^\beta\right) = \frac{\partial}{\partial x}\left(\tilde{D}_{32}\frac{\partial N_2^\alpha}{\partial x}\right) + \frac{\partial}{\partial x}\left(\tilde{D}_{33}\frac{\partial N_3^\alpha}{\partial x}\right), \quad (3)$$

where $N_i^j$ is an molar ratio of *i*-th components in *j*-th phase and $\varphi$ is a volume fraction of α-phase (matrix).

In the considered phase diagram, Fig. 2, the third component has constant concentration in the two-phase zone, and the consequently cross interdiffusion coefficients ware omitted by authors, $\tilde{D}_{32} = \tilde{D}_{23} = 0$. Consequently Eq. (3) takes form:

$$\frac{\partial}{\partial t}\left(\varphi N_3^\alpha + (1-\varphi) N_3^\beta\right) = 0. \quad (4)$$

The constant concentration of third component causes that the volume fraction of α-phase, $\varphi$, has to be constant as well. Above simplification together with $\tilde{D}_{23} = 0$ allow reducing equation (2) to the final form:

$$\frac{\partial}{\partial t} N_2 = \frac{\partial}{\partial x}\left(\tilde{D}_{22}\frac{\partial N_2^\alpha}{\partial x}\right), \quad (5)$$

which is identical with Fick's second law for diffusion in a binary single-phase system. Thus, assuming that the tie-lines radiate regularly from the single-phase β-region, the diffusion path resembles that shown in Fig. 2.

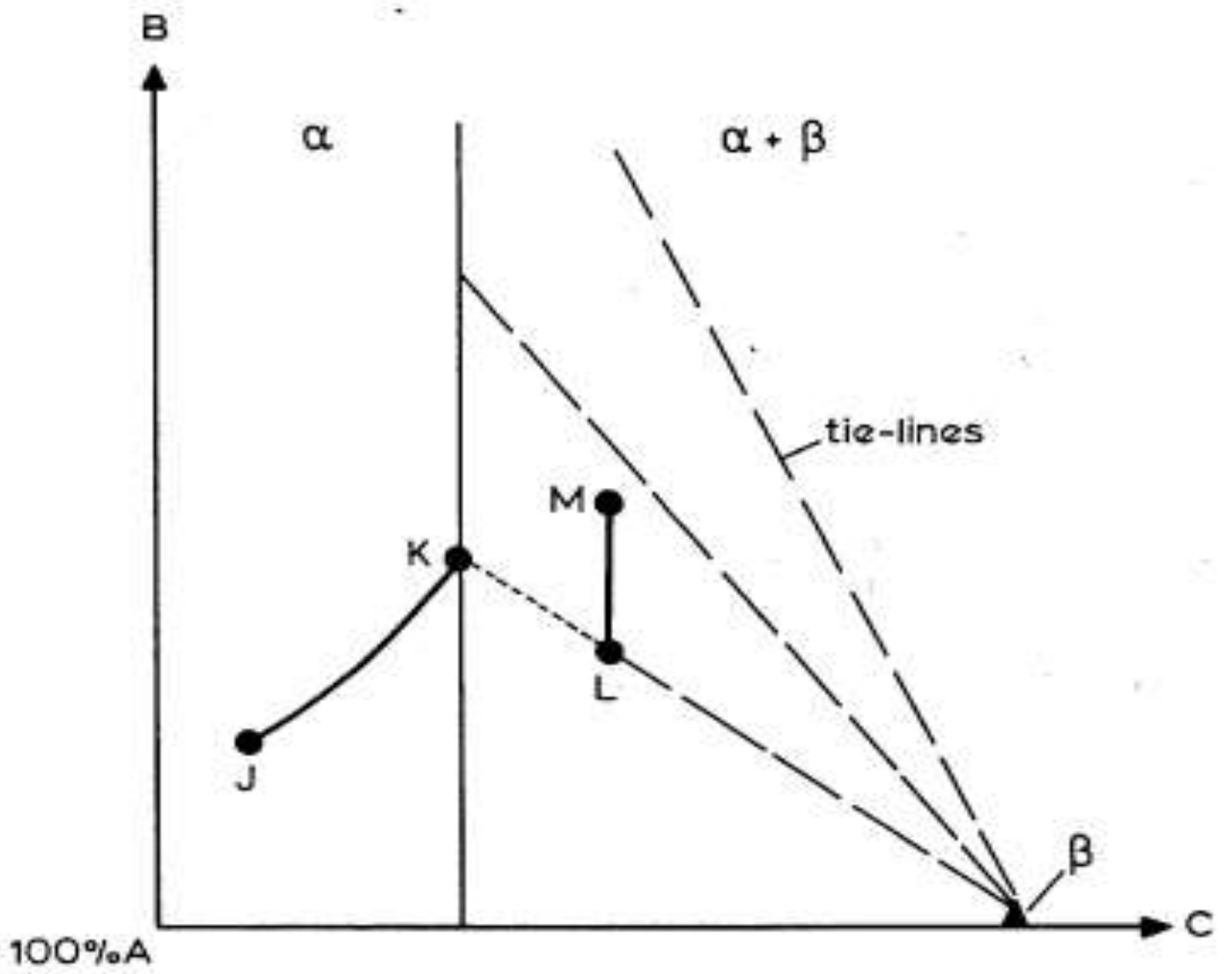

**Fig. 2. Diffusion path (thick line) and schema of the A-component rich part of phase diagram analyzed by Roper et al. [27].**

### 1.2.2 Model by Gusak and Lyashenko (1991)

The advanced model describing diffusion in two-phase zone was proposed by Gusak and Lyshayenko in 1991 [30]. The authors base on the hypothesis of local quasi-equilibrium in the diffusion process. Quasi-equilibrium implies that chemical potential of the species in both phases are locally (in a small volume) equal:

$$\mu_i^\alpha(c_1^\alpha, c_2^\alpha) = \mu_i^\beta(c_1^\beta, c_2^\beta) \quad i = 1,2,3. \tag{6}$$

where: $c_1^\alpha, c_2^\alpha$ and $c_1^\beta, c_2^\beta$ are the concentrations of the components in the contacting α and β phases.

The above three conditions imply that there is only one independent parameter, and all parameters can be described as a function of a single parameter, e.g. “k” (tie-lines parameter). In the case presented in Fig. 3, when phase boundaries are parallel to the 1-2 components line, and conodes radiate from the third specie corner, the following equations can be used to describe the conodes [30]:

$$c_1^\alpha / e^\alpha = c_1^\beta / \left(1 - e^\beta\right), \quad c_2^\alpha / e^\alpha = c_2^\beta / \left(1 - e^\beta\right), \tag{7}$$

where: $e^\alpha = c_1^\alpha + c_2^\alpha$ and $1 - e^\beta = c_1^\beta + c_2^\beta \ / e^\alpha$.

The parameters $e^\alpha$ and $e^\beta$ characterize the phase boundaries. If one defines $c_1^\beta$ as the conode parameter, then the following relations are true:

$$c_1^{\alpha}(k)=\frac{e^{\alpha}}{1-e^{\beta}}k,\;\; c_1^{\beta}(k)=k,$$
$$c_2^{\alpha}(k)=\left(1-\frac{k}{1-e^{\beta}}\right)e^{\alpha},\; c_2^{\beta}(k)=1-e^{\beta}-k. \tag{8}$$

Diffusion fluxes are expressed in Onsager linear form:

$$J_i=-\sum_{j=1}^{2}\tilde{L}_{ij}\frac{\partial\left(\mu_j-\mu_3\right)}{\partial x}\quad i=1,2\,, \tag{9}$$

where: $\tilde{L}_{ij}$ are a kinetic coefficients.

Taking into account that chemical potential of the components depends only on the k parameter, Eq. (9) takes a form:

$$J_i=-\tilde{M}_i\frac{\partial k}{\partial x}\quad i=1,2, \tag{10}$$

where mutual diffusion coefficients in two-phase zone, $\tilde{M}_i$, are given by:

$$\tilde{M}_1=\tilde{L}_{11}\frac{\partial\left(\mu_1-\mu_3\right)}{\partial k}+\tilde{L}_{12}\frac{\partial\left(\mu_2-\mu_3\right)}{\partial k},$$
$$\tilde{M}_2=\tilde{L}_{21}\frac{\partial\left(\mu_1-\mu_3\right)}{\partial k}+\tilde{L}_{22}\frac{\partial\left(\mu_2-\mu_3\right)}{\partial k}. \tag{11}$$

Consequently mass conservation equation takes the form:

$$\frac{\partial c_i}{\partial t}=-\frac{\partial}{\partial x}\left(\tilde{M}_i\frac{\partial k}{\partial x}\right)\quad i=1,2, \tag{12}$$

where the mutual diffusion coefficients in two-phase zone, $\tilde{M}_i$, can be computed from the mutual diffusivities in the both phases, ($\tilde{M}_i^{\alpha}$ and $\tilde{M}_i^{\beta}$) :

$$\tilde{M}_i=\tilde{M}_i^{\alpha}\varphi^{\alpha}+\tilde{M}_i^{\beta}\varphi^{\beta}\,, \tag{13}$$

where:

$$\tilde{M}_i^{j}=D_{i1}^{j}\frac{\partial c_1^{j}}{\partial k}+D_{i2}^{j}\frac{\partial c_2^{j}}{\partial k}\;\text{ For } i=1,2 \text{ and } j=\alpha,\beta\,. \tag{14}$$

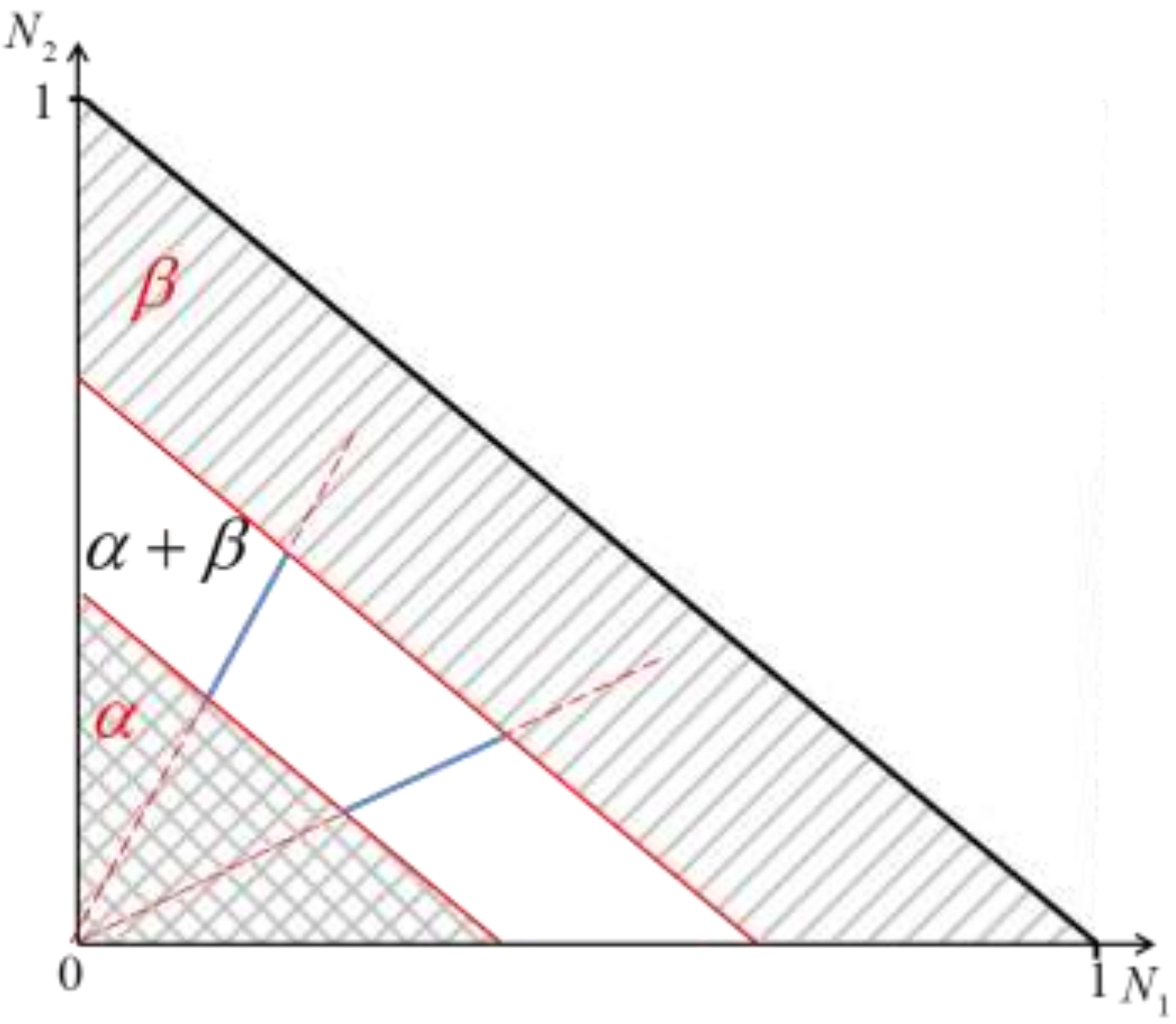

**Fig. 3. Schema of phase diagram analyzed by Gusak and Lyshayenko, after [30].**

### 1.2.3 Model by Hopfe and Morral (1994)

Gusak concept was followed by Hopfe and Morral. Their model was initially formulated for α+β|α+β diffusion couple with type 0 interphase boundary [2]. The model was characterized by following assumption:

a) diffusion between α+β and α+β terminal alloys is considered;
b) there is local quasi-equilibrium (LE) in the two-phase zone[1];
c) the β-phase is present as the isolated precipitates;
d) diffusion occurs only through the dominant continuous α-phase (matrix);
e) the overall concentration in dominant α-phase is constant: $\sum c_i^{\alpha} = \sum c_i = c$ .

The last assumption allows reducing the numbers of components from three to two and the mass conservation law in the matrix form becomes:

$$\frac{\partial (N)}{\partial t} = -\frac{\partial}{\partial x}\left(\tilde{J}\right), \tag{15}$$

where $(N)$ is a 2x1 matrix of overall molar rations (overall concentrations) and $(\tilde{J})$ is 2x1 matrix of interdiffusion fluxes.

By assuming that the diffusion occurs only in α-phase, the overall flux, $\tilde{J}$, is given by:

[1] Described by Morral as local equilibrium, (LE).

$$\left(\tilde{J}\right)=\left(\tilde{J}^{\alpha}\right)=-\left[\tilde{D}^{\alpha}\right]\frac{\partial\left(N^{\alpha}\right)}{\partial x}, \tag{16}$$

where $\left[\tilde{D}^{\alpha}\right]$ is the 2x2 diffusivity matrix and $\left(N^{\alpha}\right)$ is the 2x1 column matrix of molar ratios in α-phase.

In order to eliminate the α-phase composition, the singular transformation matrix, $\left[N^{TM}\right]$, given by following equation is introduced:

$$\left[N^{TM}\right]=\begin{bmatrix}\dfrac{\partial N_1^{\alpha}}{\partial N_1} & \dfrac{\partial N_1^{\alpha}}{\partial N_2}\\ \dfrac{\partial N_2^{\alpha}}{\partial N_1} & \dfrac{\partial N_2^{\alpha}}{\partial N_2}\end{bmatrix}. \tag{17}$$

The equations (15), (16) and (17) can be rewritten in the form:

$$\begin{aligned}\frac{\partial N_1}{\partial t}&=\left(\tilde{D}_{11}^{\alpha}\frac{\partial N_1^{\alpha}}{\partial N_1}+\tilde{D}_{12}^{\alpha}\frac{\partial N_2^{\alpha}}{\partial N_1}\right)\frac{\partial^2 N_1}{\partial x^2}+\left(\tilde{D}_{11}^{\alpha}\frac{\partial N_1^{\alpha}}{\partial N_2}+\tilde{D}_{12}^{\alpha}\frac{\partial N_2^{\alpha}}{\partial N_2}\right)\frac{\partial^2 N_2}{\partial x^2},\\ \frac{\partial N_2}{\partial t}&=\left(\tilde{D}_{21}^{\alpha}\frac{\partial N_1^{\alpha}}{\partial N_1}+\tilde{D}_{22}^{\alpha}\frac{\partial N_2^{\alpha}}{\partial N_1}\right)\frac{\partial^2 N_1}{\partial x^2}+\left(\tilde{D}_{21}^{\alpha}\frac{\partial N_1^{\alpha}}{\partial N_2}+\tilde{D}_{22}^{\alpha}\frac{\partial N_2^{\alpha}}{\partial N_2}\right)\frac{\partial^2 N_2}{\partial x^2}.\end{aligned} \tag{18}$$

The authors assumed that diffusivities are invariant and consequently equation (18) reduces to the trivial form.[2]

The schema of dependence between change in average molar rations and change of molar rations in α-phase is shown in Fig. 4.

---

[2] Morral introduces matrix form: $\frac{\partial\left(N\right)}{\partial t}=\left[D^{eff}\right]\frac{\partial^2\left(N\right)}{\partial x^2}$, where $\left[D^{eff}\right]$ is an effective diffusivity matrix $\left[D^{eff}\right]=\left[\tilde{D}^{\alpha}\right]\left[N^{TM}\right]$.

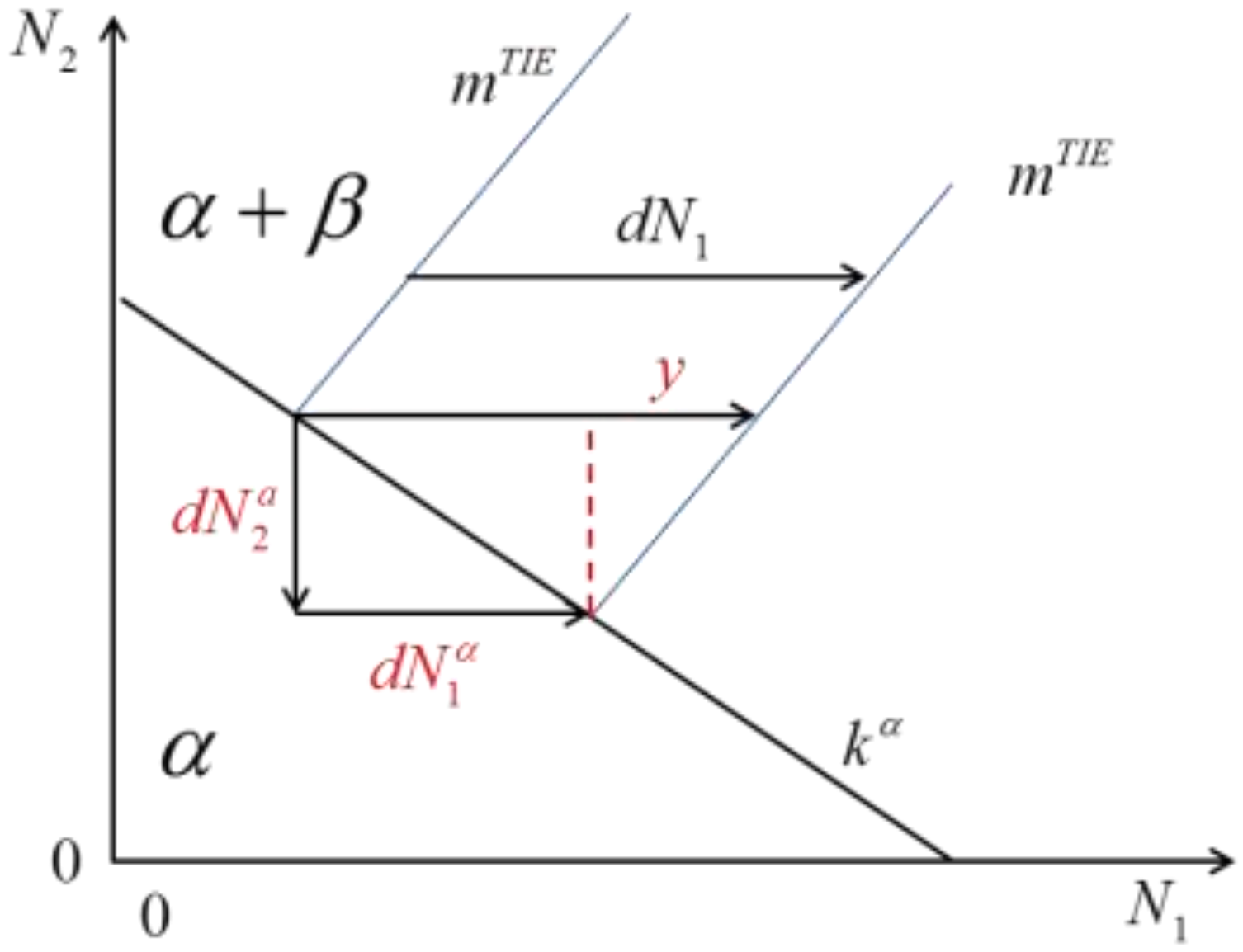

**Fig. 4. Shema of phase diagram showing a relation between a change in overall molar ratio and a change of molar ratio in α-phase. Blue lines are tie-lines.**

The Hopfe and Morral model was extended over the years. In 1999 [31], method of calculating transformation matrix for the system with nonlinear phase boundary and nonparallel tie-lines was presented. The modifications imply the concentration dependent effective diffusivity matrix and more complex form of the mass conservation law. The model was implemented into DICTRA software [32], combined with thermodynamic data and multicomponent diffusivities. The model can be applied independently of the numbers of the phases in the ternary system [33].

Hopfe and Morral model fulfils Kirkaldy theorems. It has been used by those authors to define types of interphase boundaries in multiphase interdiffusion [34-37].

### 1.2.4 Phase-field method

There is a growing interest in simulating diffusion in multiphase, multicomponent system using phase field method [37-59]. This method allows to predict evolution and morphology of precipitates during crystallization. As noted by Qin [60] the method allows obtaining only quantitative results because not all parameters necessary for qualitative calculations are known. Some parameters are arbitrary, adjustable or phenomenological and therefore there is no satisfactory agreement between theoretical predictions and experiment. Consequently very few papers combine diffusion with phase field method and compare the results with experiment [44, 49, 57]. Despite low agreement, the phase field method combined with diffusion is one of the key targets in non-reversible thermodynamic and metallurgy. In this section I will present the phase field method following Wu at. all [38].

Qualitative description of interdiffusion, as given in phase field method, is based on the Cahn-Hilliard diffusion equation [38]:

$$\frac{\partial N_i}{\partial t} = -D_i \nabla^2 \left( \mu_i - \mu_3 - 2\kappa_i \nabla^2 N_i \right) \quad i = 1, 2, 3, \tag{19}$$

where $N_i$ is a molar ratio of the *i*-th component, $\kappa_i$ is a constant coefficient related to the gradient of energy due to exchange enthalpies between components. It contributes to the interfacial energy of the interface of the phases. $\mu_i$ is chemical potentials of *i*-th component. The chemical potentials of the components are computed, e.g. from the regular solid solution model:

$$G_m = RT\left(N_1 \ln N_1 + N_2 \ln N_2 + N_3 \ln N_3\right) + I\left(N_1 N_3 + N_2 N_3\right), \tag{20}$$

where $I$ is the regular solution mixing parameter.

For the ternary system the Cahn-Hiliard equation, can be reduced to two components and becomes:

$$\begin{aligned} \frac{\partial N_1}{\partial t} &= \nabla\left[M_{11} \nabla\left(\mu_1 - \mu_3 - 2\kappa_{11}\nabla^2 N_1 - 2\kappa_{12}\nabla^2 N_2\right)\right], \\ &+ \nabla\left[M_{12} \nabla\left(\mu_2 - \mu_3 - 2\kappa_{12}\nabla^2 N_1 - 2\kappa_{22}\nabla^2 N_2\right)\right], \end{aligned} \tag{21}$$

$$\begin{aligned} \frac{\partial N_2}{\partial t} &= \nabla\left[M_{21} \nabla\left(\mu_1 - \mu_3 - 2\kappa_{11}\nabla^2 N_1 - 2\kappa_{12}\nabla^2 N_2\right)\right], \\ &+ \nabla\left[M_{22} \nabla\left(\mu_2 - \mu_3 - 2\kappa_{12}\nabla^2 N_1 - 2\kappa_{22}\nabla^2 N_2\right)\right]. \end{aligned} \tag{22}$$

In the simplified case, the mobility $M_{ij}$ can be calculated following Morral schema [2, 31]:

$$\left[M\right] = \left[D^m\right]\left[N^{TM}\right], \tag{23}$$

where $\left[N^{TM}\right]$ is a transformation matrix, and $\left[D^m\right]$ is the diffusivity matrix calculated from the mobilities of the components, $B_i$:

$$D_{11}^m = D_{12}^m = RT\left[B_1 + N_1\left(B_3 - B_1\right)\right], \tag{24}$$

$$D_{21}^{m} = D_{22}^{m} = RTc^{m}\left[B_3 - B_1\right],$$

where $c^m$ is the concentration of component 2 in the continuous phase (matrix phase).

## 1.3 Status of experimental investigation of diffusion in Ni-Cr-Al system

The diffusion experiments in Ni-Cr-Al were started in 1978 in NASA Lewis Research Center [61]. Over 150 Ni-rich Ni-Cr-Al diffusion couples were annealed at 1000, 1095, 1150 and 1205°C for 100, 300 and 500 hours and examined by optical microscopy. The studies were limited to metallography and microstructural observations. Concentration profiles were not measured (published).

Also in NASA Lewis Research Center the first diffusion//concentration profiles in Ni-Cr-Al system were measured [62]. Eleven diffusion couples between four γ+β, (C1, C2, C3, C4), and three γ+γ′ alloys, (S1, S2, S3), were annealed at 1200°C for 200 h (Figs. 5 -7). The concentration profiles were measured by the electron microprobe (EMP). The overall concentrations in γ+β zone were calculated based on the volume fractions of the phases and concentrations measured separately for each phase. This technique can be used only when the precipitates are bigger than the area of interaction of the electrons with the material. Therefore, the overall concentrations in the two-phase γ+γ′ zone in [62] were not measured, and the diffusion paths in those regions were approximated. The obtained diffusion paths are shown in the Figs. 5 - 7.

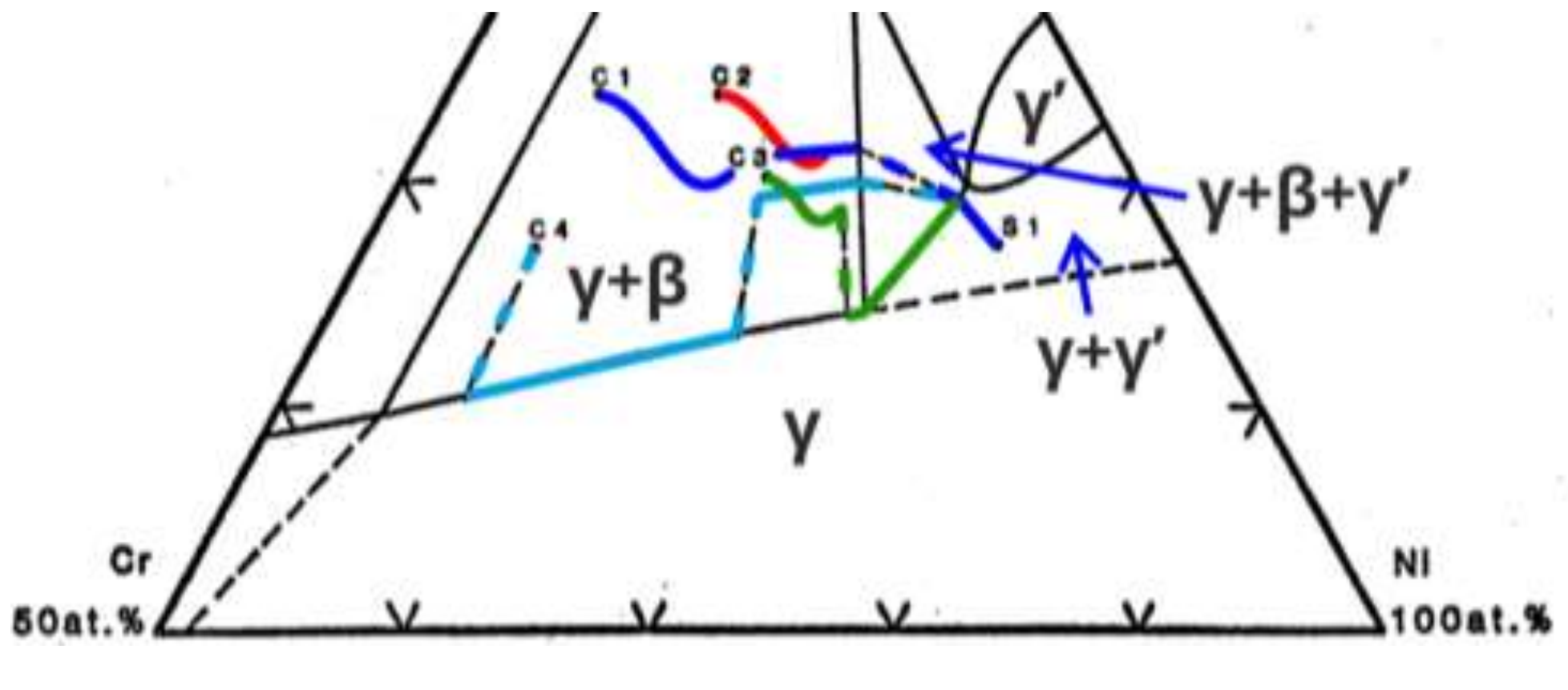

**Fig. 5. Diffusion paths for couples C1, C2, C3, C4|S1 [62].**

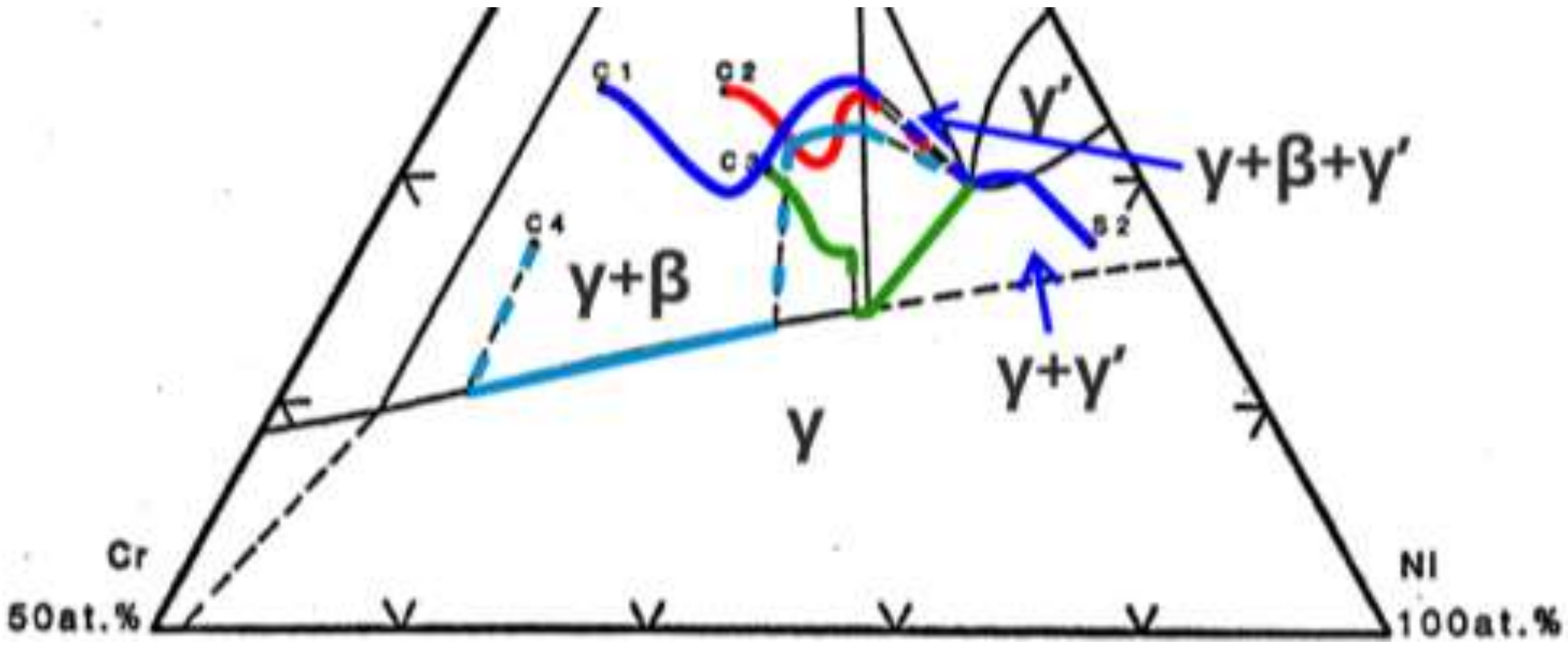

**Fig. 6. Diffusion paths for couples C1, C2, C3, C4|S2 [62].**

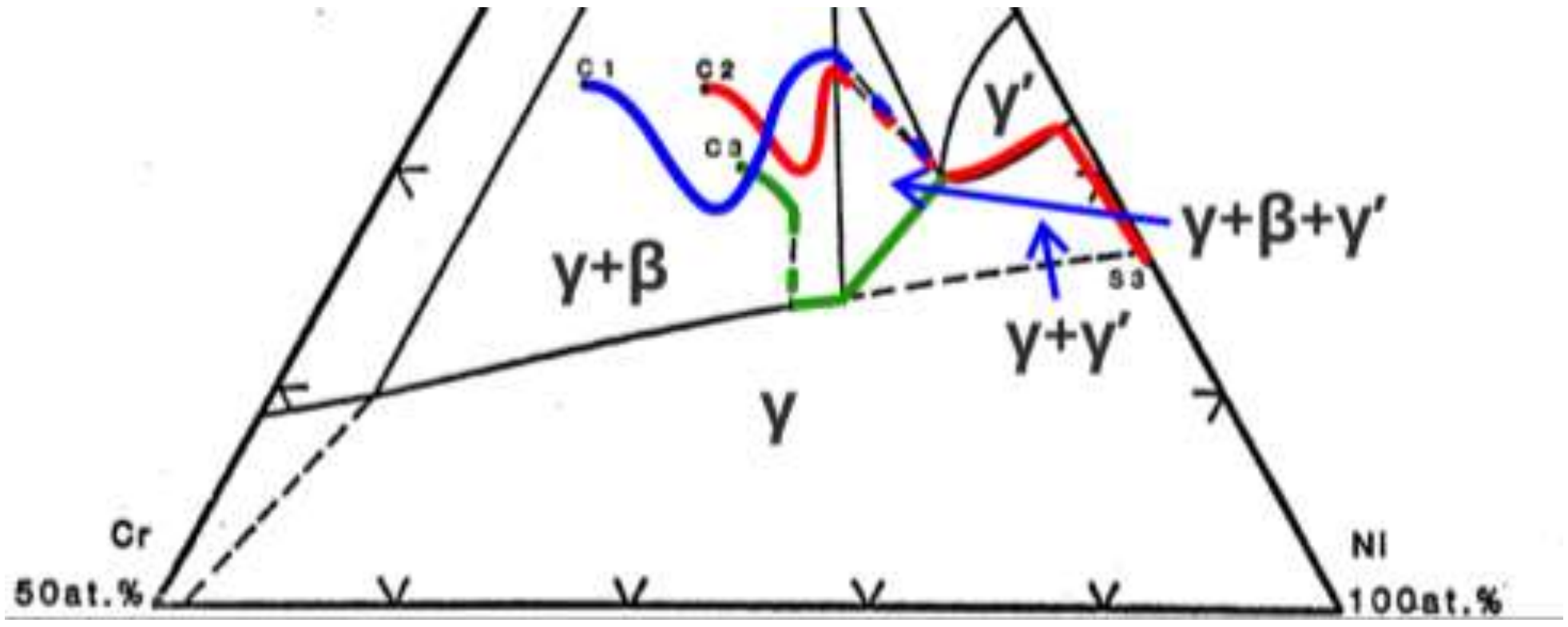

**Fig. 7. Diffusion paths for couples C1, C2, C3|S3 [62].**

This method is limited by the errors in the determination of volume fraction of the phases, (as illustrated in Fig. 8). The volume fraction of γ-phase, $\varphi^{\gamma}$, was measured as the surface fraction of in ~8x450 μm rectangles on the SEM image of γ+β|γ+β diffusion couple [41]. It is seen that the determination of volume fraction is distorted by the microstructural inhomogeneity. These distortions are reduced when the sizes of precipitates are small, relative to the size of analyzed area, however not smaller than SEM resolution.

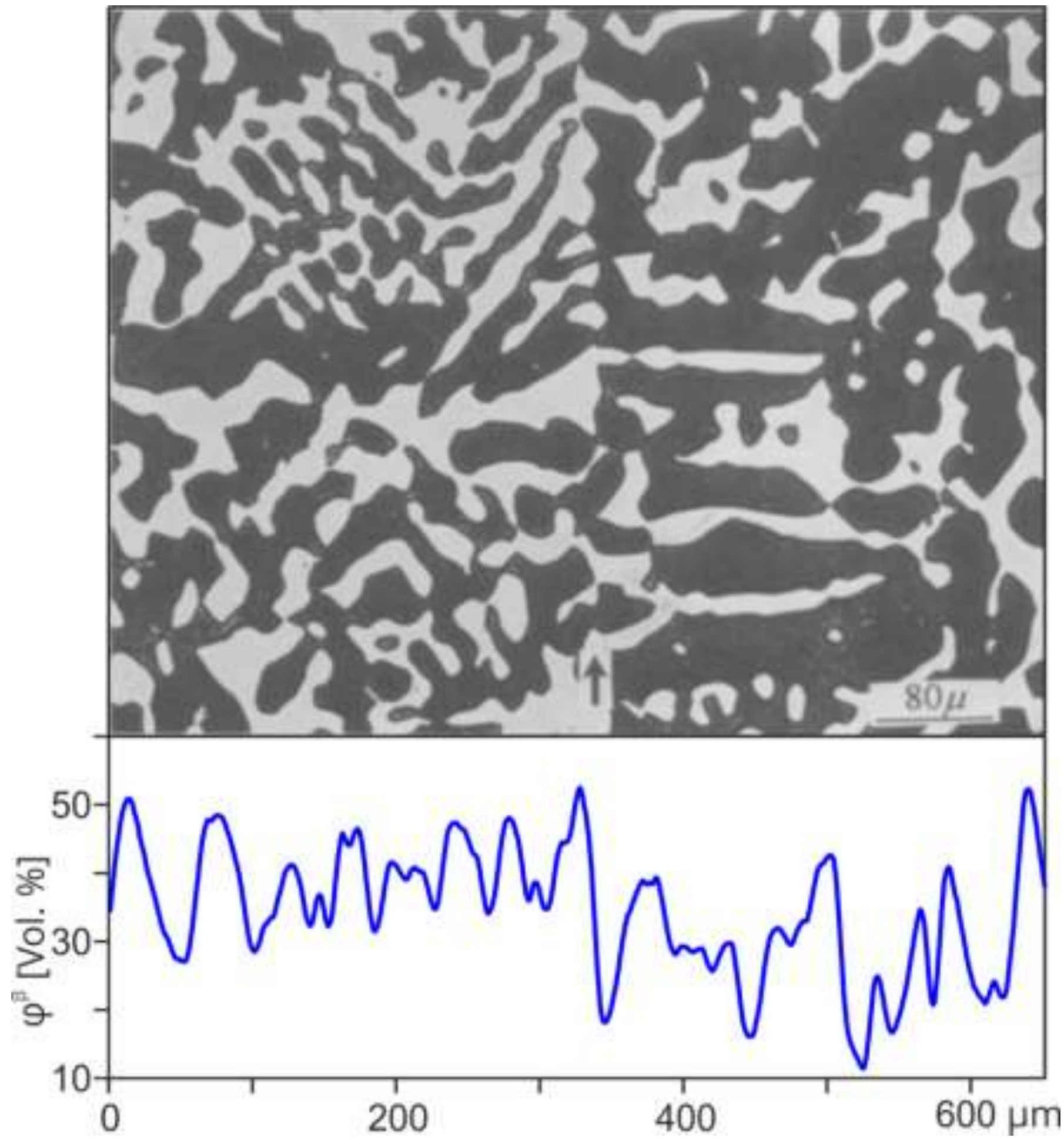

**Fig. 8. Interdiffusion in Ni-Cr-Al alloys [61]. Crosssection of the β+γ|β+γ diffusion couple annealed at 1150°C for 500 hours and measured volume fraction of β-phase, $\varphi^{\beta}$.**

Diffusion in Ni-Cr-Al system was experimentally investigated by Nesbitt and Heckel [63, 64]. Over 40 diffusion couple, including γ|γ+β multi-phase couples, were annealed at 1100 or 1200°C for 100 hours. The concentration profiles were measured by the electron microprobe. However, the measurements were limited only to the γ-phase. The concentration profiles and diffusion paths in two-phase zone were approximated from the two terminal compositions. Obtained data were used to calculate the concentration dependent interdiffusion coefficients in γ-phase. The exemplary results are shown in Figs. 9 and 10 [65].

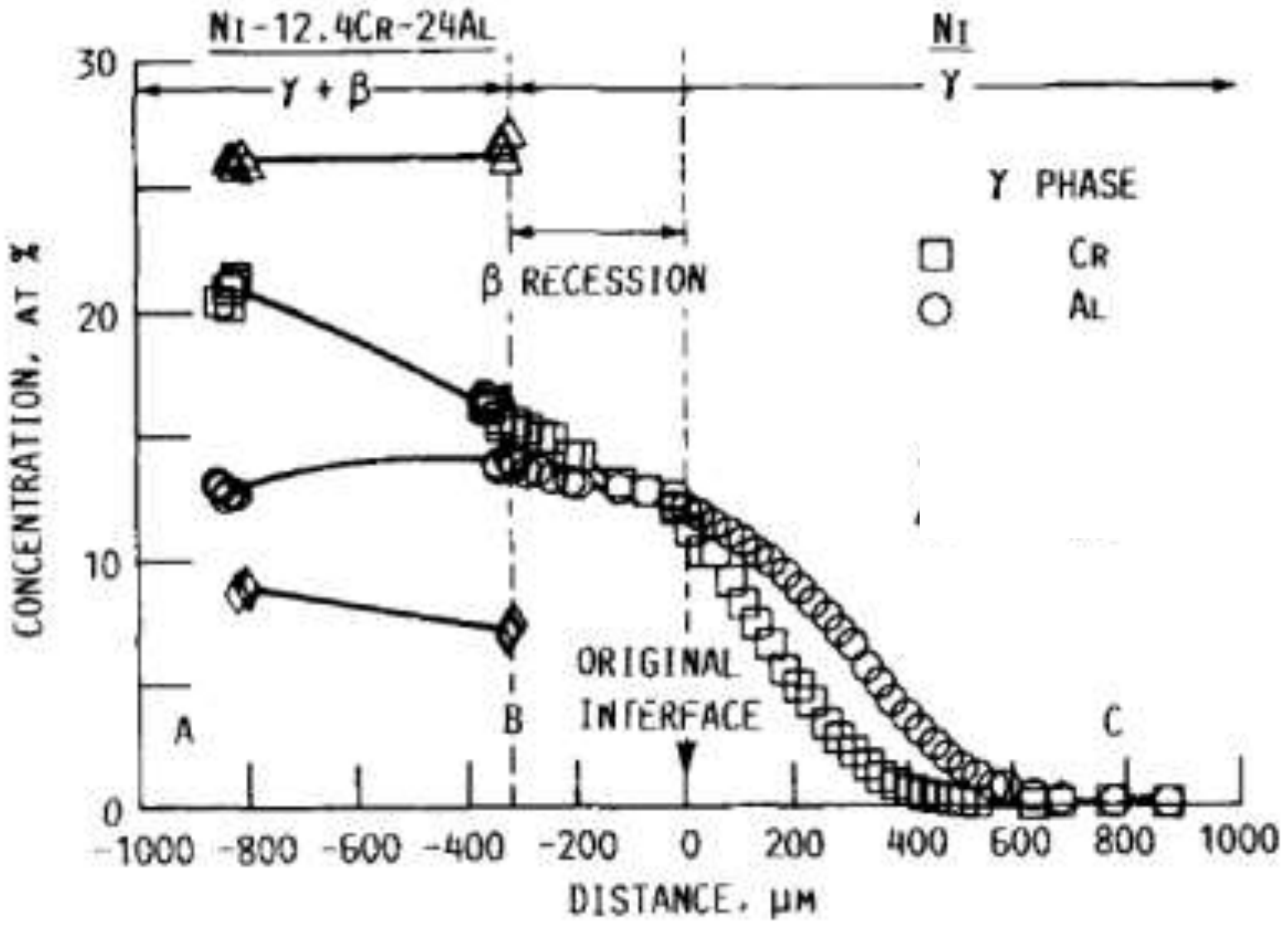

**Fig. 9. Interdiffusion in Ni-Cr-Al alloys [65]; the concentrations profiles measured in γ-phase (dots) and approximated in two-phase zone (lines).**

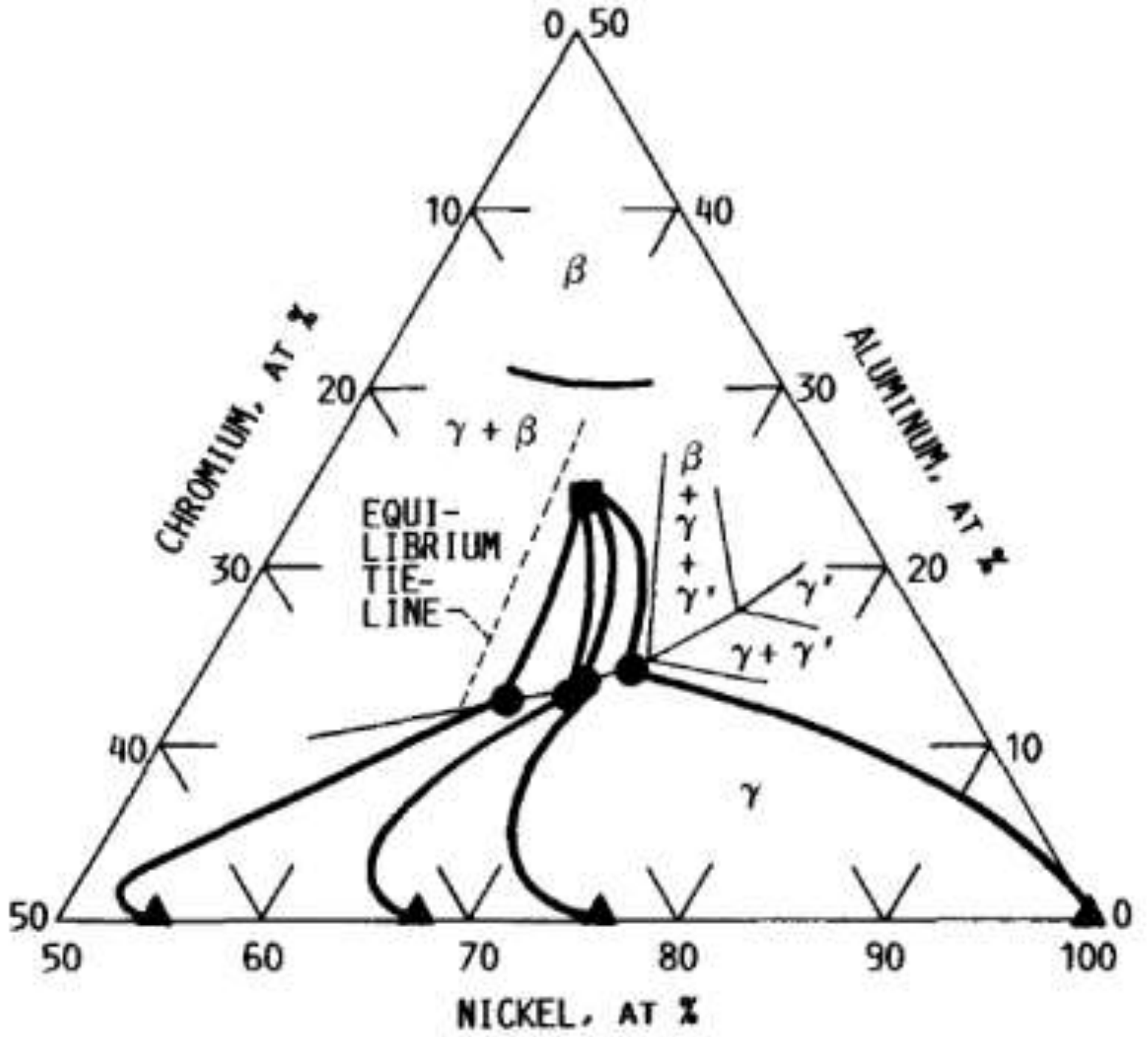

**Fig. 10. Interdiffusion paths in Ni-Cr-Al alloys [65]. Only diffusion paths in single-phase γ region were drawn in accordance with the measured concentration profiles.**

Diffusion between β-NiAl and γ-NiCr alloys was studied by Merchant et al. [66, 67]. Several diffusion couples consisting of a common terminal β-NiAl end-member and a series of binary γ-alloys containing 10, 20, 30, 35, 38, 40 Cr at. % and ternary, 40Cr6Al, 40Cr10Al at. %, balanced with Ni, as well as pure Ni subjected to diffusion at 1150°C for 49 hours were investigated. Chosen examples of the diffusion paths are shown in Fig. 11. The difference between diffusion paths starting from the following terminal compositions: 38Cr, 40Cr, 40Cr6Al, and 40Cr10Al (balanced by Nickel) are too small, to present on this picture.

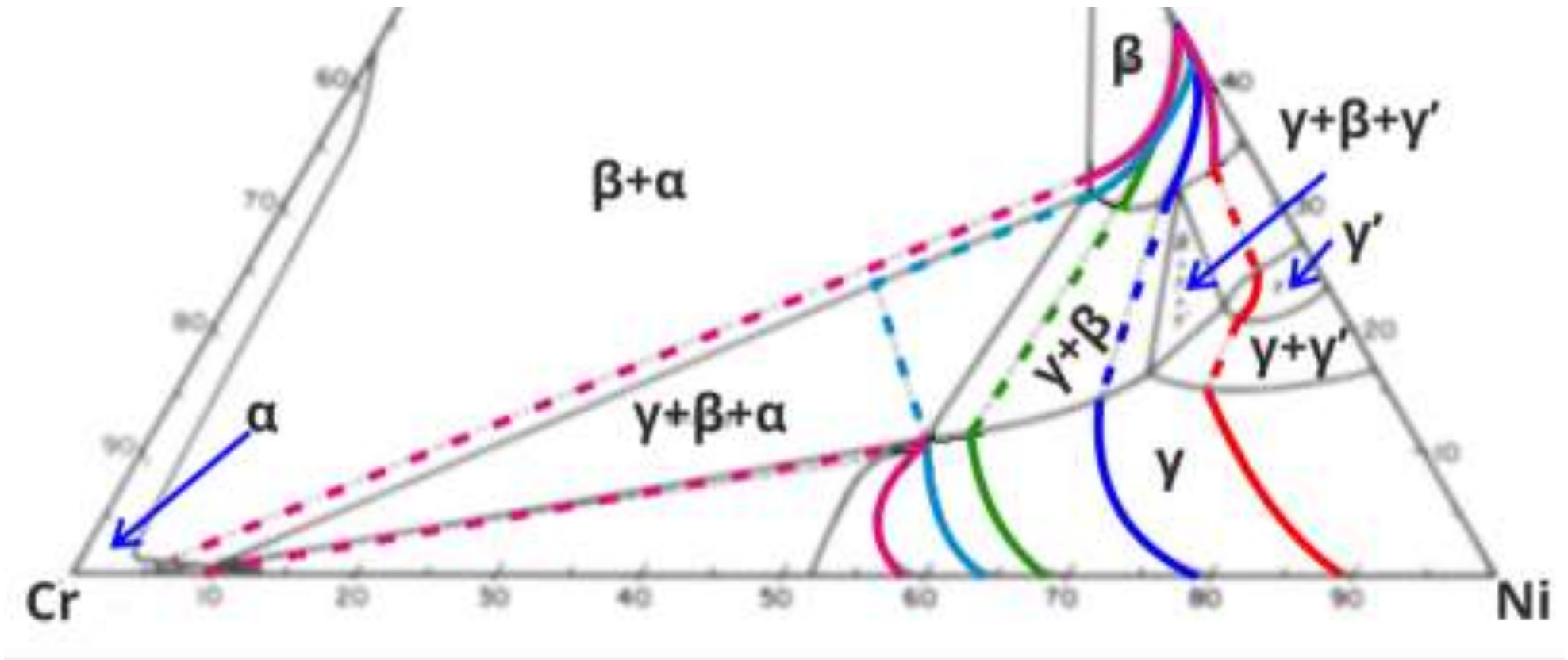

**Fig. 11. Interdiffusion in Ni-Cr-Al alloys [66]. Diffusion paths for β|γ terminal composition.**

Diffusion in Ni-Cr-Al system was examined by Xin [68]. He measured the overall concentration in multiphase zone as average values for several square areas within the diffusion zone. This technique improves the accuracy of the measured concentration profiles in two-phase zone. Diffusion zone in the two-phase region was divided into the 15 to 20 square areas. Such method increases precision of determination of overall concentration at the cost of lower spatial resolution. The diffusion paths obtained by Xin are shown in Fig. 12.

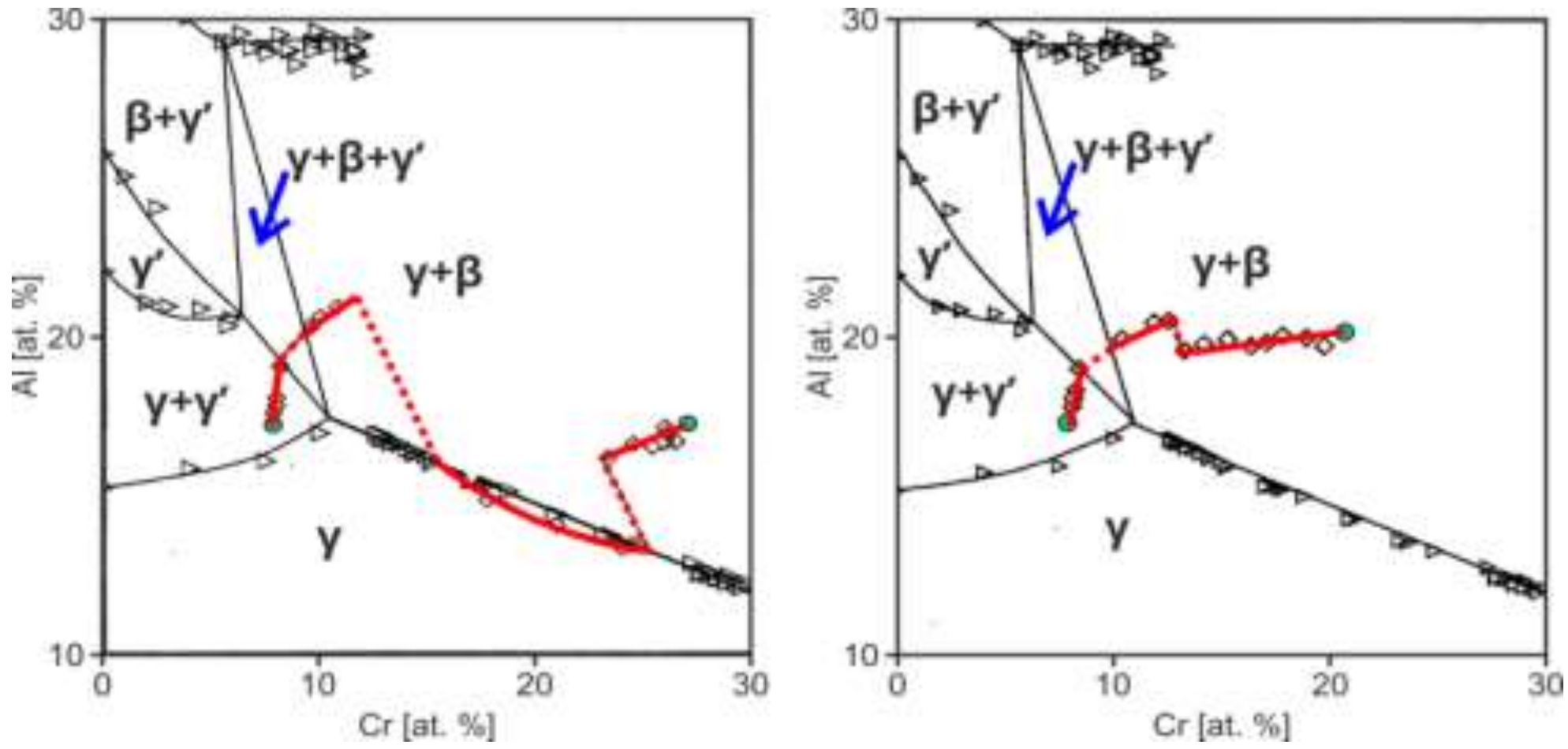

**Fig. 12. Diffusion paths in Ni-Cr-Al alloys [66].**

## 1.4 Summary

In this section the progress in the models of diffusion in ternary multi-phase systems was shown, and the theorems describing behavior of the diffusion path were presented. The theorems describe possible ways in which the diffusion path can cross the phase boundary lines.

Presented short review indicates unresolved problems:

- Why do some diffusion paths go along phase boundaries?
- Why some experimental results (Fig. 6) seem to contradict[3] the diffusion theorems?

And targets:

- Adequate experimental methods of measuring the overall concentrations in the multiphase zones (advanced multipoint integral methods, e.g. wide-line EDS analysis (WL-EDS)).
- Generalization of the models of quantitative description of interdiffusion, namely:
    - considering different partial molar volumes of components;
    - verifying of the results by quantitative computing of the entropy production due to diffusive mass transport;
    - multidimensional interdiffusion formalism at macro-scale with reaction diffusion at micro-scale.

[3] No jump of concentration is observed on a type 1a interphase boundary in Fig. 6.

# Chapter 2. Objectives and motivation

The interdiffusion in metals, especially in multicomponent systems, is still a source of inspirations and challenges for researchers. This issue is also very important from technical point of view. A big part of materials in constructions working at high temperatures are made of multiphase alloys that remain in contact with other materials, e.g.: protective layers, thermal barrier coatings or welds. The lifetime of materials is closely related to interdiffusion processes that occur at working conditions. The diffusion controlled processes are also widely applied to improve surface properties of the materials, e.g., carburizing, nitriding, aluminizing and other.

The Ni-base superalloys show excellent mechanical properties at high temperatures, and are widely used in elements working at extremal conditions in the gas turbines, jet and rocket engines, and often as parts of chemical reactors [6, 7]. The superalloys may have eight or more components and the studies of the diffusional interactions in such systems are generally very complex, time consuming and expensive. Here the alloy is simplified to the model ternary Ni-Cr-Al system.

The computer simulations significantly reduce the time and cost of research and speed up the development process. They present very useful tool which links theory and experiment. Models give the answers to many questions and allow verifying the present understanding of the complex phenomena.

The objectives of my thesis are both theoretical and experimental:

a) Development of a generalized model that allows simulating interdiffusion in the multi-component multi-phase system. Entirely new elements in model are:
   - consideration of different partial molar volumes of components;
   - consideration of changing of sample dimension due to reaction;
   - calculations of the entropy production due to diffusive mass transport in multiphase systems.
b) Modeling of interdiffusion in two-phase zone in macro-scale with reaction diffusion in the micro-scale.
c) Experimental studies of interdiffusion in the ternary multi-phase Ni-Cr-Al alloys. In particular:

- development of the effective method of samples preparation (diffusion multiples using HIP method);
- increasing of the accuracy of concentration measurements by using WL-EDS technique;
- measurements of the Kirkendall effect in multiphase alloys.

Finally by combining the generalized theoretical model and advanced experimental techniques I will explain the following phenomena:

- existence of diffusion paths that go along phase boundaries;
- absence of jump of concentration in the experimental results at the type 1a interphase boundary;
- existence of the diffusion paths with segment tangent to the phase boundary;
- condition for the local maxima in local entropy production rate at interphase boundaries.

# Chapter 3. Generalized models of the diffusion in two-phase ternary system

## 3.1 Multi-phase multi-component model in R1

The models presented in this section are extension of the bi-velocity method by Holly and Danielewski [69-71]. Bi-velocity method bases on the postulate that interdiffusion flux, $\tilde{J}_i = c_i \tilde{\upsilon}_i$, is a result of two velocities: diffusion velocity, $\upsilon_i$, which depends on the diffusion potential gradient and is independent of the choice of the reference frame, and drift velocity, $\upsilon$, that is common for all components and depends on the choice of reference frame. In very recent paper we showed, that the bi-velocity method is fully consistent with Onsager phenomenology [72].

The first multi-phase modification of the bi-velocity method [73] was used to model diffusive coating formation [73-79]. Here the method is generalized to include the effect of different molar volumes of the phases present in the ternary alloys.

### 3.1.1 Model assumptions

The method bases on the following assumption:

a) two-phase may occurs simultaneously (two-phase zones exist);
b) in the two-phase zone, the geometry of the precipitates and grain boundary mass transport are neglected;
c) diffusion between the phases (e.g., reactions between matrix and precipitates in the local scale) is faster than interdiffusion in the global scale (the whole diffusion zone) [80]. Consequently, the local quasi-equilibrium[4] (LE) between phases is provided.
d) the concentrations at the interfaces follow from the phase diagram (from LE);
e) formation of voids and stresses are neglected;
f) diffusion of the components occurs within all phases present;
g) intrinsic diffusion coefficients, $D_i^j$, are constant but differ for each specie and phase;

[4] Precisely the local ortho-equilibrium

h) partial molar volumes of the components, $\Omega_i^j$, are constant but differ for each specie and phase;

i) the Vegard's law, $\sum_i^R \Omega_i^j c_i^j = 1$ holds for every phase: j=α, β…;

j) there is no diffusion through the external boundaries of the diffusion couple, i.e., system is closed;

k) the overall molar volume depends on the molar fractions of all phases.

### 3.1.2 Laws and constitutive relations

The following laws and constitutive relations form the model (initial boundary-value problem):

**Laws:**

a) mas conservation law:

$$\frac{\partial c_i}{\partial t} = -\frac{\partial \tilde{J}_i}{\partial x} \quad \text{for } i = 1,2,3, \tag{25}$$

where $c_i$ and $\tilde{J}_i$ are overall concentration and overall flux of the *i*-th components;

b) volume continuity equation [71]:

$$\frac{\partial}{\partial x}\left(\sum_i \Omega_i^j J_i^j - \upsilon^j\right) = 0, \qquad j = \alpha, \beta, \tag{26}$$

where $J_i^j$ and $\upsilon^j$ are the diffusion flux and drift velocity in the *j*-th phase;

c) volume continuity equation in the two-phase zone[5]:

$$\frac{\partial\left(\upsilon c^{\alpha}\right)}{\partial x} = \frac{\Omega^{\alpha} - \Omega^{\beta}}{3\Omega}\frac{\partial\left(\varphi c^{\alpha}\right)}{\partial t}, \tag{27}$$

where $\Omega^{\alpha}$ and $\Omega^{\beta}$ are molar volumes of α and β phases, $\Omega$ is overall molar volume, $\varphi$ is volume fraction of the α-phase[6], $c^{\alpha}$ is overall concentration in the α-phase and $\upsilon$ is drift velocity due to reaction in the two-phase zone;

[5] Volume change as a results of reactions, in two-phase zone.
[6] It follows from LE assumption and is computed from thermodynamic phase diagram.

**Constitutive relations:**

d) Vegard's formula:

$$\Omega^j = 1/c^j = \sum_i \Omega_i^j N_i^j \,, \tag{28}$$

where $N_i^j$ is molar fraction of the *i*-th component in the *j*-th phase.

e) diffusion flux, Nernst-Planck equation [81]

$$J_i^j = c_i^j \upsilon_i^j = c_i^j B_i^j F_i^j \,, \quad j = \alpha, \beta, \; i = 1, 2, 3, \tag{29}$$

where $B_i^j$ denotes a mobility of the *i*-th component in *j*-th phase, and $F_i^j$ is a sum of total forces acting on atoms;

f) interdiffusion flux:

$$\tilde{J}_i^j = J_i^j + c_i^j \upsilon^j \,, \quad j = \alpha, \beta, \; i = 1, 2, 3; \tag{30}$$

g) overall interdiffusion flux:

$$\tilde{J}_i = \varphi \tilde{J}_i^\alpha + (1-\varphi) \tilde{J}_i^\beta + c_i \upsilon, \quad i = 1, 2, 3; \tag{31}$$

h) molar ratios:

$$\sum_i N_i^j = 1, \tag{32}$$

i) overall molar volume of two-phase system

$$\Omega = 1/c = \varphi^* \Omega^\alpha + \left(1-\varphi^*\right) \Omega^\beta \,, \tag{33}$$

where $\varphi^*$ is a molar fraction of α-phase;

j) lever rule as follows from Eqs. (28) and (33):

$$\begin{matrix} c_i = \varphi c_i^\alpha + (1-\varphi) c_i^\beta \\ N_i = \varphi^* N_i^\alpha + \left(1-\varphi^*\right) N_i^\beta \end{matrix} \qquad i = 1, 2, 3 \,, \tag{34}$$

k) the relationship between molar phase fraction, $\varphi^*$ and volume phase fraction, $\varphi$, as follows from Eqs. (28), (33) and (34):

$$\varphi^* = \varphi \frac{\Omega^\alpha}{\Omega} \,. \tag{35}$$

### 3.1.3 Local equilibrium in two-phase zone

The overall molar ratios of components in the two-phase zone, can be calculated by the lever rule, Eqs. (34). Calculation of concentrations base on the assumption of the local equilibrium that allows using thermodynamic data. The simplest way to present equilibrium parameters for the ternary system is the phase diagram. It is used here to obtain equilibrium parameters for the system in the two-phase region.

Below an example of the calculations of the concentrations under the local equilibrium conditions is shown for the system presented in Fig. 13. The phase boundary lines are described by the functions: $k^{\alpha}$ and $k^{\beta}$, the tie-lines are represented as the function $k\left(N_1, N_2\right) = k\left(\mathbf{N}\right)$, where $\mathbf{N}$ denotes a vector of the overall molar ratio, Fig. 13. All tie-lines pass through the common point, $\mathbf{p}^{tie} = \left(N_1^{tie}, N_2^{tie}\right)$.

In two-phase region, the molar ratios in coexisting phases are determined by the position of intersection, $\mathbf{N}^{\alpha|\beta}$ of the tie-line, $k$, with the phase boundaries, $k^{\alpha|\beta}$:

$$\mathbf{N}^{\alpha} = k^{\alpha} \cap k\left(N_1, N_2\right), \tag{36}$$

$$\mathbf{N}^{\beta} = k^{\alpha} \cap k\left(N_1, N_2\right). \tag{37}$$

Following Vegard's law, Eq. (28), the calculated concentrations are:

$$c_i^j = \frac{N_i^j}{\Omega^j} = \frac{N_i^j}{\sum_{i=1}^{3} \Omega_i^j N_i^j} \qquad j = \alpha, \beta,\ i = 1,2,3. \tag{38}$$

The volume fraction of α phase, follows from lever rule (34):

$$\varphi = \frac{c_i - c_i^{\beta}}{c_i^{\alpha} - c_i^{\beta}}. \tag{39}$$

In single-phase region, ($\varphi = \varphi^* = 1$ or $\varphi = \varphi^* = 0$), the concentrations of components equal to their overall concentration.

$$\begin{aligned} \alpha: &\quad \varphi = 1,\ N_i^{\alpha} = N_i \ \text{ and } \ c_i^{\alpha} = c_i; \\ \beta: &\quad \varphi = 0,\ N_i^{\beta} = N_i \ \text{ and } \ c_i^{\beta} = c_i. \end{aligned} \tag{40}$$

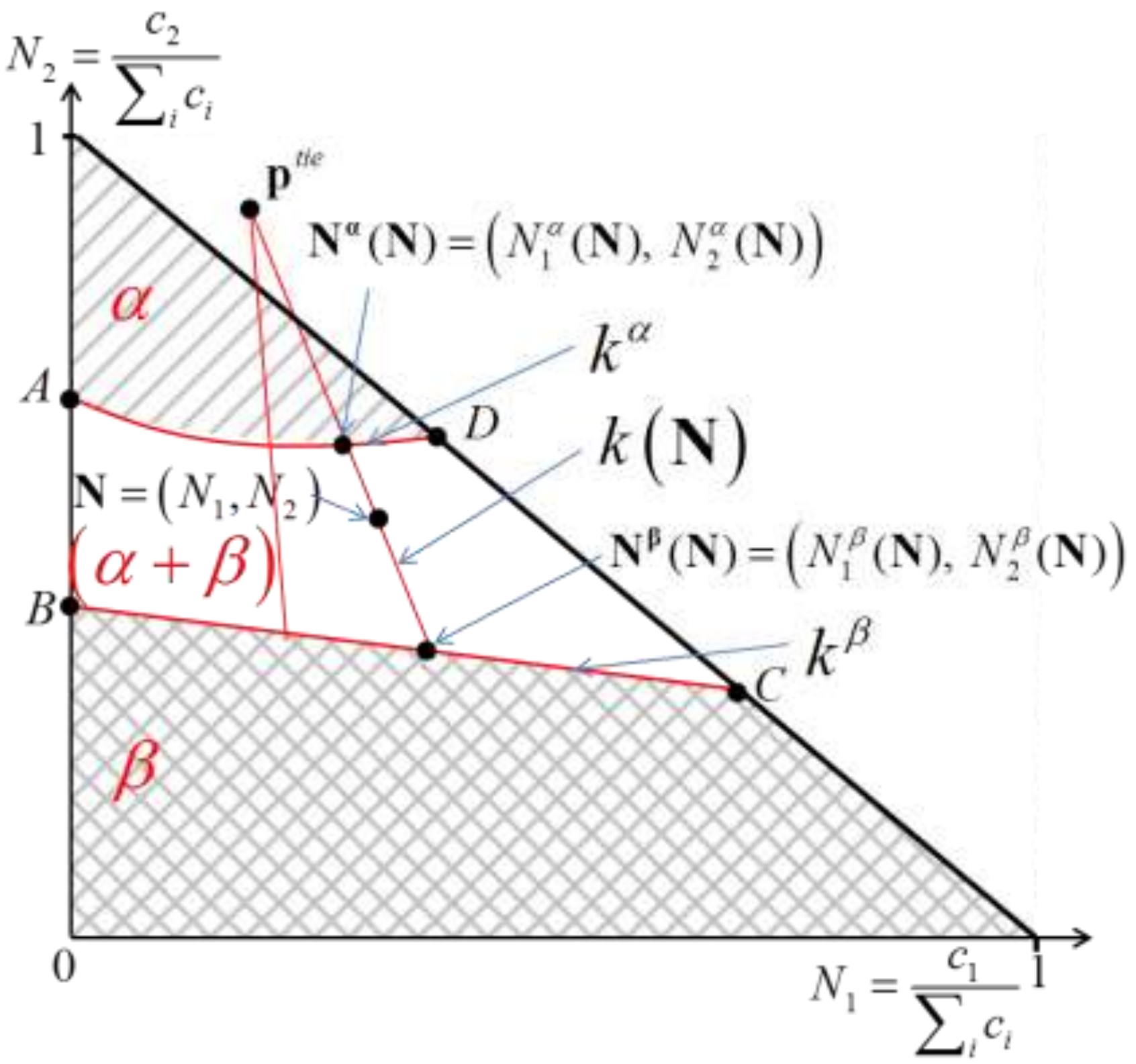

**Fig. 13. An example of phase diagram for the two-phase ternary system. α and β denote single-phase regions, α+β two-phase region, $k^{\alpha}$ and $k^{\beta}$ are the α and β phase boundaries, $k(\mathbf{N})$ the tie-line (conode) and $\mathbf{N} = (N_1, N_2)$ denotes the overal molar ratio vector.**

### 3.1.4 Evolution of concentration

The interdiffusion flux in each phase, $\tilde{J}_i^j$, consists two components:

a) diffusion flux, $J_i^j$, which is calculated in material reference frame for each components, Eq. (29);

   and

b) drift velocity, $\upsilon^j$, that is common for all components in the *j*-phase and depends on the choice of the external reference frame, Eq. (26).

The diffusion flux can be calculated from Nernst-Planck flux formula (29) [81]:

$$J_i^j = c_i^j \upsilon_i^j = c_i^j B_i^j F_i^j, \quad j = \alpha, \beta, \; i = 1, 2, 3, \tag{29}$$

where $B_i^j$ denotes a mobility of the *i*-th component in *j*-th phase, and $F_i^j$ is a sum of all forces acting on diffusing atoms. In general case, there can be many types of forces, other than the gradient of chemical potential, e.g., gravitational force, gradient of pressure, etc.:

$$F_i^j = -\nabla\mu_i^j + M_i G - \Omega_i^j \nabla p + \dots . \tag{41}$$

In the case, when chemical potential is the only driving force and the *j*-phase can be treated as an ideal solid solution, Eq. (29) reduces to the first Fick law:

$$J_i^j = -D_i^j \nabla c_i^j, \quad j = \alpha, \beta, \; i = 1, 2, 3. \tag{42}$$

In this approach, the unbalanced diffusion fluxes cause drift (stress free deformation in the material), without creation of voids [72]. Drift velocity in *j*-th phase can be calculated from volume continuity equation (26):

$$\upsilon^j = -\sum_i \Omega_i^j J_i^j, \qquad j = \alpha, \beta . \tag{43}$$

Thus, the interdiffusion flux in the phase is given by Eq. (30):

$$\tilde{J}_i^j = J_i^j + c_i^j \upsilon^j, \quad j = \alpha, \beta, \; i = 1, 2, 3 . \tag{30}$$

In the multiphase system such that partial molar volumes differ, the reaction between phases can lead to a change of the overall volumes of the system and cause additional drift $\upsilon$. Its velocity follows from Eq. (27):

$$\frac{\partial\left(\upsilon c^\alpha\right)}{\partial x} = \frac{\Omega^\alpha - \Omega^\beta}{3\Omega} \frac{\partial\left(\varphi c^\alpha\right)}{\partial t} \tag{27}$$

where the volumes $\Omega^\alpha$, $\Omega^\beta$ and $\Omega$ can be calculated from Eqs. (28) and (33).

The overall interdiffusion flux of the *i*-th component in the two-phase zone, $\tilde{J}_i$, can be calculated using lever rule (34), from a weighted average of interdiffusion fluxes in the phases, $\tilde{J}_i^\alpha$ and $\tilde{J}_i^\beta$. In the expression for $\tilde{J}_i$ the drift velocity resulting from different molar volumes of coexisting phases, Eq. (27), is also included:

$$\tilde{J}_i = \varphi \tilde{J}_i^\alpha + \left(1 - \varphi\right) \tilde{J}_i^\beta + c_i \upsilon, \quad i = 1, 2, 3 . \tag{31}$$

The evolution of the overall concentration of the component is given by the mass conservation law Eq. (25):

$$\frac{\partial c_i}{\partial t} = -\frac{\partial \tilde{J}_i}{\partial x} \;\; \text{for} \;\; i = 1, 2, 3 . \tag{25}$$

### 3.1.5 Entropy production and Kirkendall plane movement

**The Kirkendall plane position**, $x_k = x_k(t)$, is calculated by trajectory method [82], which predicts that velocity of Kirkendall plane equals to overall drift velocity, $\upsilon^{Drifft}$:

$$\frac{\partial x_k}{\partial t} = \upsilon^{Drifft}(x_k, t), \tag{44}$$

where $\upsilon^{Drifft} = \upsilon^{Drifft}(x_k, t)$ is approximated by:

$$\upsilon^{Drifft} = \varphi\upsilon^{\alpha} + (1-\varphi)\upsilon^{\beta} + \upsilon . \tag{45}$$

**The local entropy production** rate during diffusion process in ternary system, is within linear thermodynamics formalism represented in bilinear form [83]:

$$\dot{\sigma}(t,x) = \sum_i \sum_j L_{ij} X_j X_i, \quad i, j = 1,2, \tag{46}$$

where $L_{ij}$ form Onsager diffusivity matrix and $X_{i|j}$ is a force acting on the specie:

$$X_i = -\frac{\nabla \mu_i}{T} \overset{\text{ideal solid solution}}{=} R\frac{\nabla c_i}{c_i}. \tag{47}$$

In the recent work [72] it has been demonstrated, that the entropy production rate is independent on the reference frame choice:

$$\dot{\sigma}(t,x) = \sum_{i=i}^{R} J_i^j X_i^j = \sum_{i=i}^{R} \tilde{J}_i^j X_i^j, \quad j = \alpha, \beta, \tag{48}$$

in [72] the equivalence of Eqs. (46) and (48) was proved.

To calculate entropy production in two-phase zone the assumption of additivity of entropy production has been used:

$$\dot{\sigma}(t,x) = \varphi \sum_{i=1}^{3} \left(-\tilde{J}_i^{\alpha} X_i^{\alpha}\right) + (1-\varphi) \sum_{i=1}^{3} \left(-\tilde{J}_i^{\beta} X_i^{\beta}\right), \tag{49}$$

where $\tilde{J}_i^{\alpha}$ and $\tilde{J}_i^{\beta}$ are the interdiffusion fluxes in α and β phases, calculated by Eq. (31) and $X_i^j$ is the force acting for the *i*-th specie in *j*-th phase.

In this work the entropy production due to reaction is not considered.

### 3.1.6 Boundary and initial conditions

In the present model the 1-dimensional closed diffusion couple, of the length $2\Lambda$ is considered. There is no material flux through the couple boundary, thus the Neumann boundary condition can be assumed:

$$\begin{aligned} \nabla c_i(t,-\Lambda) &= 0, \\ \nabla c_i(t,\Lambda) &= 0. \end{aligned} \tag{50}$$

Initial concentrations of components are arbitrary, e.g., given by Heaviside function:

$$\begin{aligned} c_i(0,x) &= c_i^{left} \quad \text{for } x \in (-\Lambda, 0), \\ c_i(0,x) &= c_i^{right} \quad \text{for } x \in \langle 0, \Lambda ). \end{aligned} \quad i = 1,2,3 \tag{51}$$

The initial position and velocity of the Kirkendall plane $x_k = x_k(t)$ are arbitrary, eg., in the center of the sample: $x_k(0) = 0$, and $\partial x_k / \partial t = \upsilon^{Drifft}(x_k, 0) = 0$.

The presented model has been discretized with finite difference method, and the resulting equations were solved using a MATLAB program. The numerical schema is shown in Appendix 1.

## 3.2 Diffusion in ternary two-phase system – numerical experiments in R1

In the papers by Kirkaldy and Morral the main rules (theorems) concerning diffusion in multiphase systems are presented. The authors classified various types of the interphase boundaries and characterized their properties and related diffusion path shapes [3, 24] (section 1.1.2). In this section I'm presenting the numerical experiments which verify the Kirkaldy and Morral rules and allow formatting new rules that follow from generalized model and concern:

- behavior of diffusion path during the single-phase zone growth (theorem no A9 section 1.1.2 [3]);
- conditions for diffusion path on the phase boundary line (theorem no A9 section 1.1.2 [3]);

Moreover the appearance of singularities of the local entropy production at IB is verified, and the impact of the phase-boundary shape on the diffusion path is demonstrated. The type 3 interphase boundary (theorems A11) is not considered in this work.

### 3.2.1 Type 0 interphase boundary in α+β|α+β diffusion couple

During the interdiffusion between two alloys that differ only with the phase concentrations the interphase boundary type 0 is formed. It is the most characteristic example of the diffusion in two-phase systems which results in the diffusion path of zigzag shape. In Fig. 14 and 15 the example is shown. In this particular case, type 0 IB with one inward and one outward horns is evident. The outward horn relates to the formation of narrow zone of the pure β-phase $\varphi^{\alpha} = 0$. Such effect was before found experimentally [35, 84].

Fig. 16 shows the "time evolution" of the type 0 IB shown in Fig. 14. The results of numerical experiments for various processing times, 1, 3, and 10 hour, are compared. It is seen that the diffusion paths are similar. The only differences are duo to applied numerical method.

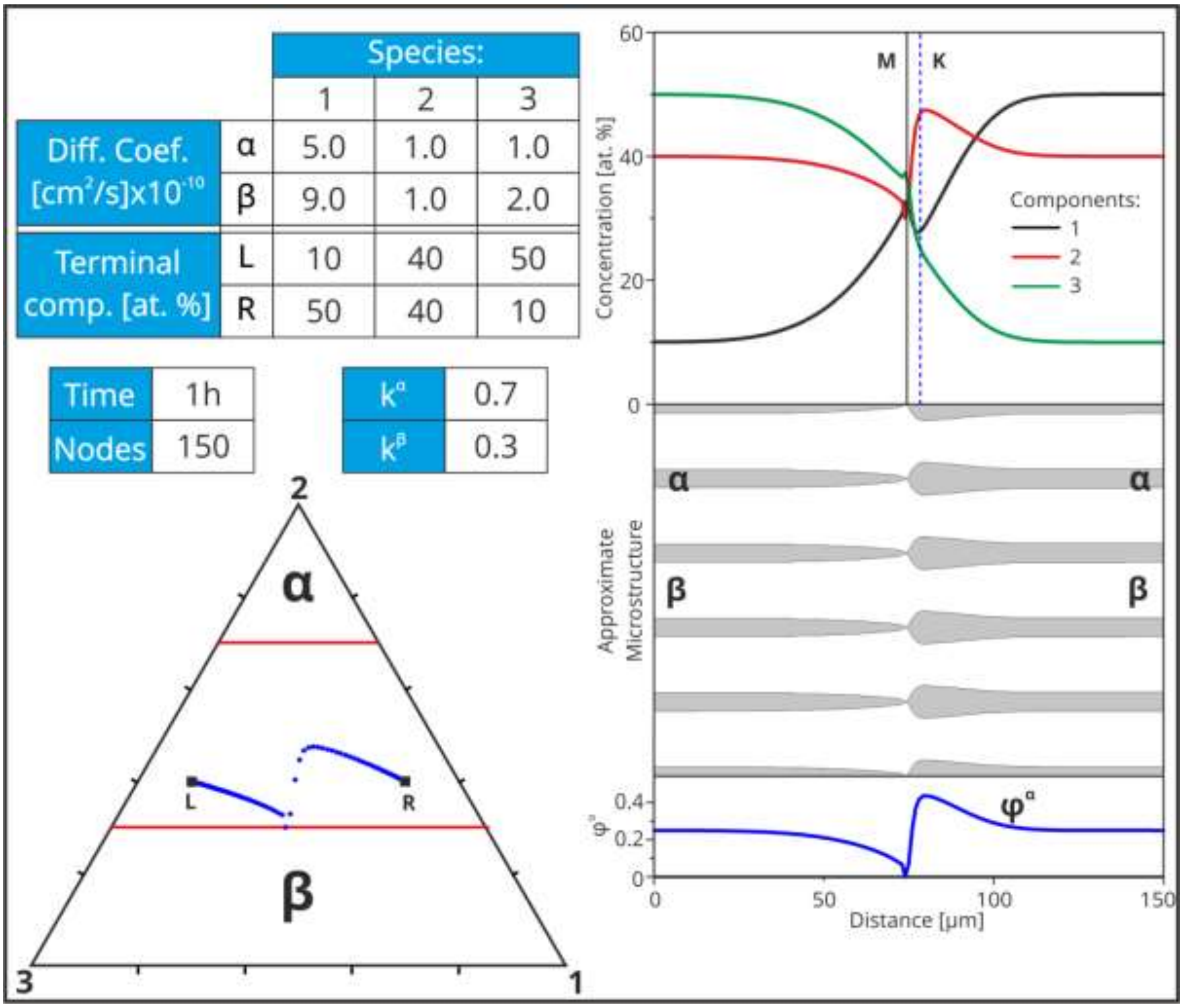

**Fig. 14. Diffusion in α+β|α+β couple with type 0 IB, the results of numerical experiment. Left, from the top: data (diffusion coefficients, terminal compositions, annealing time, number of grid points, α and β phase boundaries ($k^{\alpha}$ and $k^{\beta}$), phase diagram with the diffusion path. Right, from the top: concentration profiles with Matano (M) and Kirkendall (K) planes, draft of an arbitrary microstructure, and volume fraction of α-phase.**

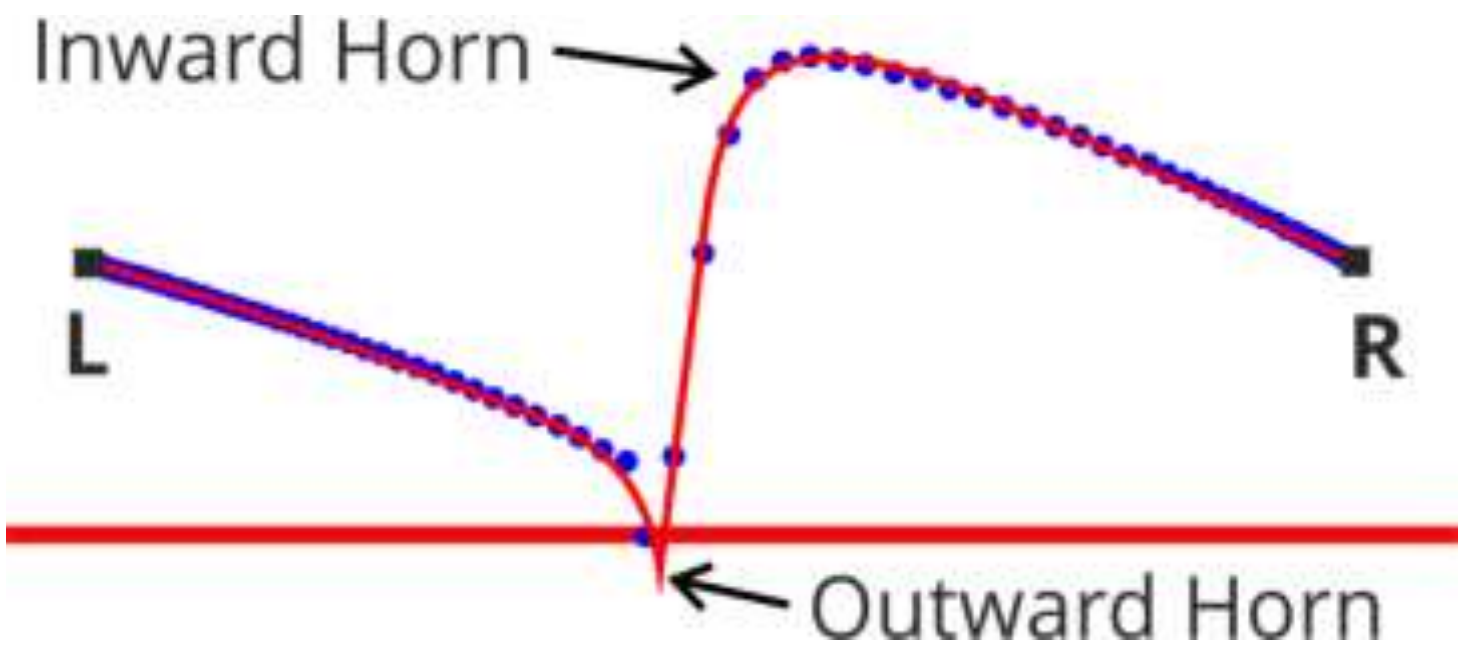

**Fig. 15. Diffusion in α+β|α+β couple with type 0 IB. Magnification of diffusion path shown in Fig. 14 displaying the outward and inward horns.**

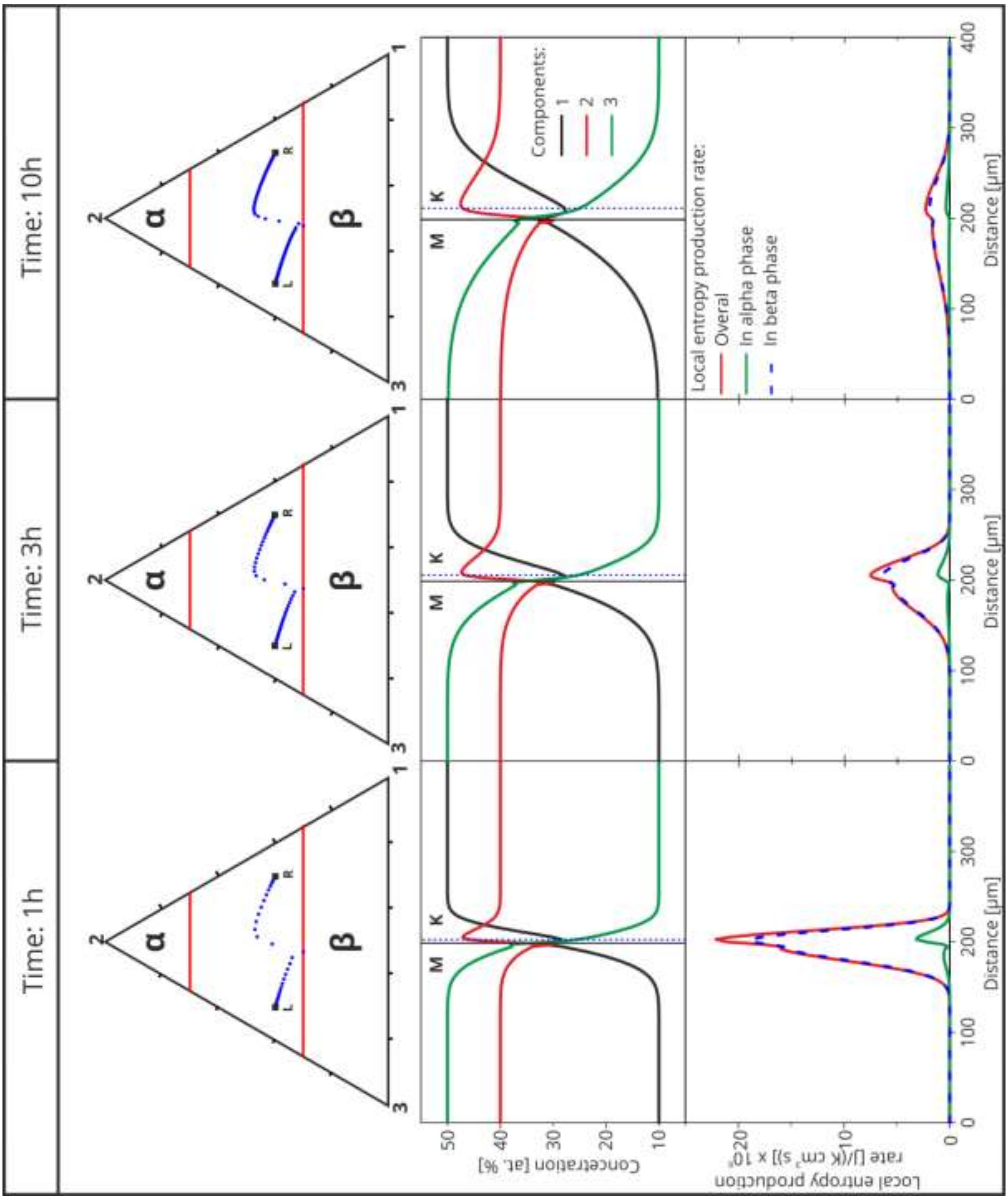

**Fig. 16. Time evolution of the diffusion path shown in Fig. 14. The results for 1, 3 and 10 hours. In the rows, from the top: Processing time, phase diagram with diffusion path, the concentration profiles with Matano and Kirkendall planes and the local entropy production rate.**

In Fig. 17 the paths after 10 and 1000h are compared. It is seen that after very long time, the terminal compositions change and the system reaches thermodynamic equilibrium. The two-phase alloy formed by long diffusional intermixing consists of alpha and beta phases of the compositions laying on the same conode, i.e., the diffusion path converge to multi dot path on the conode. The only driving force here is the concentration gradient in the coexisting phases. Consequently, full homogenization is stopped and overall concentrations and phase volume fractions vary with the position. In the real systems, the of stresses and/or the action of surface tension may result in the full homogenization of the microstructure.

The position and movement of the type 0 IB are difficult to determine experimentally. Numerical experiments shown in Fig. 17 that this interface is moving during the diffusion process.

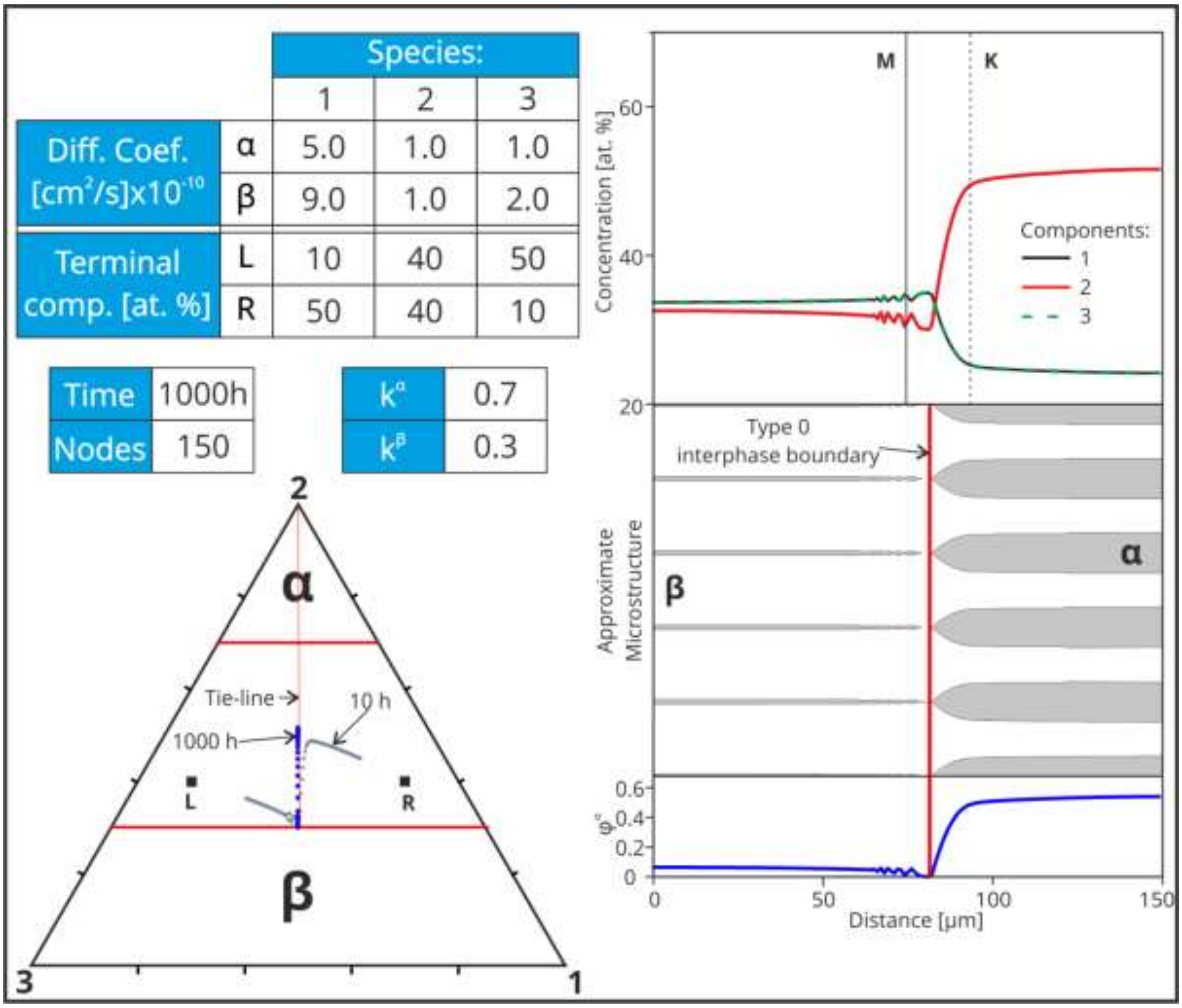

| | | 1 | 2 | 3 |
|---|---|---|---|---|
| Diff. Coef. [cm²/s]x10⁻¹⁰ | α | 5.0 | 1.0 | 1.0 |
| | β | 9.0 | 1.0 | 2.0 |
| Terminal comp. [at. %] | L | 10 | 40 | 50 |
| | R | 50 | 40 | 10 |

| | |
|---|---|
| Time | 1000h |
| Nodes | 150 |
| $k^{\alpha}$ | 0.7 |
| $k^{\beta}$ | 0.3 |

**Fig. 17. Long lasting (1000h) diffusion in α+β|α+β diffusion couple with type 0 IB. The representative parts of the presentation like in Fig. 14. On the phase diagram the diffusion path after 10 and 1000h are compared.**

The following conclusions based on the obtained results can be drawn based on the numerical experiments:

a) The zigzag diffusion path that develops horns has two different horns (inward and outward), in agreement with the result by Wu [36].
b) Outward horn (singularity) may cause a growth of the single-phase zone, Figs. 15 and 16.
c) As long as the terminal compositions are not affected by diffusion then:
    - the diffusion paths are time invariant;
    - diffusion process is parabolic.
d) The solutions of multi-multi equations are unstable with respect to disruptions of the diffusion paths at the conodes, Fig. 17, t=1000 h.
e) In the present model the thermodynamic equilibrium has been achieved after long time of annealing, but the gradients of phase volume fractions and the gradients of overall concentrations have not vanished. Full homogenization can be predicted only in the case of further generalization of the model by introducing stochastisation of diffusion, stresses and/or surface tension effects.
f) Contrary to the Morral theorem, the presented simulations shows that the type 0 IBs can move.

### 3.2.2 Type 0 interphase boundary in α+β|α+β diffusion couple; an effect of thermodynamics

In most papers dedicated to interdiffusion in ternary systems the phase boundaries on phase diagram are given as straight lines, e.g., represents constant concentration of one specie [2, 26, 30, 31]. Such assumption allows acceptable model simplification, like in sections 3.3 and 3.4 of this thesis. However the real phase boundary is seldom straight line. In this section I will show how a shape of the calculated diffusion paths in α+β|α+β diffusion couple depends on the shape of phase boundary, say β|α+β. To demonstrate this, I made calculations of diffusion path for three little different shapes of phase boundaries: straight, concave and convex, Fig. 18. The maximal deflection of convex and concave of the β-phase boundary is 1.7 at. %. The results are shown in Fig.

18. They demonstrate that even relatively small deviations of the phase boundary considerably affect diffusion path.

The results allow formulating following conclusions:

- even small change of the shape of phase boundary may result in the big change of diffusion path shape.

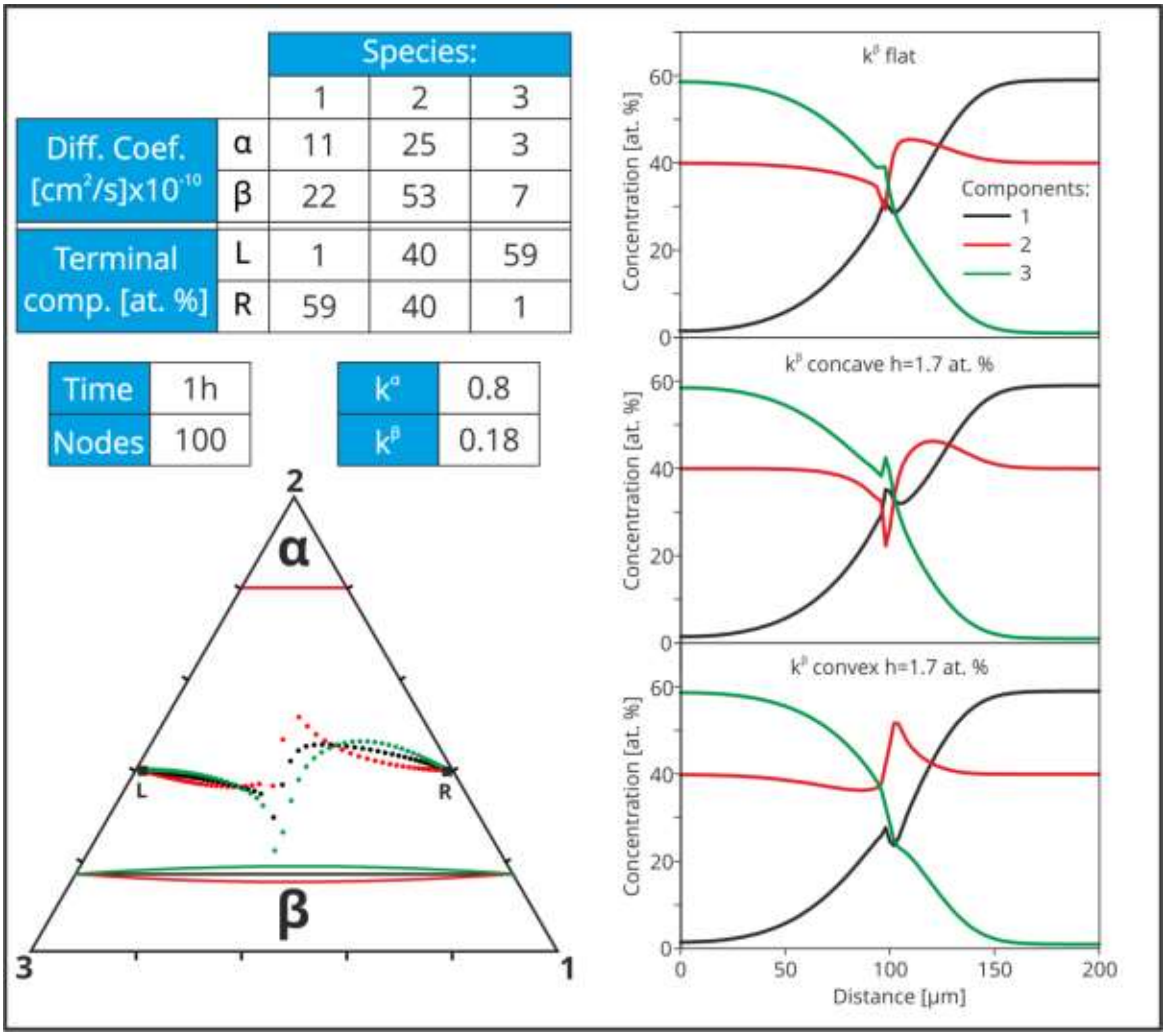

**Fig. 18. Influence of β-phase boundary shape to the process of diffusion in two-phase zone. Left, from the top: data (diffusion coefficients, terminal composition, process time, number of grid points, positions of α and β phase boundaries, phase diagram with three diffusion paths for three shapes of β-phase boundary: Flat - black, concave – green and convex – red. Right: the concentration profiles for corresponding diffusion paths.**

### 3.2.3 Type 1a interface boundaries in α+β|α+β diffusion couple and formation of single-phase zone

There are both experimental [35, 84] and numerical evidences that outward horn (singularity) on the diffusion path with type 0 IB may results in a growth of a thin single-phase zone. However when compositions of the terminal alloys are close to the phase boundary, say α+β|β, then the thick single phase zone grows between two type 1a IBs. It was reported in the past [18, 57, 62, 68] that the corresponding diffusion path goes, in such case along the phase boundary.

In the following section I will show that the diffusion path can exhibit different behaviour. The simulations were made for three only little different compositions, all lying very close to the α+β|β phase boundary ($k^{\beta}$). The results are shown in Fig. 19 and they confirm that the diffusion path enters the β region not tangentially to the phase boundary line. Simultaneously the IB moves opposite to the initial contact interface (Matano plane). The distributions of local entropy production rate are semi-similar.

The obtained results confirm that when at least one of the couple alloys has the composition close to the phase boundary. Then:

- the diffusion path enters the single β-phase region;
- single β-phase grow in the diffusion zone;
- two type 1a IB are formed;
- velocities and directions of velocities of both IB differ.

The concentration jump on type 1a IB can be very small, then can be not observed during the experiment (section 1.3 Fig. 6, [62]).

Note that the local entropy production rate in the left α-phase zone is very low but not equal to zero.

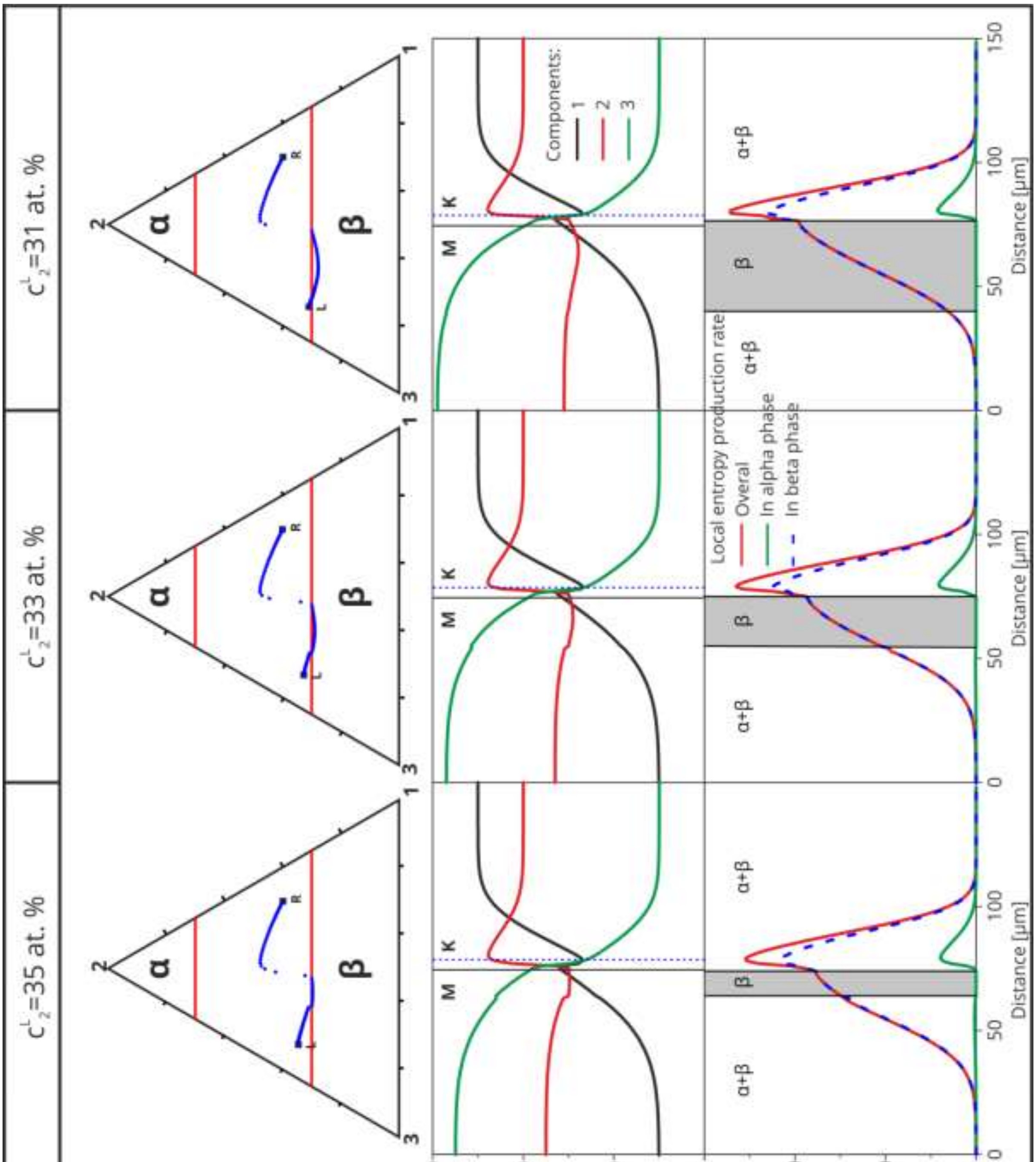

**Fig. 19. Single-phase zone formation in the diffusion couple with one terminal concentration (here left) lying close to the phase boundary. The results for three diffeent concentrations of the second components (35, 33 and 31 at.%). From the top: Diffusion path, concentration profiles and local entropy production rate. Processing parameters, except composition of left alloys, are the same as in Fig. 14.**

### 3.2.4 Type 1a and 1b interphase boundaries in single-phase β|β diffusion couple and formation of α precipitates

The interphase boundary type 1b and two-phase zone are formed in initially single-phase diffusion couple when the diffusion path leaves the single-phase region, say β, enters the two-phase region, say α+β, and the goes back to the single β-phase region. Such behavior of the diffusion path is related to the growth of the α-precipitates in β-matrix, Fig. 20. Such behavior was predicted by Morral [3]. The characteristic feature of type 1b IB is that at the β side the fragment of the diffusion path is tangent to the β|α+β phase boundary line and that after entering the α+β region the diffusion path changes direction, Fig. 20. On the contrary, the type 1a IB is when the diffusion path crosses the phase boundary line on the β-region side at any non-zero angle and the jump of the concentration at IB is generated. Consequently, the fluxes at two sides of interface differ (there is jump of the flux at IB) and IB is moving, Fig. 21.

In single-phase β|β diffusion couple two type 1 interphase boundaries are formed: 1a and 1b. The main difference between 1a and 1b IB is no concentration jump at 1b IB while such jump is observed at 1a IB. Both interphases, 1a and 1b, move in the same direction but with different velocities. Thus the precipitates move as well.

The subsequent conclusions follow from the above:

- the growth of precipitates in the two-phase zone in β|β diffusion couple leads to a formation of two different type 1 IB: 1a and 1b;
- there is no jump of concentration at 1b IB;
- there is jump of concentration at 1a IB;
- there is jump of the flux at the 1a IB and no jump at 1b IB;
- the precipitates move (drift) during diffusion.

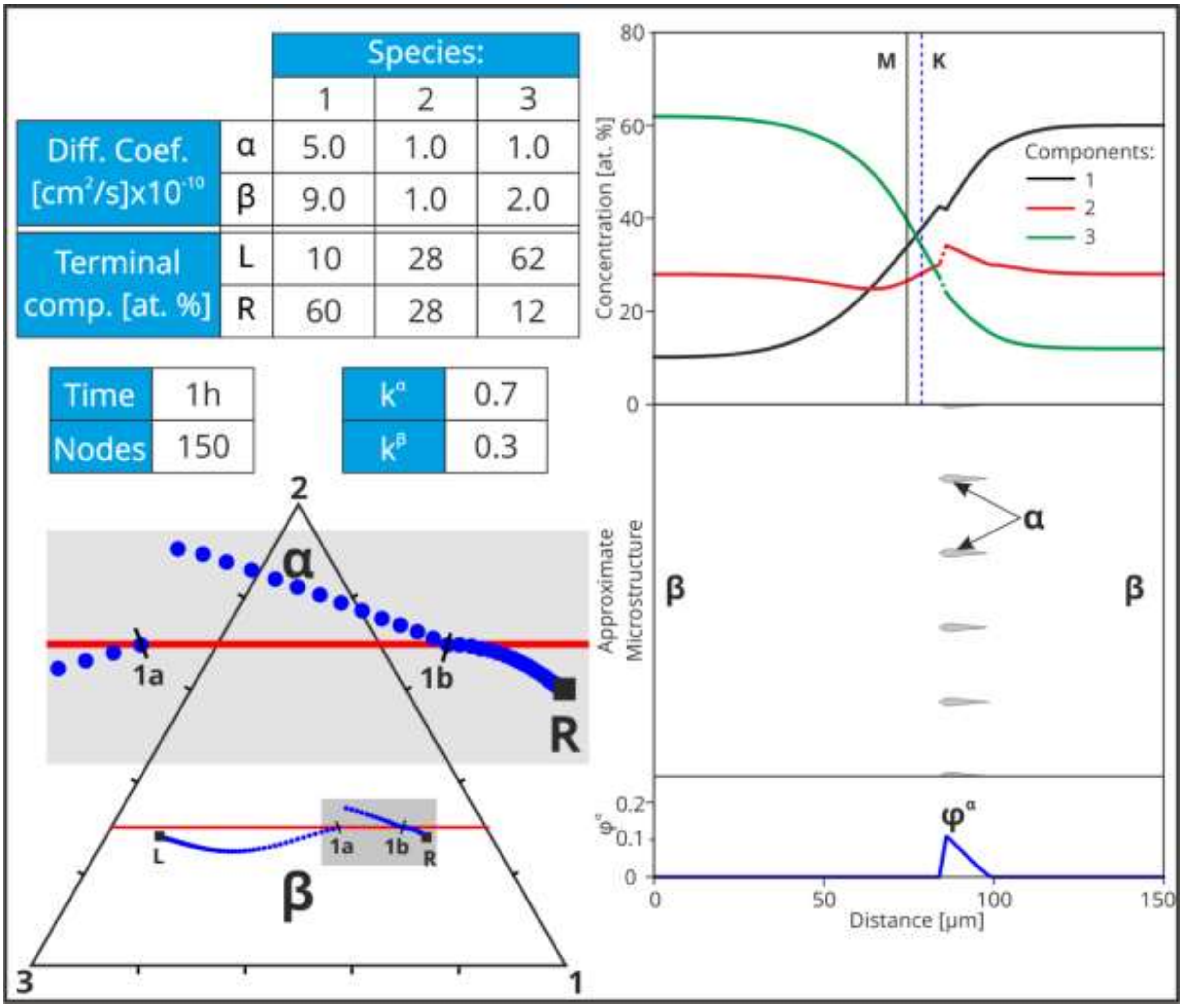

**Fig. 20. Formation of two-phase zone in single-phase diffusion couple. The sequence of the figures and markings like in Fig. 14.**

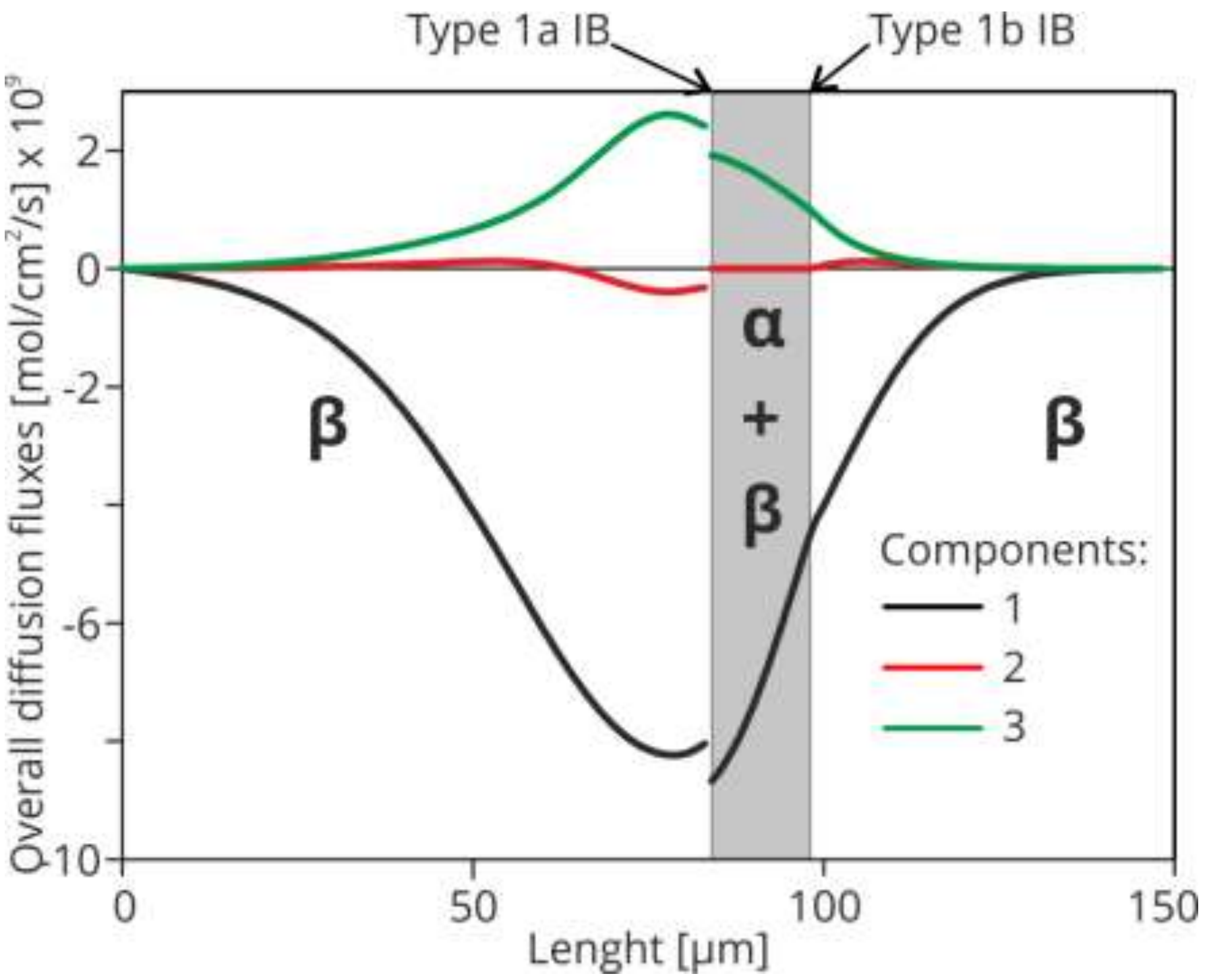

**Fig. 21. Overall diffusion fluxes for the couple shown in Fig. 20. The positions of interphase boundaries are marked. In gray the two-phase zone.**

### 3.2.5 Type 1a interphase boundary in α+β|β diffusion couple and regression of α-phase

During diffusion in the couple composed of single and two-phase alloys, a single IB type 1a can be formed, Fig. 22. In this case, the jump of concentration at the interphase boundary exists. The IB moves toward the two-phase zone and its velocity depends on the volume fraction of α-phase.

The conclusions are as follows:

- the IB type 1a implies, as in the previous case, the concentration jump, Fig. 23;
- the growth of beta-phase occurs at the cost of the two-phase zone.

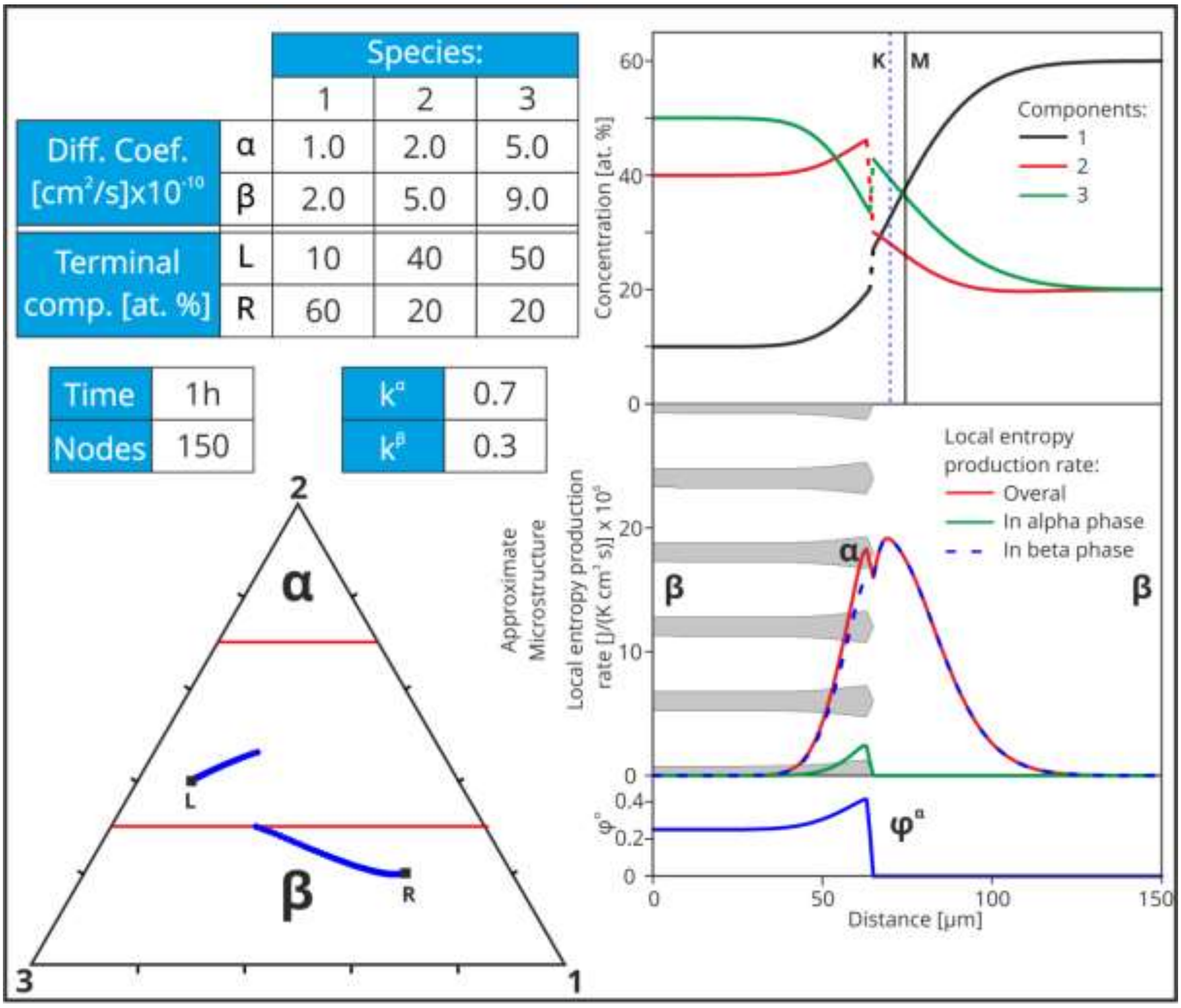

**Fig. 22. Diffusion in α+β|β couple and regression of α-phase. The sequence of the figures like in Fig. 14.**

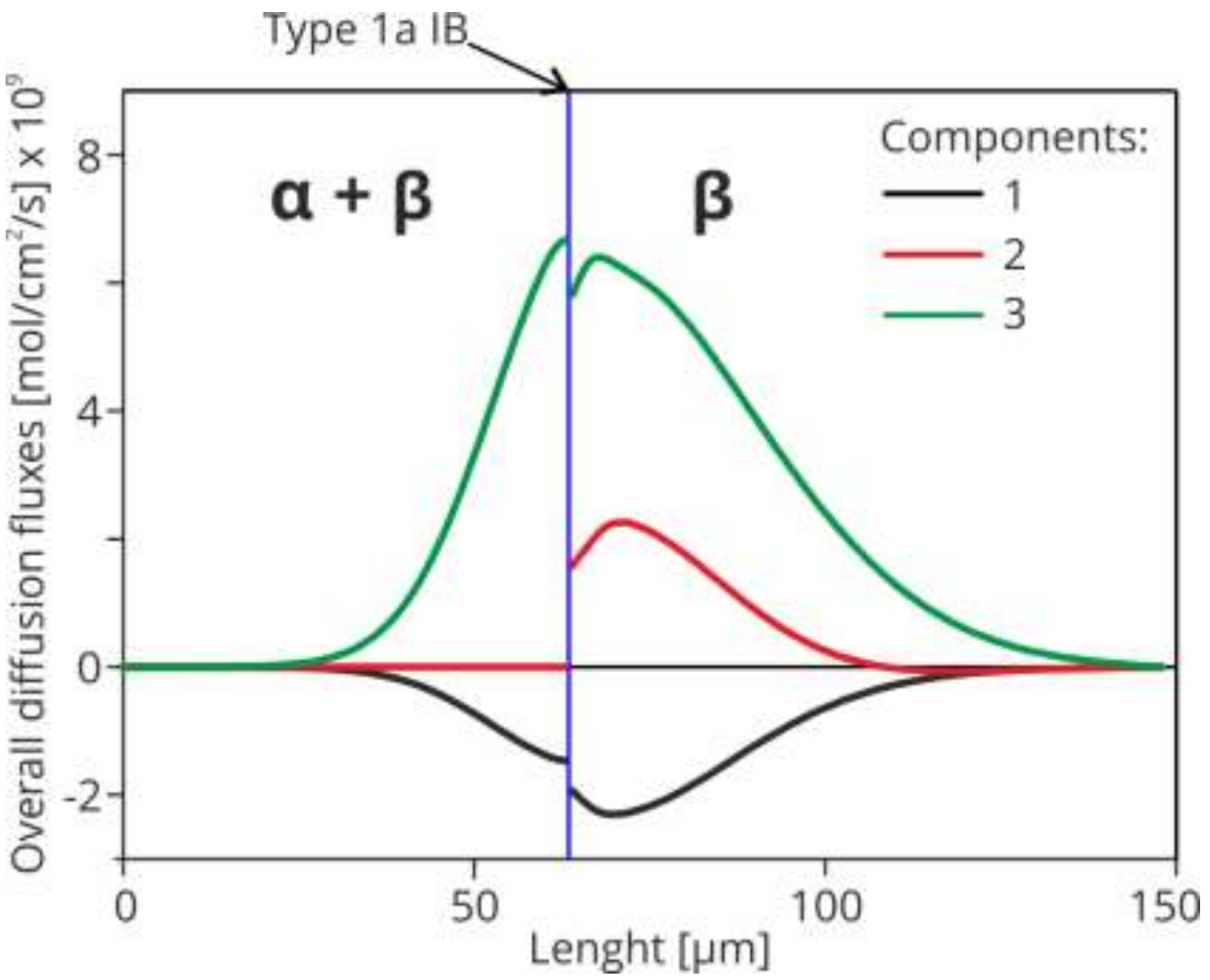

**Fig. 23. Overall diffusion fluxes in α+β|β experiment shown in Fig. 22. The location of interphase boundary is marked on the top.**

### 3.2.6 Type 1a and 1b multiple interface boundaries in α+β|β diffusion couple

When both compositions of the terminal alloys are located close to the phase boundary then the interdiffusion can result in formation of the multiple, single and two-phase zones, Fig. 24. Accordingly, the diffusion path can be divided into fragments that can be analyzed independently (theorem A3) [3]. Such cases of the diffusion paths have been already analyzed in sections 3.2.1 (α+β|β|α+β) and 3.2.3 (β|α+β|β) and adequate conclusions from those sections are valid in the present case as well:

- a complex diffusion paths can be formed in even in "simple" two-phase ternary system;
- all theorems are valid regardless of the complexity level of the diffusion path.

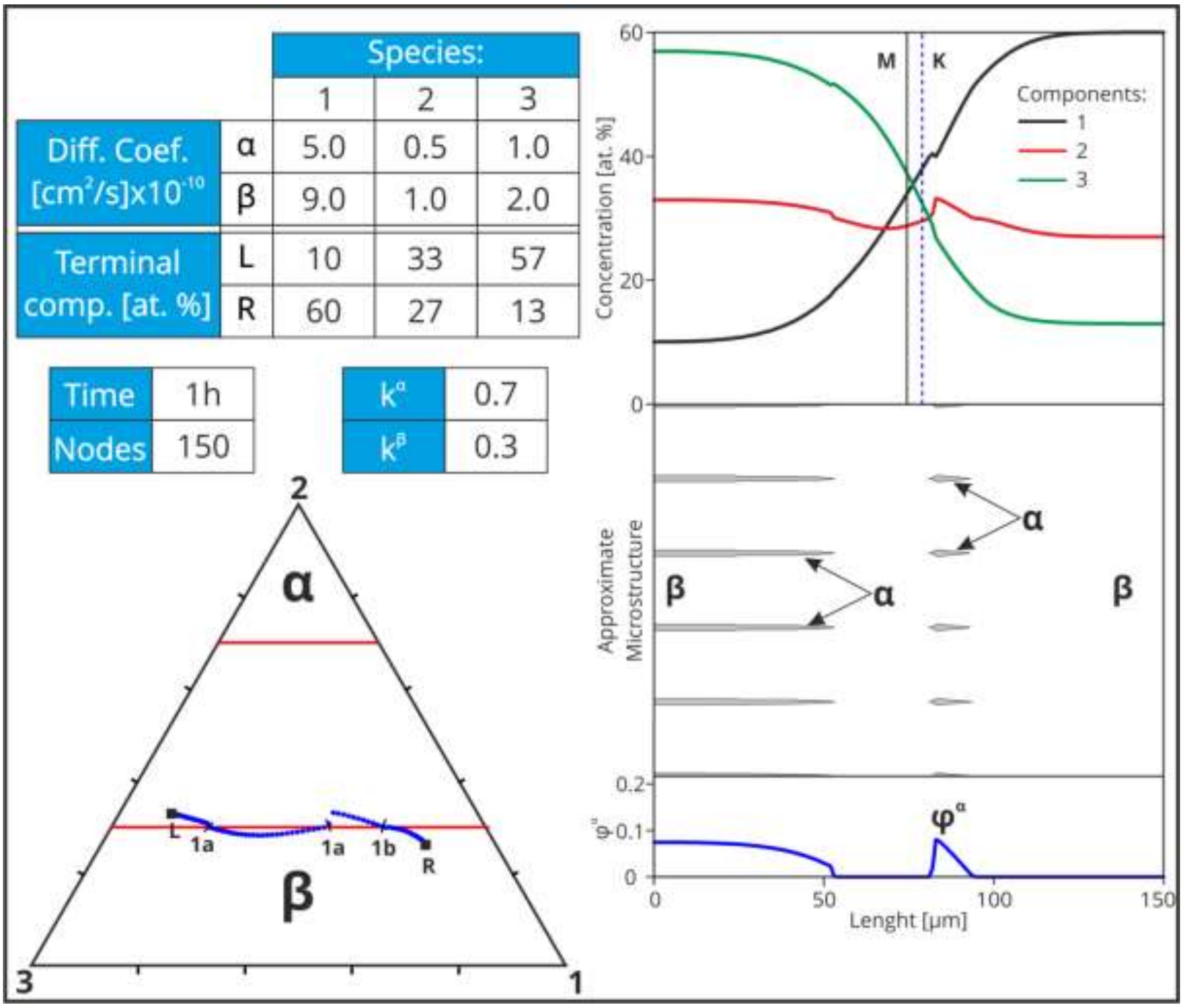

**Fig. 24. Diffusion path with two diffusion zones formed between terminal alloys in α+β|β couple. The sequence of the figures like in Fig. 14.**

### 3.2.7 Type 2 interphase boundary in α|β diffusion couple

Diffusion between two single-phase, different phase alloys is shown in Fig. 25. The jump of the concentration occurs along the tie-line and two-phase zone is not formed. The left and right alloys have at the IB the same chemical potential.

These sample can be divided along the IB into two semi-infinite diffusion zones with constant concentration at the interface. When the transport properties of phases differ, the bifurcation of the Kirkendall plane can be observed.

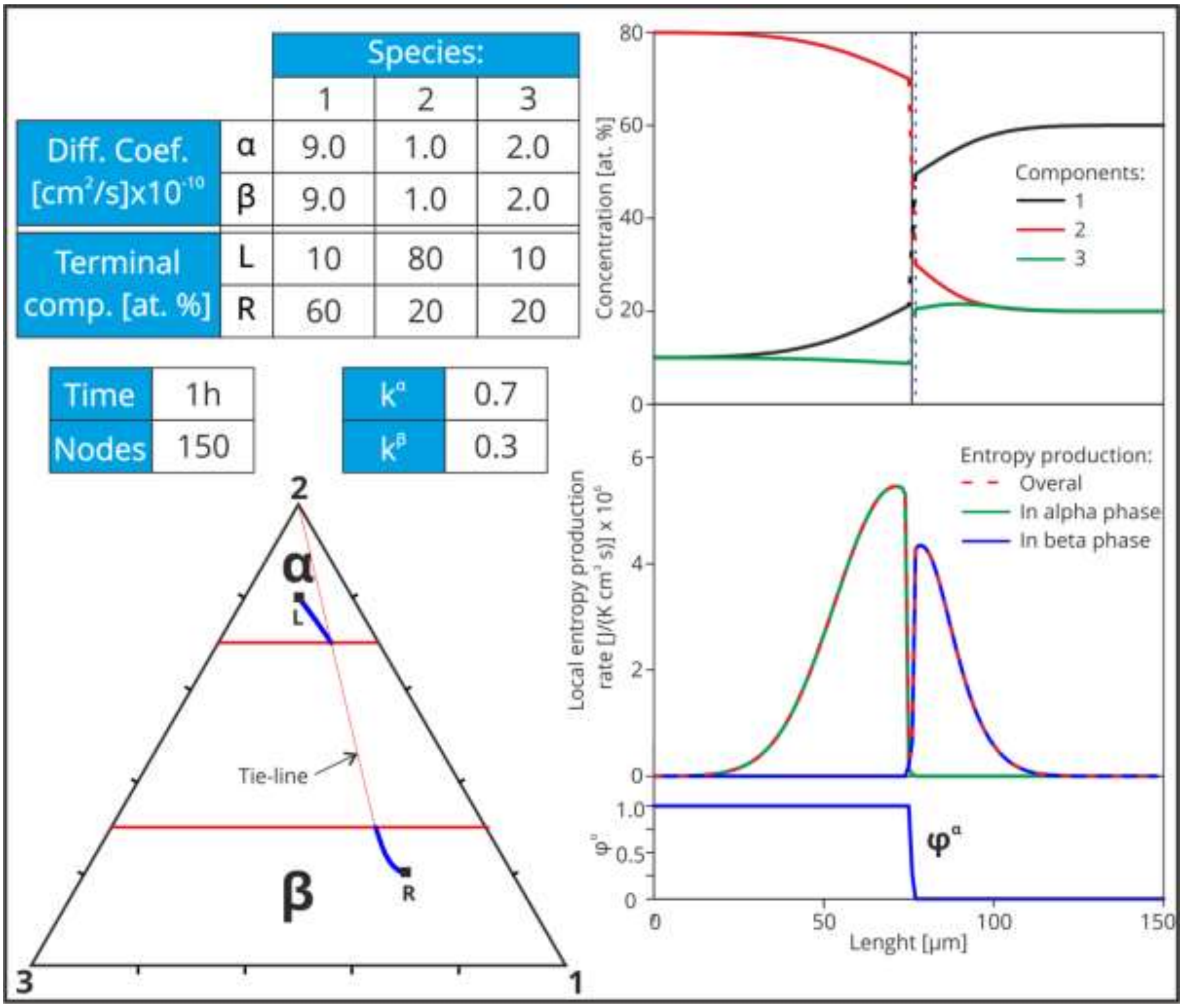

**Fig. 25. Diffusion in α|β couple. Left, from the top: data (diffusion coefficients, terminal composition, process time, number of grid points, positions of α and β phase boundaries ($k^{\alpha}$ and $k^{\beta}$), phase diagram with diffusion path. Right, from the top: concentration profiles with Matano and Kirkendall planes, the local entropy production rate, and volume fraction of α-phase.**

The following conclusions results from the simulation:

- the two-phase zone is not formed;
- the concentration of components on both sites of the type 2 IB are invariant with time;
- the diffusion couple can be divided into two semi-infinite parts with constant concentrations at the boundary (Dirichlet boundary condition);

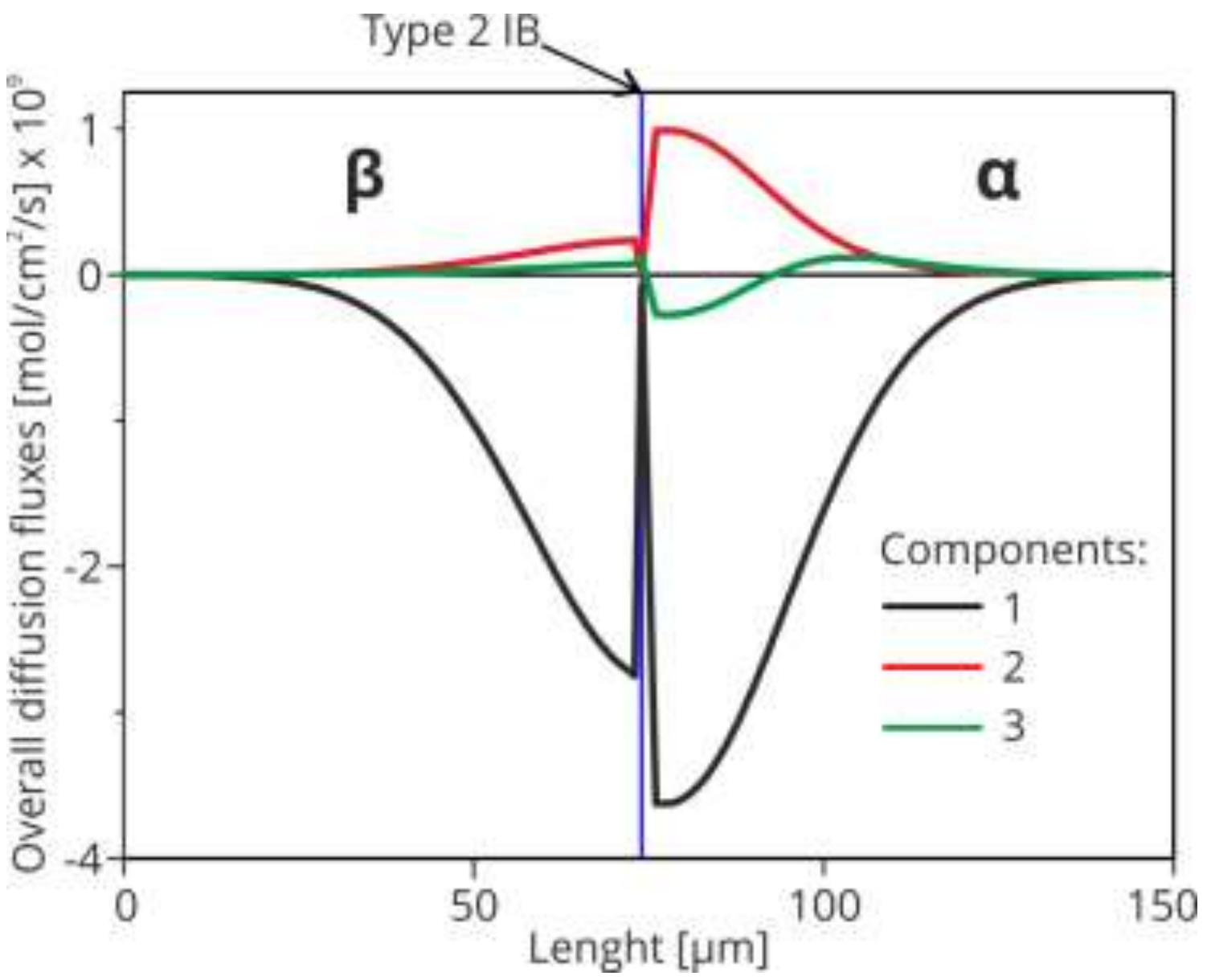

**Fig. 26. Overall diffusion fluxes for the experiment shown in Fig. 25. The position of type 2 IB is marked by blue line.**

### 3.2.8 Singularities in entropy production at the interphase boundaries

Wierzba et al. reported that entropy production density exhibits maxima (peeks) at the interphase boundaries. Consequently, the positions of IBs and thickness of the two-phase zone can be determined from the positions of the representative peeks [78]. The present results indicate that such conclusion is not justified. To verify the problem the simulations for the formation of two-phase zone in single-phase diffusion couple shown in Fig. 20 have been repeated using various grids. The results are shown in Fig. 27.

It is seen that the change of the number of grid points affects neither the shape of diffusion path nor the concentrations profiles. Also the profiles of local entropy production rate are similar, except the maxima (peeks) at the IBs. These singularities (peaks) increase with increasing grid density.

The examination of the finite difference method applied here shows that the singularity (maxima) originates in α-phase when the α-phase volume fraction is near to zero, i.e., when precipitates density is near to zero. In this case the numerical procedure generates big gradients of concentrations. As the entropy production depends on the square of the concentration gradient, the “pseudo-singularities” emerge at the local entropy production rate distribution.

The pseudo-singularities are the consequence of numerical method. Thus, they should be removed from the simulated distributions.

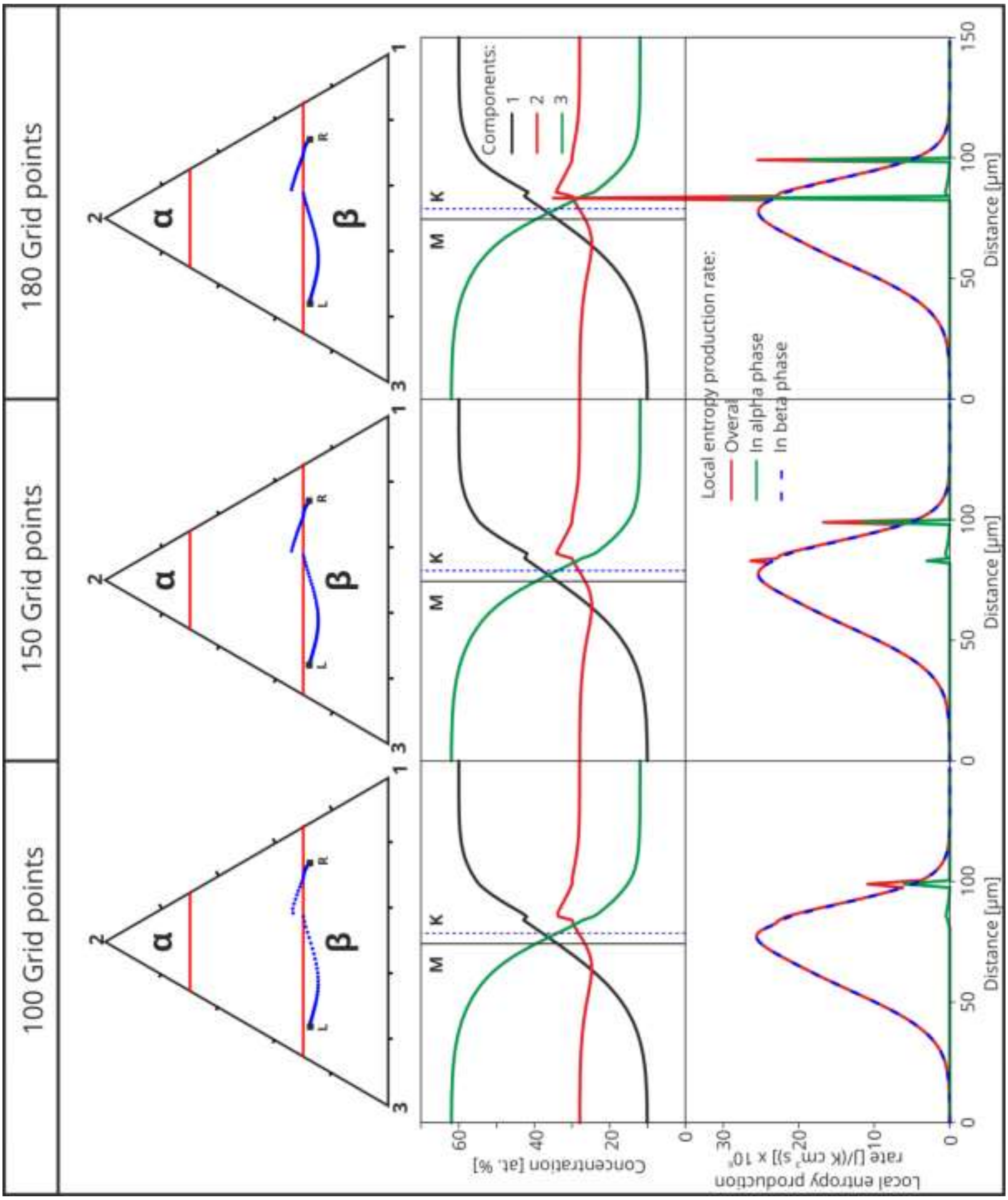

**Fig. 27. The results of numerical simulation shown in Fig. 20, for various numbers of grids points: 100, 150 and 180. In the row, from the top: number of grid points, phase diagram with diffusion path, the concentration profiles with Matano and Kirkendall planes and the local entropy production rate: overall and in individual phases.**

## 3.3 Dual-scale two-phase model in R3

The entirely new approach describing diffusion in the two-phase zone with consideration evolution in morphology was developed recently [80]. The diffusion couple is formed by two alloys, divided by the pseudo-interface, *y(t)*, (Fig. 28). The left alloy is a α-phase matrix with inclusions of small β-phase precipitates ($\varphi^{\beta}$<10 Vol. %). The right alloy is β-phase matrix with inclusions of small α-phase precipitates ($\varphi^{\alpha}$<10 Vol. %). In this approach the diffusion is described in two scales: (Fig. 30):

a) **Coarsened:** one-dimensional in the matrix phase, perpendicular to the alloys interface (R1).
b) **Local:** mass transport occurs between matrix and spherical precipitates, and produces the growth or shrinkage of precipitates (R3).

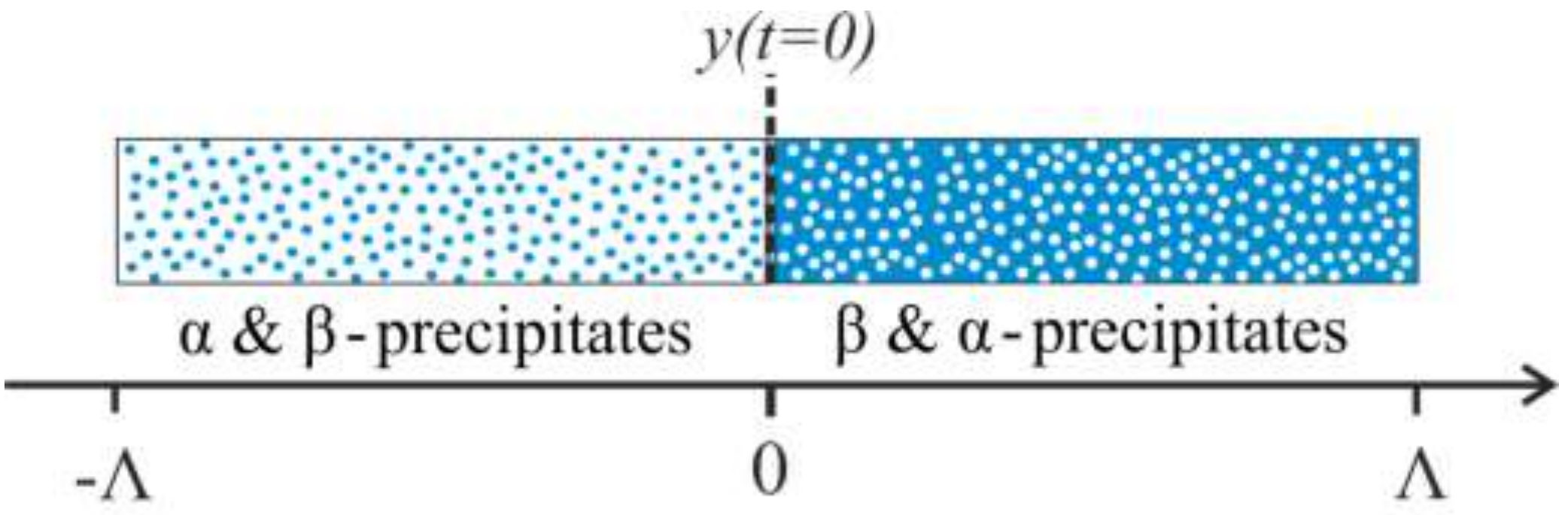

**Fig. 28. Two-phase diffusion couple formed by ternary alloys.**

In this model the local equilibrium between phases is assumed only at the boundaries between phases, shown in local scale in Fig. 30 and at the pseudo-interface in Fig. 31. Concentration in volume of matrix can deviate from equilibrium. The precipitates are small, thus the equilibrium concentration in whole volume of precipitate is almost immediately. The precipitates move only with the matrix lattice (Darken drift), can grow or shrink as a results of diffusion with matrix, however the creation or annihilation of is not assumed.

In this approach the following phase diagram is considered (Fig. 29):

a) phase boundary lines of α and β phase are horizontal (equilibrium concentration od second component is constant);

b) the extension of all tie-lines pas through 2-nd component corner;

c) the partial molar volumes of components in α and β phase, are all equal. Consequently diffusion does not led to change of volume of the diffusion couple.

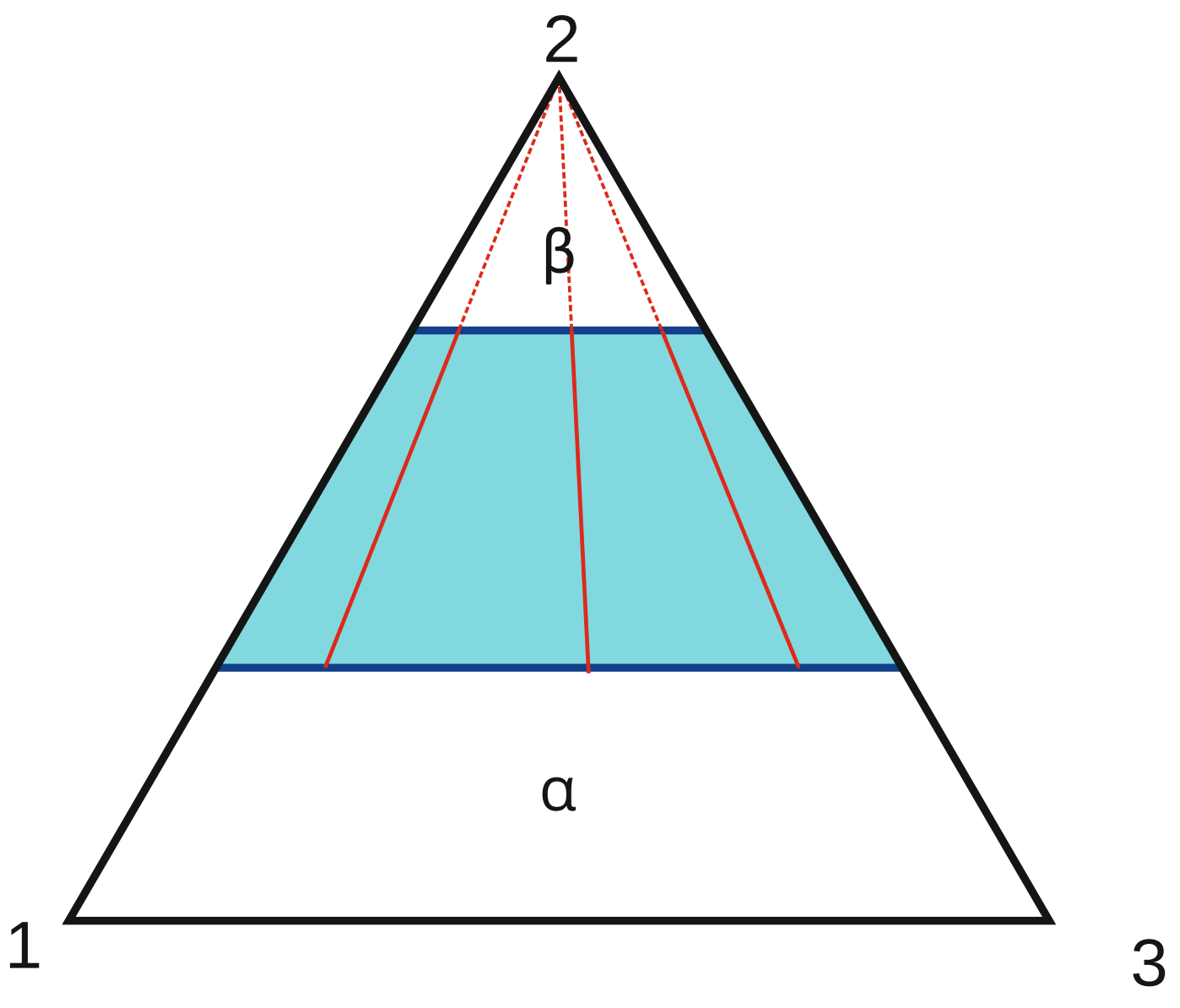

**Fig. 29. Regular solution approximation of the ternary alloy phase diagram.**

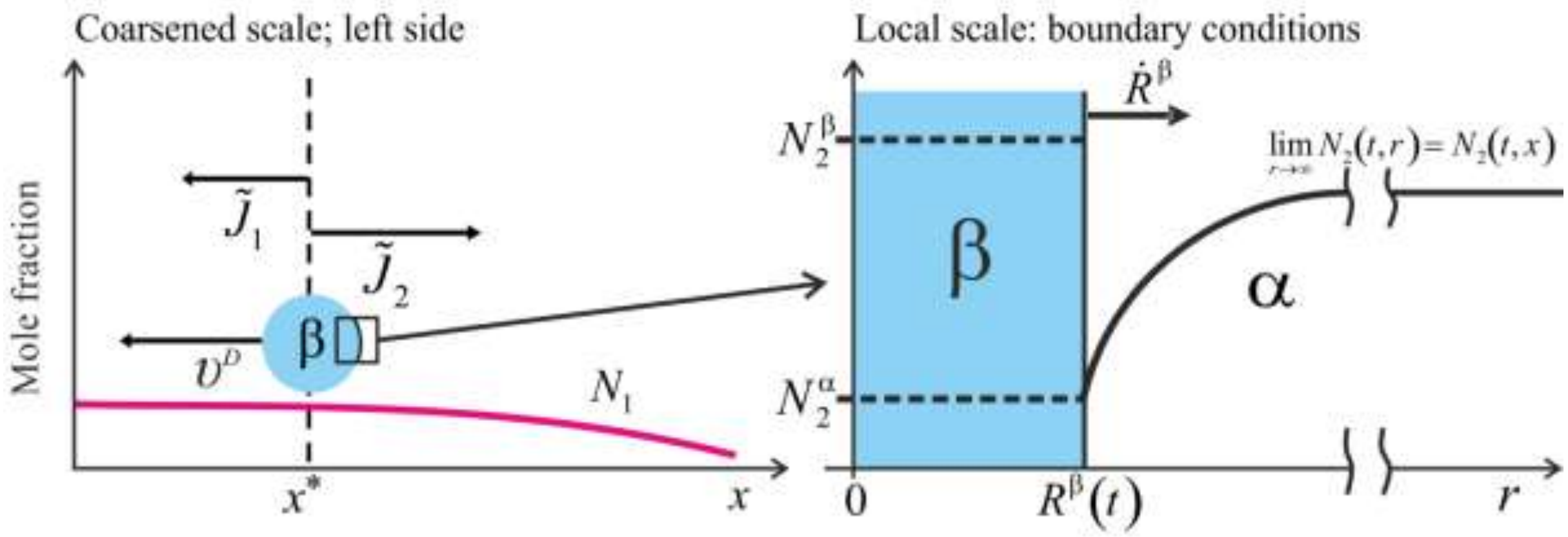

**Fig. 30. Schema of the mass transport in the two-phase diffusion couple: a) fluxes and concentration profile in coarsened scale, b) concentration profile around β-precipitate in α-matrix in local micro-scale.**

Diffusion in global scale, like in Hopfe-Morral model [2] occurs only through the matrix phase, however the process of reaching of equilibrium is controlled by the diffusion in local scale. All following equations described for left side (α-phase matrix), has analogical form for right side, only the superscripts must be changes.

The mass conservation law, for the left side (α-phase matrix) is given by:

$$\begin{aligned}
\frac{\partial N_1}{\partial t} &= \frac{\partial}{\partial x}\left(\Omega\,\tilde{J}_1\right) + \Omega\,4\pi\left(R^\beta\right)^2 n^\beta\,\tilde{J}_1^\alpha,\\
\frac{\partial N_2}{\partial t} &= \frac{\partial}{\partial x}\left(\Omega\,\tilde{J}_2\right) + \Omega\,4\pi\left(R^\beta\right)^2 n^\beta\,\tilde{J}_2^\alpha,\\
\begin{bmatrix}\text{mass change}\\ \text{in global}\\ \text{scale = labo-}\\ \text{ratory RF}\end{bmatrix} &= \begin{bmatrix}\text{interdiffussion fluxes}\\ \text{in global scale}\end{bmatrix} + \begin{bmatrix}\text{Intake/withdrawal}\\ \text{from the precipitate}\\ \text{in local scale =}\\ \text{material RF}\end{bmatrix},
\end{aligned} \tag{52}$$

where: $N_1 = N_1^\alpha(t,x)$ and $N_2 = N_2^\alpha(t,x)$ denote the molar ratio of first and second components in α-matrix (not average ratio), $\tilde{J}_1^\alpha = \tilde{J}_1^\alpha\left(R^\beta\right)$ and $\tilde{J}_2^\alpha = \tilde{J}_2^\alpha\left(R^\beta\right)$ are the interdiffusion fluxes in local scale through the β-precipitate external surface, and $R^\beta = R^\beta\left(t,x\right)$ and $n^\beta = n^\beta\left(t,x\right)$ are the radius of the precipitate and the precipitates density in global scale.

The average molar ratio can be calculated as:

$$\bar{N}_i = \left(1-\varphi^\beta\right)N_i + \varphi^\beta N_i^\beta, \tag{53}$$

where $N_i^\beta$ is a (equilibrium) molar ratio in the precipitate, $\varphi^\beta$ is the volume fraction of β-phase in α-matrix, depends on the precipitates density, $n^\beta = n^\beta\left(t,x\right)$, and their radius, $R^\beta = R^\beta\left(t,x\right)$. Thus we have:

$$\varphi^\beta = \frac{4}{3}\pi\left(R^\beta\right)^3 n^\beta \quad \text{for} \quad x < y\left(t\right). \tag{54}$$

### 3.3.1 Model of the regular solution

In this approach the diffusion driving force is a gradient of chemical potential. The following parameters describe the thermodynamics of two-phase system shown in Fig. 29:

$$\Phi_{11} = \Phi_{22} = \Phi_{33} = 0, \qquad E_{13}^{mix} = 0, \qquad E_{23}^{mix} = E_{12}^{mix} = E^{mix} > 0\,. \tag{55}$$

where $\Phi_{ij}$ is a pair interaction energy between nearest neighboring *i* and *j* atoms and $E_{ij}^{mix}$ is a mixing energy.

From the theory of regular solution:

$$g = ZE^{mix} N_2 (1 - N_2) + kT (N_1 \ln N_1 + N_2 \ln N_2 + N_3 \ln N_3), \tag{56}$$

where $g$ is the Gibbs free energy per atom, $Z$ is number of nearest neighbors and:

$$ZE^{mix} = \frac{kT}{1 - 2N_2^{\alpha}} \ln \frac{1 - N_2^{\alpha}}{N_2^{\alpha}}. \tag{57}$$

The concentrations of species 1 and 2 are threat as independent $(N_3 = 1 - N_1 - N_2)$:

$$g = N_1\mu_1 + N_2\mu_2 + N_3\mu_3, \quad \frac{\partial g}{\partial N_1} = \mu_1 - \mu_3, \quad \frac{\partial g}{\partial N_2} = \mu_2 - \mu_3. \tag{58}$$

The chemical potentials in such regular solution depend on composition:

$$\begin{aligned} &\frac{\partial \mu_1}{\partial N_1} = kT\left(\frac{1}{N_1}\right); &&\frac{\partial \mu_1}{\partial N_2} = kT\left(\frac{1}{N_1}\right) + 2ZE^{mix} N_2; \\ &\frac{\partial \mu_2}{\partial N_1} = 0; &&\frac{\partial \mu_2}{\partial N_2} = kT\left(\frac{1}{N_2}\right) - 2ZE^{mix}(1 - N_2);, \\ &\frac{\partial \mu_3}{\partial N_1} = -kT\left(\frac{1}{N_3}\right); &&\frac{\partial \mu_3}{\partial N_2} = -kT\left(\frac{1}{N_3}\right) + 2ZE^{mix} N_2. \end{aligned} \tag{59}$$

The concentrations in the two-phase zone obey relations:

$$N_2^{\alpha} = const, \quad N_2^{\beta} = const, \quad N_1^{\beta} = N_1^{\alpha} \frac{1 - N_2^{\beta}}{1 - N_2^{\alpha}}. \tag{60}$$

### 3.3.2 Diffusion in global scale

Consistency of the Darken and Onsager methods [72], allows writing the interdiffusion presented in Darken form:

$$\Omega J_i = \Omega J_i^d + N_i \upsilon^D \quad \text{for} \quad i = 1,2,3, \tag{61}$$

where: $J_i^d$ is the diffusion flux of *i*-th spices in material reference frame and $\upsilon^D$ is the drift (Darken) velocity calculated as:

$$\upsilon^D = -\sum_i \Omega J_i^d \tag{62}$$

The volume diffusion fluxes in the material reference frame are given by the Nernst-Planck diffusive flux formula where chemical potential gradients are given by Eqs. (59):

$$-\Omega J_1^d = L_{11}\nabla\mu_1 = D_1^* \operatorname{grad} N_1 + D_1^* \frac{2ZE^{mix}}{kT} N_1 N_2 \operatorname{grad} N_2, \tag{63}$$

$$-\Omega J_2^d = L_{22}\nabla\mu_2 = D_2^* \left(1 - \frac{2ZE^{mix}}{kT} N_2 \left(1 - N_2\right)\right) \operatorname{grad} N_2, \tag{64}$$

$$-\Omega J_3^d = L_{33}\nabla\mu_3 = -D_3^* \operatorname{grad} N_1 + D_3^* \left(\frac{2ZE^{mix}}{kT} N_2 N_3 - 1\right) \operatorname{grad} N_2. \tag{65}$$

Equations (63)–(65) in „component reference frame” have form[7]:

$$\Omega \tilde{J}_i = -\sum_{k=1}^{2} \tilde{D}_{ik} \nabla N_k \quad \text{for} \quad i = 1, 2, \tag{66}$$

where: $\tilde{D}_{ik}$ is an elements of 2x2 interdiffusivity matrix [72, 85]:

$$\tilde{D}_{11} = \left(1 - N_1\right) D_1^* + N_1 D_3^*, \; \tilde{D}_{12} = \left( D_3^* - D_2^* + \left[ D_1^* \left(1 - N_1\right) + D_2^* \left(1 - N_2\right) + D_3^* N_3 \right] N_2 \frac{2ZE^{mix}}{kT} \right) N_1,$$

$$\tilde{D}_{21} = \left(D_3^* - D_1^*\right) N_2, \qquad \tilde{D}_{22} = D_2^* + \left(D_3^* - D_2^*\right) N_2 - \left[ D_2^* \left(1 - N_2\right)^2 + \left(D_1^* N_1 + D_3^* N_3\right) N_2 \right] N_2 \frac{2ZE^{mix}}{kT}.$$

(67)

Drift velocity becomes:

$$\upsilon^D = \left(D_1^* - D_3^*\right) \nabla N_1 + \left\{ D_2^* - D_3^* + \left[ D_3^* - D_2^* + \left(D_1^* - D_3^*\right) N_1 + \left(D_2^* - D_3^*\right) N_2 \right] N_2 \frac{2ZE^{mix}}{kT} \right\} \nabla N_2.$$

(68)

Equations (66) - (68) represent interdiffusion in α as well as in β-matrix, therefore next sections were reduced only to left side.

### 3.3.3 Diffusion in local scale

Simplifications in local scale:

a) quasi steady-state interdiffusion around precipitate;

---

[7] Component frame of reference, i.e., frame where the r-diffusional fluxes are transformed to (r-1) interdiffusion fluxes [72].

b) local changes of matrix phase composition are small enabling to consider the interdiffusion coefficients, $\tilde{D}_{ij}$, in the vicinity of precipitate as constant, and given by Eqs. (70) and (71);

c) redistribution time (equalization) of incoming atoms inside small precipitates is considered as faster than redistribution around precipitate;

d) the low volume fraction of the precipitates allows to neglect the impact of the quasi-stationary interdiffusion around precipitate - impact of fluxes in local RF, $\tilde{J}_i^{\alpha} = \tilde{J}_i^{\alpha}(t,r)$ and $\tilde{J}_i^{\beta} = \tilde{J}_i^{\beta}(t,r)$ - on the interdiffusion fluxes in matrix phase: $\tilde{J}_i := \tilde{J}_i(t,x)$;

e) in this particular case of the phase diagram, at the straight horizontal phase boundaries: α|(α+β) and β|(α+β) the concentrations of *2*-nd component do not change: $N_2^{\beta}, N_2^{\alpha} = const$. Consequently in the whole two-phase zone at the precipitate|matrix interface the local equilibrium implies:

$$\mathrm{d}N_2^{\beta} = \mathrm{d}N_2^{\alpha} = 0 \quad \text{and} \quad N_1^{\alpha}\left(R^j\right) = N_1^{\beta}\left(R^j\right)\frac{1-N_2^{\alpha}}{1-N_2^{\beta}} \quad \text{for } j = \alpha, \beta. \tag{69}$$

In the local scale the concentrations in the matrix phase (on left site) are denoted as $N_1^{\alpha} := N_1^{\alpha}(t,r)$. The equilibrium concentrations at the every precipitate|matrix interface at its given position in global scale are denoted as: $N_1^{\alpha}\left(R^{\beta}\right) := N_1^{\alpha}\left(R^{\beta}(t,x)\right)$ and $N_1^{\beta}\left(R^{\beta}\right) := N_1^{\beta}\left(R^{\beta}(t,x)\right)$.

The mass conservation in α-matrix local scale, Fig. 30b, is formulated in spherical coordinates:

$$\frac{\partial N_1^{\alpha}}{\partial t} = \frac{1}{r^2}\frac{\partial}{\partial r}\left(r^2\left(\tilde{D}_{11}\frac{\partial N_1^{\alpha}}{\partial r} + \tilde{D}_{12}\frac{\partial N_2^{\alpha}}{\partial r}\right)\right), \qquad R^{\beta}(t,x) < r < \infty, \tag{70}$$

$$\frac{\partial N_2^{\alpha}}{\partial t} = \frac{1}{r^2}\frac{\partial}{\partial r}\left(r^2\left(\tilde{D}_{21}\frac{\partial N_1^{\alpha}}{\partial r} + \tilde{D}_{22}\frac{\partial N_2^{\alpha}}{\partial r}\right)\right), \qquad R^{\beta}(t,x) < r < \infty. \tag{71}$$

With boundary conditions:

$$N_1^{\alpha}\left(R^{\beta}\right) = N_1^{\beta}\left(R^{\beta}\right)\frac{1-N_2^{\alpha}}{1-N_2^{\beta}}, \quad N_1^{\alpha}(t,r) = N_1(t,x) \quad \text{for } r \to \infty, \tag{72}$$

$$N_2^{\alpha}\left(R^{\beta}\right) = N_2^{\alpha}, \qquad\qquad N_2^{\alpha}(t,r) = N_2^{\alpha}(t,x) \quad \text{for } r \to \infty. \tag{73}$$

The quasi-steady solution of Eqs. (70) and (71) was found by Ham for the case of precipitate grow [86]. Using this result the fluxes in α-matrix at the α|β interface are given by:

$$\Omega\,\tilde{J}_1^{\alpha}\left(R^{\beta}\right)=-\frac{1}{R^{\beta}}\left[\tilde{D}_{11}\left(N_1-N_1^{\beta}\left(R^{\beta}\right)\frac{1-N_2^{\alpha}}{1-N_2^{\beta}}\right)+\tilde{D}_{12}\left(N_2-N_2^{\alpha}\right)\right], \tag{74}$$

$$\Omega\,\tilde{J}_2^{\alpha}\left(R^{\beta}\right)=-\frac{1}{R^{\beta}}\left[\tilde{D}_{21}\left(N_1-N_1^{\beta}\left(R^{\beta}\right)\frac{1-N_2^{\alpha}}{1-N_2^{\beta}}\right)+\tilde{D}_{22}\left(N_2-N_2^{\alpha}\right)\right], \tag{75}$$

where $N_1=N_1(t,x)$ and $N_2=N_2(t,x)$ are concentrations in α-matrix in global scale.

### 3.3.4 Evolution and movement of precipitates

The model allows to calculate the microstructural change as the change of precipitates radius $R^j := R^j(t,x)$ and concentration $n_1^j := n_1^j(t,x)$. It be realized by two processes:

1) Precipitates can move as a result of drift together with matrix lattice (as “markers” in matrix) which leads to change of concentration of precipitates according to the equation:

$$\frac{\partial n^{\beta}}{\partial t}=-\frac{\partial}{\partial x}\left(n^{\beta}\,\upsilon^{D}\right). \tag{76}$$

2) Can grow and shrink due to exchange of mas with matrix (local scale diffusion). This second process lead to change concentration in precipitates and is inversely proportional to the precipitate size. The change of precipitates radius, $R^{\beta}$, and concentration of first component, $N^{\beta}$, (the concentration of second component are constant), in laboratory reference frame:

$$\begin{array}{ccccc}\dfrac{\partial R^{\beta}}{\partial t} & = & \left.\dfrac{\mathrm{D}R^{\beta}}{\mathrm{D}t}\right|_{\upsilon^{D}} & - & \upsilon^{D}\dfrac{\partial R^{\beta}}{\partial x}\\ \begin{bmatrix}\text{changes at given}\\ \text{position in}\\ \text{global scale =}\\ \text{laboratory RF}\end{bmatrix} & = & \begin{bmatrix}\text{changes in}\\ \text{local scale}\\ =\\ \text{material RF}\end{bmatrix} & + & \begin{bmatrix}\text{changes due to}\\ \text{material drift}\\ \text{in global scale}\\ \text{= laboratory RF}\end{bmatrix},\end{array} \tag{77}$$

$$\frac{\partial N_1^{\beta}}{\partial t}=\left.\frac{\mathrm{D}N_1^{\beta}}{\mathrm{D}t}\right|_{\upsilon^{D}}-\upsilon^{D}\frac{\partial N_1^{\beta}}{\partial x}. \tag{78}$$

The change of precipitates radius and concentration inside it, can be calculated from the mass balance at precipitate|matrix interface (boundary conditions) relate rate of the radius and composition changes with interdiffusion fluxes in local scale:

$$
\begin{array}{ccccc}
-4\pi\left(R^{\beta}\right)^{2}\Omega\,\tilde{J}_1^{\alpha}\left(R^{\beta}\right)\mathrm{d}t & = & 4\pi\left(R^{\beta}\right)^{2}\left[N_1^{\beta}\left(R^{\beta}\right)-N_1^{\alpha}\left(R^{\beta}\right)\right]\mathrm{d}R^{\beta} & + & \frac{4}{3}\pi\left(R^{\beta}\right)^{3}\mathrm{d}N_1^{\beta} \\
\begin{bmatrix}\text{Inflow through} \\ \beta\text{-precipitate} \\ \text{surface, Fig. 3}\end{bmatrix} & = & \begin{bmatrix}\text{Consumption due to the} \\ \text{precipitate radius change}\end{bmatrix} & + & \begin{bmatrix}\beta\text{-phase} \\ \text{composition} \\ \text{adjustment}\end{bmatrix},
\end{array} \tag{79}
$$

where $\tilde{J}_1^{\alpha}\left(R^{\beta}\right)$, $\tilde{J}_2^{\alpha}\left(R^{\beta}\right)$ denote the quasi steady-state interdiffusion fluxes through the surface of the precipitate in the local scale, $N_1^{\beta}\left(R^{\beta}\right)$, $N_1^{\alpha}\left(R^{\beta}\right)$ and $N_2^{\beta}$, $N_2^{\alpha}$ are local equilibrium concentrations at β-precipitate|α-matrix interface given by Eqs. (72) and (73).

From relation (79) it follows:

$$
\left.\frac{\mathrm{D}N_1^{\beta}}{\mathrm{D}t}\right|_{\upsilon^D} = -\frac{3}{R^{\beta}}\left(\Omega\,\tilde{J}_1^{\alpha}\left(R^{\beta}\right)+\left[N_1^{\beta}\left(R^{\beta}\right)-N_1^{\alpha}\left(R^{\beta}\right)\right]\left.\frac{\mathrm{D}R^{\beta}}{\mathrm{D}t}\right|_{\upsilon^D}\right) \tag{80}
$$

Because the concentration of the second component in the precipitate is constant ( it not affected by its growth or shrinking)

$$
\begin{aligned}
&-4\pi\left(R^{\beta}\right)^{2}\Omega\,\tilde{J}_2^{\alpha}\left(R^{\beta}\right)\mathrm{d}t = 4\pi\left(R^{\beta}\right)^{2}\left(N_2^{\beta}-N_2^{\alpha}\right)\mathrm{d}R^{\beta} \quad \Rightarrow \\
&\Rightarrow \qquad \left.\frac{\mathrm{D}R^{\beta}}{\mathrm{D}t}\right|_{\upsilon^D} = -\frac{1}{N_2^{\beta}-N_2^{\alpha}}\Omega\,\tilde{J}_2^{\alpha}\left(R^{\beta}\right),
\end{aligned} \tag{81}
$$

Combining Eqs. (81) and (75) we obtain the for changing of precipitates radius:

$$
\left.\frac{\mathrm{D}R^{\beta}}{\mathrm{D}t}\right|_{\upsilon^D} = \frac{1}{R^{\beta}\left(N_2^{\beta}-N_2^{\alpha}\right)}\left[\tilde{D}_{21}\left(N_1-N_1^{\beta}\left(R^{\beta}\right)\frac{1-N_2^{\alpha}}{1-N_2^{\beta}}\right)+\tilde{D}_{22}\left(N_2-N_2^{\alpha}\right)\right]. \tag{82}
$$

Combining Eqs. (79) and (71) we have:

$$
\left.\frac{\mathrm{D}N_1^{\beta}}{\mathrm{D}t}\right|_{\upsilon^D} = \frac{3}{\left(R^{\beta}\right)^{2}}\begin{bmatrix}-\left(N_1^{\beta}\right)^{2}\tilde{D}_{21}k_1k_3+N_1^{\beta}\left[\tilde{D}_{21}k_3N_1+\tilde{D}_{22}k_3\left(N_2-N_2^{\alpha}\right)-\tilde{D}_{11}k_1\right]+ \\ +\tilde{D}_{12}\left(N_2-N_2^{\alpha}\right)+\tilde{D}_{11}N_1\end{bmatrix}, \tag{83}
$$

where $N_1^{\beta}=N_1^{\beta}\left(R^{\beta}\right)$ and $k_1=\frac{1-N_2^{\alpha}}{1-N_2^{\beta}}$, $k_3=\frac{1}{1-N_2^{\beta}}$.

### 3.3.5 Mass balance at moving "pseudo-interface" between α-matrix and ß-matrix alloys

The plane where α-matrix contacts β-matrix, we call pseudo-interface, Fig. 31. We assume the local equilibrium (ortho-equilibrium) at pseudo-interface, thus the composition in matrix α-phase at the left side of the moving pseudo interface should be in local equilibrium with matrix β-phase at the right side of interface. It means that these compositions should be at the opposite sides of the some conode.

In order to have concise form of the boundary conditions we introduce the following denotations:

$$f^- = f^-(t) = \lim_{x \to y(t)-0} f(t,x), \qquad f^+ = f^+(t) = \lim_{x \to y(t)+0} f(t,x), \tag{84}$$

where $f^-$ and $f^+$ denote the values of the considered functions at the left and right side of the pseudo interface, that is: $N_1^-, N_1^+$, $\tilde{J}_1^-, \tilde{J}_1^+$, $R^-, R^+$, etc.

The local equilibrium shown in Eq. (60) holds for the every α|β interface in the whole diffusion zone, and pseudo-interface as well.

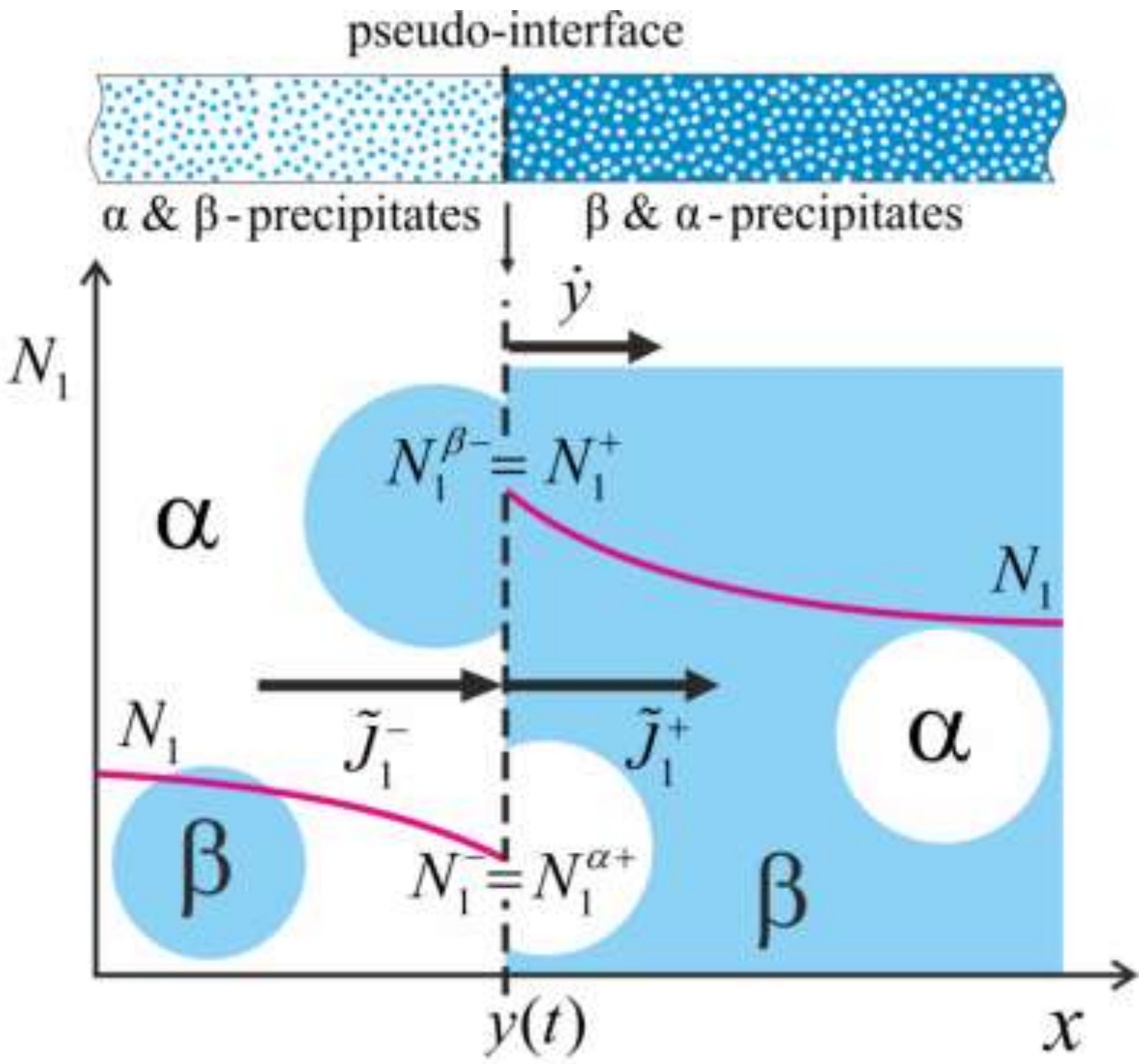

**Fig. 31. Moving "pseudo-interface", $y(t)$, between α-matrix|ß-matrix alloys.**

Equation (60), implies the following relations between compositions at both sides of the pseudo-interface shown in Fig. 31:

$$N_2^- = N_2^{\alpha+} = N_2^{\alpha}, \qquad N_2^+ = N_2^{\beta-} = N_2^{\beta},$$
$$N_1^{\beta-} = N_1^+, \qquad N_1^{\alpha+} = N_1^-, \tag{85}$$

$$N_1^+ = N_1^- \frac{1 - N_2^{\beta}}{1 - N_2^{\alpha}}. \tag{86}$$

From Eqs. (2) and (3) it follows that we have one free parameter which is $N_1^- = N_1^-(t)$ that can be computed from the mass balance at the moving pseudo interface $y(t)$, i.e., from the Stefan condition. To formulate the flux balance equations at pseudo-interface, we use the volume fractions of phases in two-phase zone. At the moving pseudo-interface we have:

$$\varphi^- = 1 - \varphi^{\beta-} = 1 - \frac{4}{3}\pi n_0^{\beta}\left(R^-\right)^3, \qquad \varphi^+ = \frac{4}{3}\pi n_0^{\alpha}\left(R^+\right)^3, \tag{87}$$

where $\varphi^- = \varphi^-(t)$ and $\varphi^+ = \varphi^+(t)$ are the volume fractions of ∝-phase at both sides of the pseudo-interface, $R^- = R^-(t)$ and $R^+ = R^+(t)$ denote the radius of the β and α-precipitates, respectively.

Main idea of flux balance at the moving boundary is that in the reference frame moving together with boundary, the incoming and out coming fluxes are equal (nothing can be accumulated at the boundary). We define an overall flux in the laboratory RF, $\bar{J}_i$, i.e. the flux due to the drift of precipitates and interdiffusion in the matrix. The fluxes at the left, $x = y(t) - 0$, and right, $x = y(t) + 0$, sides of pseudo-interface are given by:

$$\bar{J}_i^- = \varphi \tilde{J}_i^- + \left(1 - \varphi^-\right)\frac{N_i^{\beta-}}{\Omega}\upsilon^{D-},$$
$$\bar{J}_i^+ = \varphi \tilde{J}_i^+ + \left(1 - \varphi^+\right)\frac{N_i^{\beta+}}{\Omega}\upsilon^{D+} \quad \text{for } i = 1,2. \tag{88}$$

The overall fluxes relatively to the moving pseudo-interface, $\hat{J}_i(t)$, Fig. 3, are given by:

$$\hat{J}_i^-(t) = \bar{J}_i^- - \frac{\bar{N}_i^-}{\Omega}\frac{\mathrm{d}y}{\mathrm{d}t} \quad \text{where } \bar{N}_i^-(t,x) = \varphi^- N_i^- + \left(1 - \varphi^-\right)N_i^{\beta-} \quad \text{for } i = 1,2, \tag{89}$$

$$\hat{J}_i^+(t) = \bar{J}_i^+ - \frac{\bar{N}_i^+}{\Omega}\frac{\mathrm{d}y}{\mathrm{d}t} \quad \text{where } \bar{N}_i^+(t,x) = \varphi^+ N_i^+ + \left(1 - \varphi^+\right)N_i^{\beta+} \quad \text{for } i = 1,2. \tag{90}$$

The mass balance requires that $\hat{J}_i^- = \hat{J}_i^+$:

$$\bar{J}_i^- - \frac{\bar{N}_i^-}{\Omega}\frac{\mathrm{d}y}{\mathrm{d}t} = = \bar{J}_i^+ - \frac{\bar{N}_i^+}{\Omega}\frac{\mathrm{d}y}{\mathrm{d}t} \quad \text{for} \quad i = 1,2, \tag{91}$$

and velocity of pseudo-interface equals:

$$\frac{\mathrm{d}y}{\mathrm{d}t} = \Omega\frac{\bar{J}_i^+ - \bar{J}_i^-}{\bar{N}_i^+ - \bar{N}_i^-} \quad \text{for} \quad i = 1,2. \tag{92}$$

If the interface remains planar, the equations (92) must be consistent. Otherwise system will demonstrate additional two-phase zone due to transformation of planar interface into mutually penetrating phases, e.g., “tongues”. Consistency means that one can equalize the right-hand sides of Eq. (92) for both components:

$$\frac{\bar{J}_1^+ - \bar{J}_1^-}{\bar{N}_1^+ - \bar{N}_1^-} = \frac{\bar{J}_2^+ - \bar{J}_2^-}{\bar{N}_2^+ - \bar{N}_2^-}, \tag{93}$$

where fluxes and the overall concentrations are given by Eq. (88) - (90):

$$\begin{aligned}
\bar{J}_1^- &= \varphi^- \tilde{J}_1^- + \left(1-\varphi^-\right)\frac{N_1^{\beta-}}{\Omega}\upsilon^{D-}, \quad \bar{J}_1^+ = \left(1-\varphi^+\right)\tilde{J}_1^+ + \varphi^+\frac{N_1^{\alpha+}}{\Omega}\upsilon^{D+}, \\
\bar{J}_2^- &= \varphi^- \tilde{J}_2^- + \left(1-\varphi^-\right)\frac{N_2^{\beta}}{\Omega}\upsilon^{D-}, \quad \bar{J}_2^+ = \left(1-\varphi^+\right)\tilde{J}_2^+ + \varphi^+\frac{N_2^{\alpha}}{\Omega}\upsilon^{D+},
\end{aligned} \tag{94}$$

$$\begin{aligned}
\bar{N}_1^- &= \varphi^- N_1^- + \left(1-\varphi^-\right)N_1^{\beta-}, \quad \bar{N}_1^+ = \varphi^+ N_1^{\alpha+} + \left(1-\varphi^+\right)N_1^+, \\
\bar{N}_2^- &= \varphi^- N_2^{\alpha} + \left(1-\varphi^-\right)N_2^{\beta}, \quad \bar{N}_2^+ = \varphi^+ N_2^{\alpha} + \left(1-\varphi^+\right)N_2^{\beta},
\end{aligned} \tag{95}$$

and

$$\begin{aligned}
\varphi^- &= 1 - \tfrac{4}{3}\pi\left(R^{\beta-}\right)^3 n^{\beta-}, \\
\varphi^+ &= \tfrac{4}{3}\pi\left(R^{\alpha+}\right)^3 n^{\alpha-}.
\end{aligned} \tag{96}$$

Equations (2) and (3) express the local equilibrium conditions between phases at different sides of the pseudo-interface as well as within phases at the same its sides. Thus, for two concentrations at the left side of the boundary plus for two concentrations for the right side of the same boundary (in total 4 variables) we have Eq. (93) and three constraints of quasi-equilibrium, Eqs. (2), (3). After substituting Eqs. (2), (3), (94) and (95) into Eq. (93) we have only one unknown variable, $N_1^- = N_1^-(t)$

$$N_1^- = \frac{\Omega\left(1-N_2^{\alpha}\right)\left[\varphi^-\tilde{J}_1^- - \left(1-\varphi^+\right)\tilde{J}_1^+\right]}{\varphi^+\upsilon^{D+} - \left(1-\varphi^-\right)\upsilon^{D-} - \Omega\left[\varphi^-\tilde{J}_2^- - \left(1-\varphi^+\right)\tilde{J}_2^+\right]}. \tag{97}$$

Equation (97) is the missing condition for conode choice at the moving interface.

## 3.4 Diffusion in two-phase zone in R3

To verify the dual-scale two-phase model, presented in section 3.3 the simulations were compared with the multi-phase multi-component model presented in section 3.1.

For all simulations following parameters were used: Phase boundary parameters: $k^{\alpha} = 0.3,\ k^{\alpha} = 0.7$ [at. %], T=1000 K, $ZE^{mix} = 2.92318 \cdot 10^{-20}$, the initial precipitates radius: $R_0^{\alpha} = 3.0 \ \ and \ \ R_0^{\beta} = 2.8\,[\mu m]$. Diffusivities are shown in Tab. 1.

All simulations were made for time=1 h, 2Λ=100 μm, numbers of nods=102, The different initial concentrations of the terminal alloys, are shown in following sections.

**Tab. 1. Diffusion coefficients in α and β phase.**

| | | Component | | |
|---|---|---|---|---|
| | | 1 | 2 | 3 |
| Diffusion coefficient [cm$^2$/s] | α- phase | 5.0e-10 | 2.0e-10 | 1.0e-10 |
| | β-phase | 3.0e-10 | 1.0e-10 | 0.9e-10 |

### 3.4.1 Type 0 interphase boundary in α+β|β+α diffusion couple

The initial concentration of terminal alloys are shown in Tab. 2. The other parameters resulting from chosen initial compositions and phase diagram: concentration of 1-st species in matrix and precipitates, initial concentration of precipitates and volume fraction of precipitates, are presented in Tab. 3.

The comparisons of results obtained by multi-multi model (section 3.1) and dual-scale two-phase models were presented in the Fig. 32 and Fig. 33, respectively concentration profiles and diffusion path. In Fig. 33 shift from equilibrium as “pseudo diffusion path” in matrix phase were additionally presented (green dots). The size and concentration of precipitates after process are present in Fig. 34.

**Tab. 2. Initial overal concentrations of terminal alloys.**

| | | Component |
|---|---|---|

| | | 1 | 2 | 3 |
|---|---|---|---|---|
| $\bar{N}_i$ $[at.\,\%]$ | Left Site | 0.60 | 0.35 | 0.05 |
| | Right Site | 0.05 | 0.65 | 0.30 |

**Tab. 3. Initial data for the dual-scale two-phase model.**

| Terminal alloy / Initial data | Left site, α-matrix and β-precipitates | Right site. β-matrix and α-precipitates |
|---|---|---|
| Molar ratio in matrix $N_1$ $[at.\,\%]$ | 64.6 | 42.9 |
| Molar ratio in precipitates $N_1^j$ $[at.\,\%]$ | 27.7 | 10.0 |
| Concentration of precipitates $n^j$ $[cm^{-3}]$ | $1.36 \cdot 10^9$ | $1.11 \cdot 10^9$ |
| Volume fraction of precipitates $\varphi^j$ $[vol.\,\%]$ | 12.5 | 12.5 |

The concentration obtained by booth models are in agreement Figs. 32 and 33, the main difference were observed close to the interphase boundary. Both models show jump along similar conodes however, when simulations made by M-M model shows one inward and one outward horn, (see section 3.2.1), then the dual-scale two-phase models results in two outwards horns. This effect results from the equilibrium assumed at the pseudo-interface. The concentration in matrix phase at pseudo-interface must be in equilibrium, (the point related to pseudo-interface are not showed on the diffusion path), therefore the diffusion path consist of two arc. The observed change of precipitates size are related to the shift from equilibrium in matrix phase, the supersaturation leads to growth of the precipitates. The concentration of the precipitates are affected only from the drift velocity.

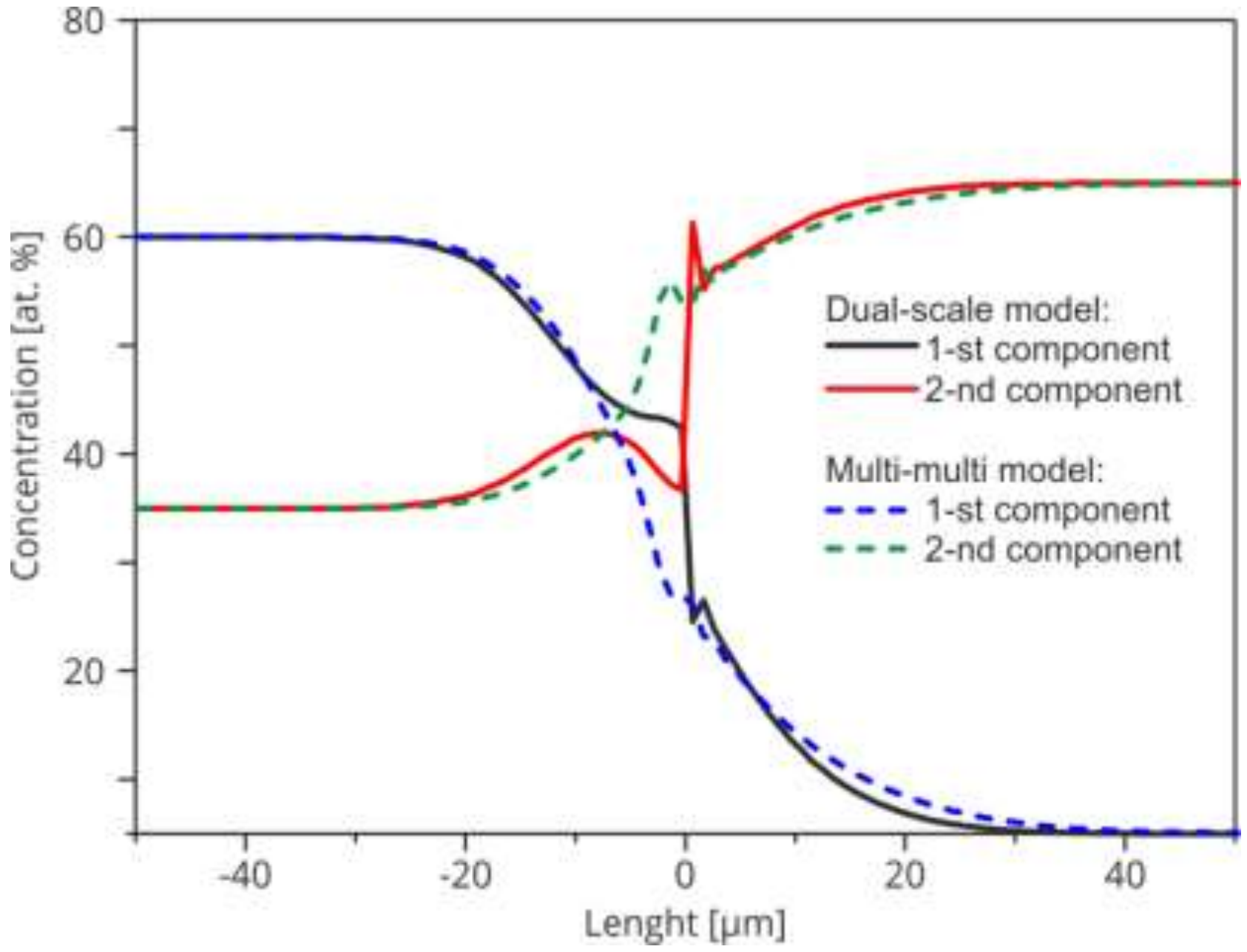

**Fig. 32. Concentration profile for initial composition presented in Tab. 2. Calculated by: a) dual-scale two-phase model (solid line), and; b) multi-phase multi-components model (dashed line).**

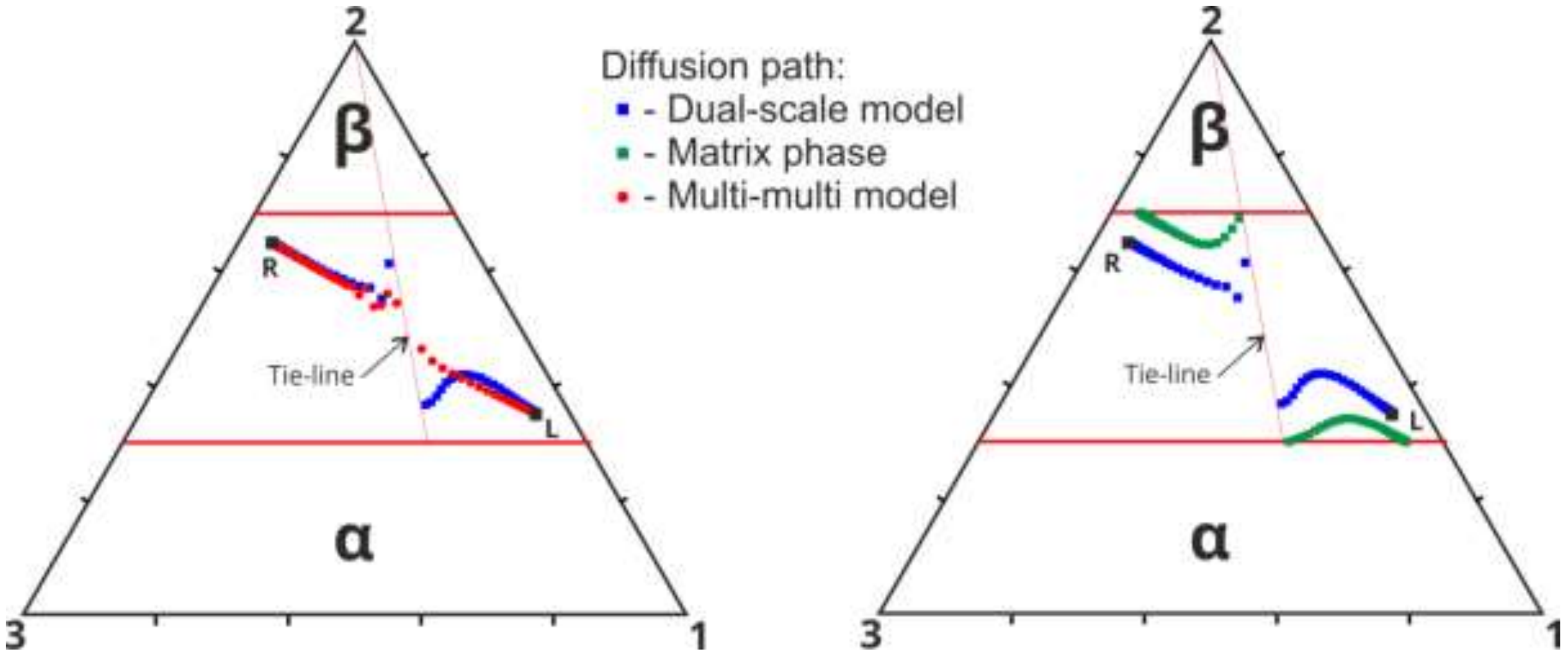

**Fig. 33. Diffusion path calculated by: dual-scale two-phase model (blue), and multi-phase multi-components model (red). On the right site deviation from the equilibrium in matrix phase (green) coresponding to the calculated diffusion path is shown. Results for initial concentrations in Tab. 2.**

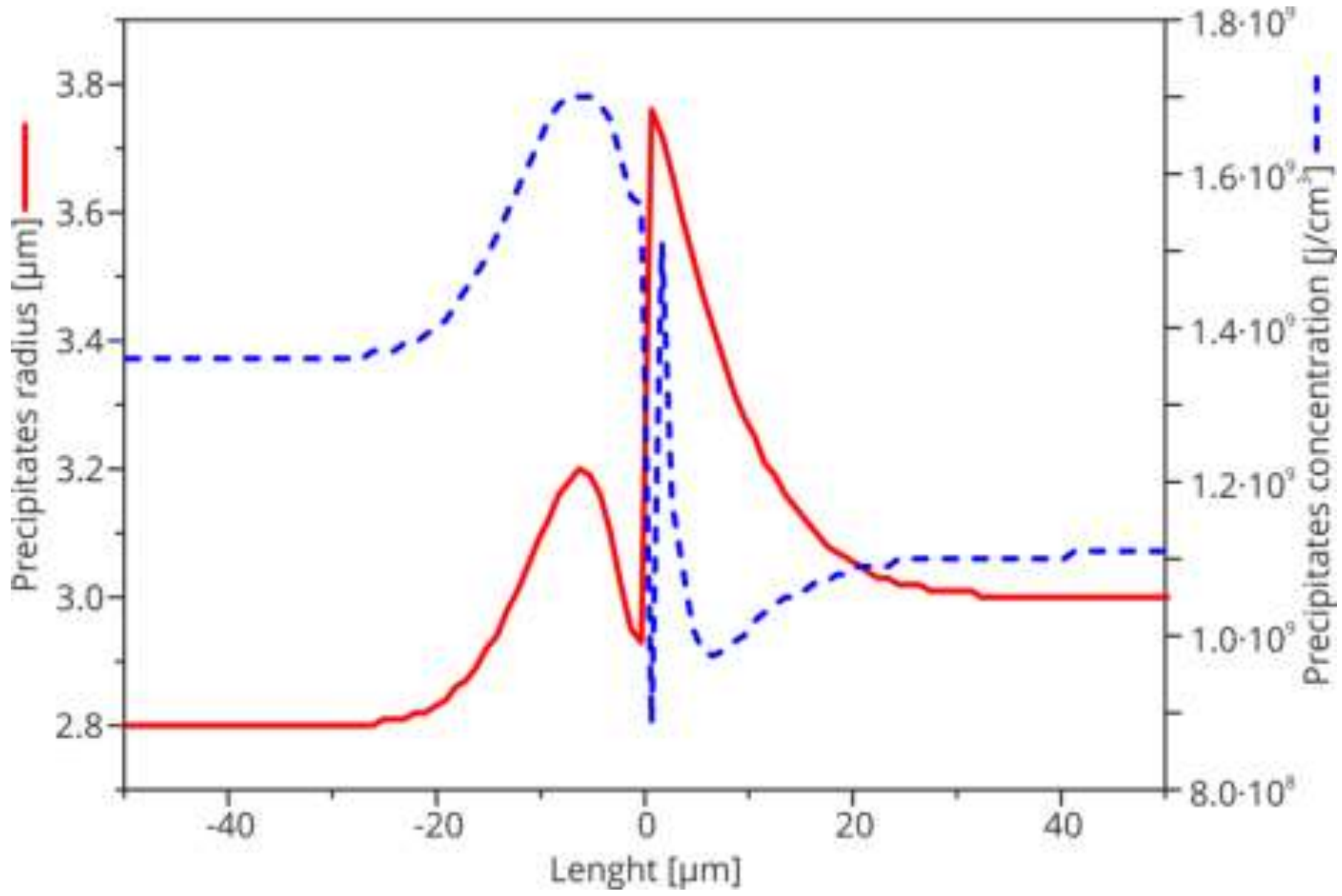

**Fig. 34. Radii (red) and concentration (blue) profiles in diffusion couple presented in Tab. 2.**

### 3.4.2 Type 1 interphase boundary in α+β|β+α diffusion couple

The initial overall concentrations are presented in Tab. 4. and Tab. 5. In the Fig. 35 the comparison of the concentration profiles obtained by muti-multi and the dual-scale two-phase models is presented. The comparison of diffusion paths is shown in Fig. 36. Figure 36 shows also the shift from equilibrium. Shift from equilibrium is shown as “pseudo diffusion path” in matrix phase, Fig. 36 (green dots). The size and concentration of precipitates after 1 hour are presented in Fig. 37.

**Tab. 4. Initial overal concentrations of terminal alloys.**

| | | Component | | |
|---|---|---|---|---|
| | | 1 | 2 | 3 |
| $\bar{N}_i$ $[at. \%]$ | Left Site | 0.05 | 0.35 | 0.60 |
| | Right Site | 0.30 | 0.65 | 0.05 |

**Tab. 5. Initial data for dual-scale two-phase model.**

| Initial data \ Terminal alloy | Left site, α-matrix and β-precipitates | Right site. β-matrix and α-precipitates |
|---|---|---|
| Molar ratio in matrix $N_1\ [at.\ \%]$ | 5.4 | 25.7 |
| Molar ratio in precipitates $N_1^j\ [at.\ \%]$ | 2.3 | 60.0 |
| Concentration of precipitates $n^j\ [cm^{-3}]$ | $1.36 \cdot 10^9$ | $1.11 \cdot 10^9$ |
| Volume fraction of precipitates $\varphi^j\ [vol.\ \%]$ | 12.5 | 12.5 |

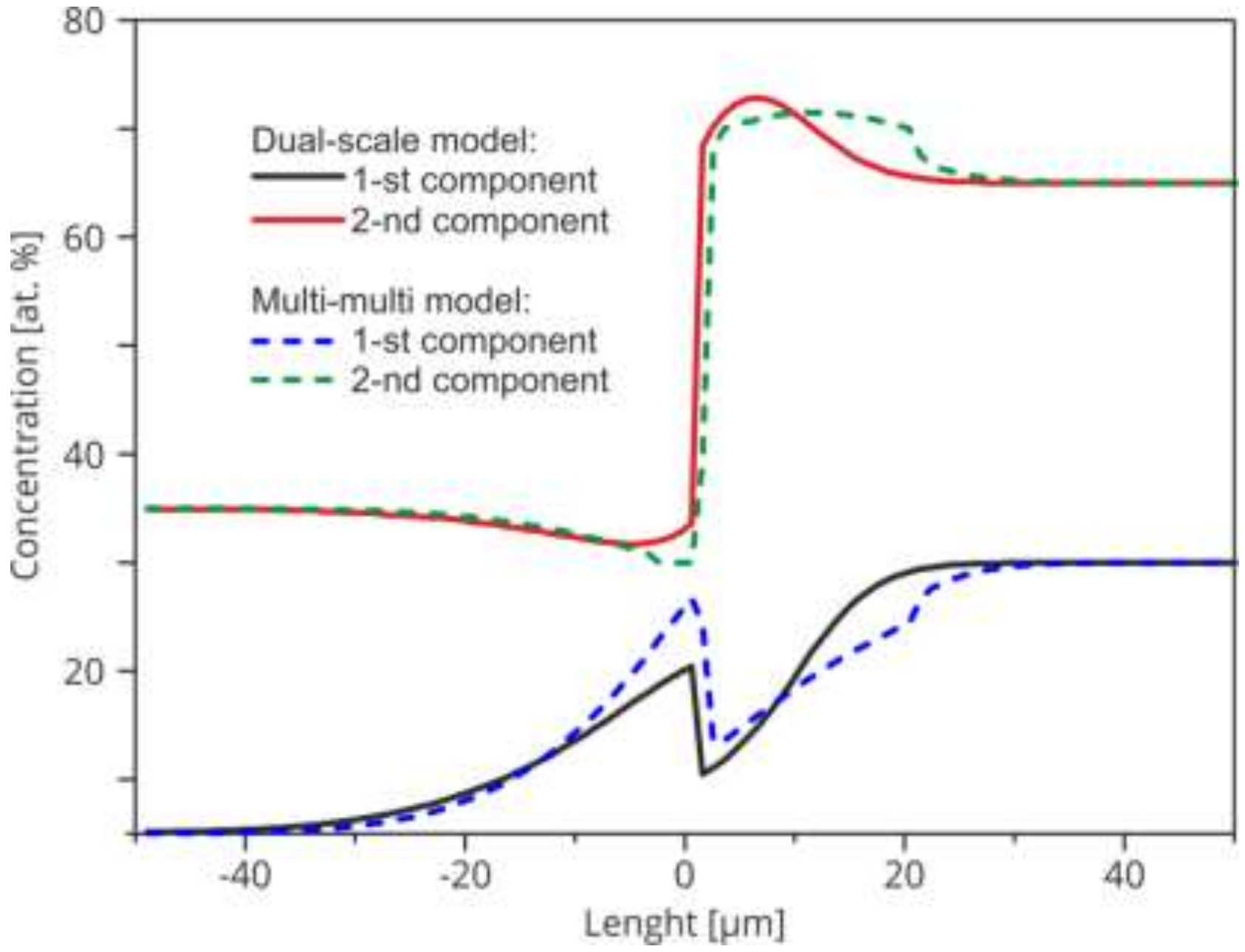

**Fig. 35. Concentration profile for initial composition presented in Tab. 4. Calculated by: a) dual-scale two-phase model (solid line), and; b) multi-phase multi-components model (dashed line).**

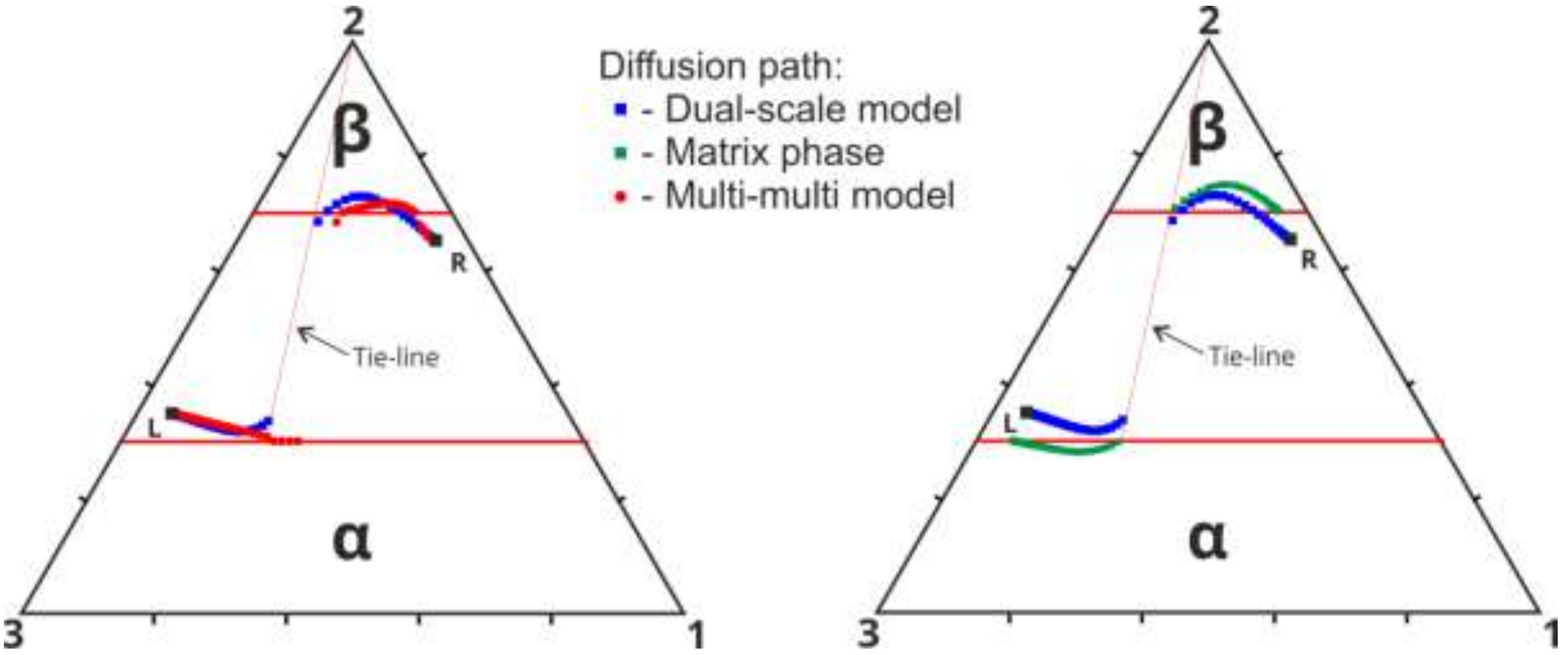

**Fig. 36. Diffusion path calculated by: dual-scale two-phase model (blue), and multi-phase multi-components model (red). On the right site defiation from the equilibrium in matrix phase (green) coresponding to the calculated diffusion path is shown. Results for initial concentrations in Tab. 4.**

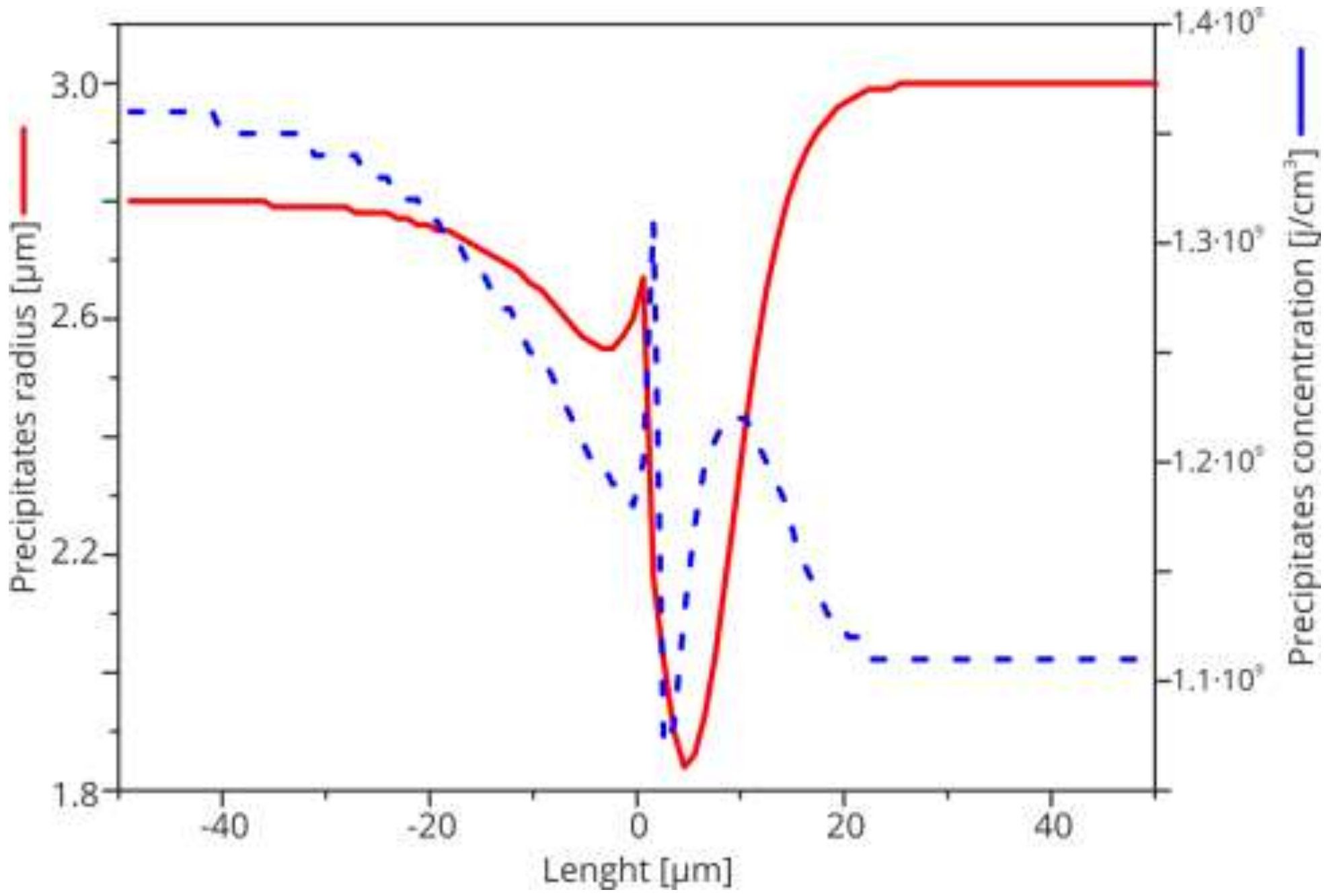

**Fig. 37. Radii (red) and concentration (blue) profiles in diffusion couple presented in Tab. 4.**

The dual-scale two-phase model considers only one interphase boundary (pseudo-interface. i.e., type 0 IB). The results obtained by booth models are in quantitative agreement, Fig. 36. The interphase boundaries go along conodes being in close proximity and the single-phase region is seen. The observed change of precipitates size is related to the shift from equilibrium in matrix phase. The undersaturation leads to shrinking of the precipitates, especially when overall concentration enters the single-phase zone. This situation can be seen on the right site of the pseudo-interface where the overall concentration enters single-phase zone and after extended time the precipitates will vanished completely [80].

## 3.5 Summary

In previous sections two models describing diffusion in multiphase system were presented. The models differ in the assumptions and the range of applications. The dual-scale two-phase model allows to calculate microstructural evolution, however can be applied only to α+β|α+β diffusion couple. The multi-multi model is less restricted by initial assumptions therefore can be applied in wide range of application, however at the cost of lower number of information. In common range of application, the models are consistent Fig. 33 and Fig. 36.

The multi-multi model was used to examine the Morral diffusion theorems A3-A9 [3]. The following conclusion can be formulated:

- the thermodynamic equilibrium can be achieved, even when the gradients of overall concentrations and phase volume fractions have not vanished;
- contrary to the Morral theorem, the presented simulations show that the type 0 IBs can move;
- the process of diffusion in two-phase zone is strongly affected by the shape of the phase boundary lines (the thermodynamics equilibrium parameters between phases);
- the diffusion path may enter the single-phase region when single-phase region grow between two two-phase regions, ie. α+β|β|α+β (section 3.2.3);
- the concentration jump on type 1a IB can be very small;
- the behavior of diffusion path at the type 1b IB when diffusion path is tangent to the phase boundary line in single-phase region can be explained. It a result of equal concentration gradients at both sides of the type 1b IB;

- the singularities of the entropy production rate are the consequence of the numerical error caused by discontinuity of the concentration at the interphase boundaries.

Presented multi-multi model does not allow to simulate diffusion in the systems with more than two phases. This limitation and the occurrence of numerical errors requires further development. These limitation and imperfections indicate the ways of the further development:

- extension the model to three and more phases;
- implementation of the equilibrium parameters in the three-phase region (tie-triangle);
- the opening of the borders of the diffusion couple, for applications like: carburizing, nitriding, and other;
- taking into account the concentration dependent diffusion coefficients;
- integration of model with the thermodynamic database to calculate the chemical potentials as a true driving force for diffusion.

The dual-scale two-phase model differs from the classical description of the diffusion in two-phase zone. It considers non-equilibrium conditions in the diffusion zone and allows to predict the changes in microstructure during diffusion. The obtained results are in quantitative agreement with M-M models, however the dual-scale two-phase model allow to obtain two horns with the same direction Figs. 33 and 36. This results required experimental proof.

# Chapter 4. Experimental investigations of diffusion in Ni-Cr-Al system

In the following sections I will summarize the experimental results of my studies on the diffusion in the ternary Ni-Cr-Al system:

- characteristic of the metals/alloys used in the experiment;
- method of preparation of diffusion multiples;
- a method used to measure concentration profiles in multi-phase regions;
- results of the concentration measurements;
- summary of the experimental results.

## 4.1 Experimental procedures

### 4.1.1 Materials

In Fig. 38. an isothermal section of Ni-rich corner of the Ni-Cr-Al phase diagram is shown for the temperature 1200°C. The diagram is drawn after [23] and presents the results calculated by Calphad. Single, two and three-phase fields, separated by phase boundaries are indicated in the drawing. Red points in the diagram represent the initial compositions of the terminal alloys used in the experiments, see Tab. 6.

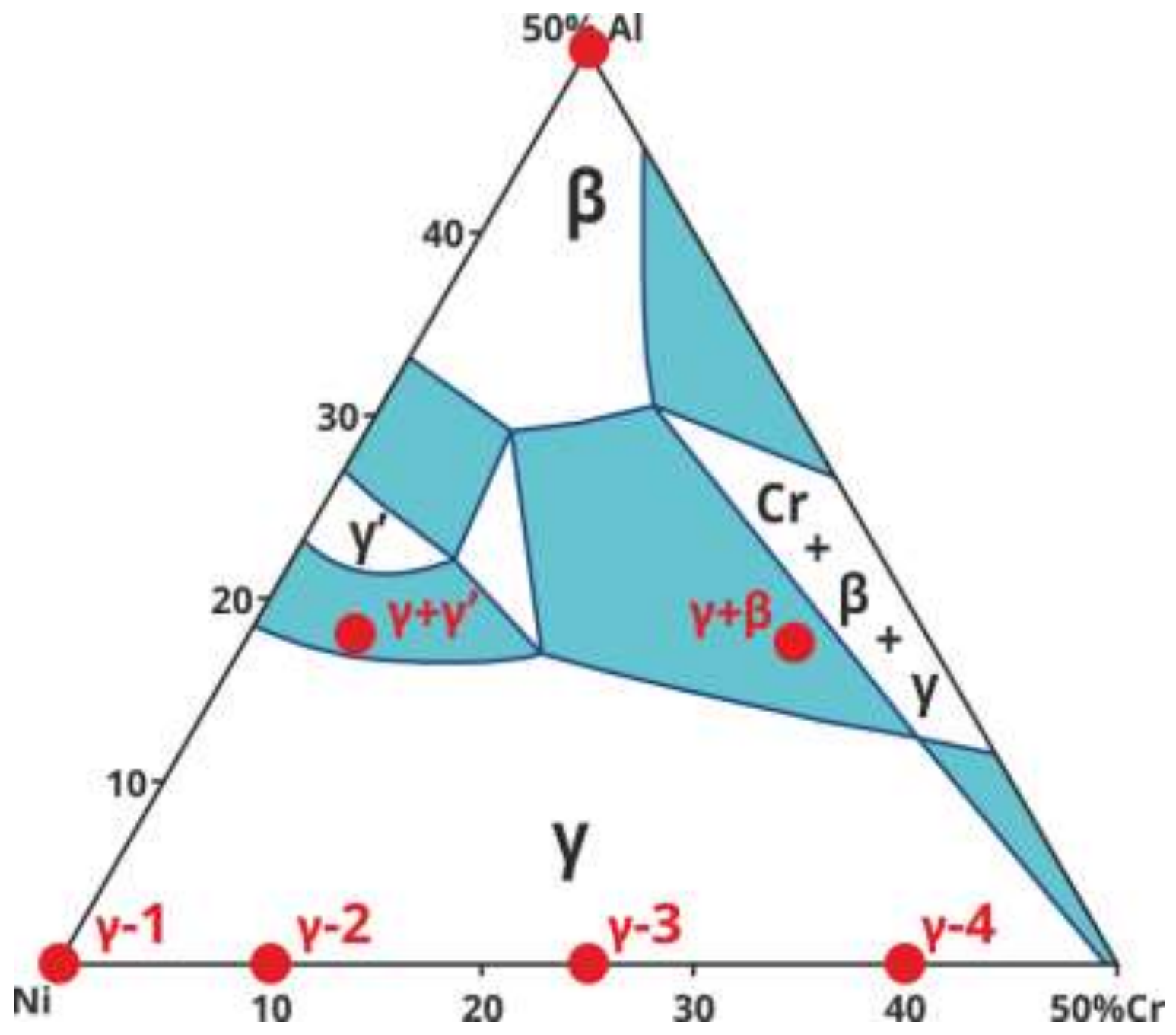

**Fig. 38. Isothermal section of the Ni-rich corner of the Ni-Cr-Al system for the temperature 1200°C and pressure 1 atm [23]. Red point relate to the compositions of the materials used in the experiments.**

The materials used in the experiment were provided by GoodFellow Inc. and MaTeck GMBH in the form of cylindrical ingots with the diameter of 10 mm. The GoodFellow materials had the purity of 99.99 % and those provided by MaTeck 99,999%. The chemical composition of the alloys used to prepare diffusion couples are specified in Tab. 6. I am presenting the data specified by the manufacturers next to the phase composition of the alloys at 1200°C – as predicted from the phase diagram.

**Tab. 6. The compositions of the alloys used in the studies of the diffusion in the ternary Ni-Cr-Al system.**

| Alloy symbol-Phase composition | Chemical composition (by the manufacturers) [% at.] | | | Manufacturer |
|---|---|---|---|---|
| | Ni | Cr | Al | |
| β | 50 | --- | 50 | MaTecK |
| γ-1 | 100 | --- | --- | GoodFellow |
| γ-2 | 90 | 10 | --- | |
| γ-3 | 75 | 25 | --- | |
| γ-4 | 60 | 40 | --- | |
| γ+γ' | 77 | 5 | 18 | |
| γ+β | 56.5 | 26 | 17.5 | |

All materials were first characterized in terms of microstructure and chemical composition, using scanning electron microscope (SEM) equipped with energy dispersive spectrometer (EDS). The results confirm that all single-phase alloys are continuous, non-porous, without precipitates. The all used alloys have average chemical composition close to that specified by manufacturer and the phase composition consistent with phase diagram.

Polycrystalline microstructure has been revealed in the case of β-NiAl alloy, which has microstructure of large-size grains, between 2 and 3 mm in the diameter. Two-phase alloys were characterized as delivered and also after annealing at 1200°C. A low-magnitude microstructure of non-annealed γ+γ′ alloy is shown in Fig. 39. The image reveals typical casting structure with longitudinal areas arranged radially from the middle of the ingot axis. Different shadow of the areas is due to various grain orientation. Grain microstructure was revealed only at higher magnifications.

The two-phase alloys were annealed at 1200°C to observe microstructure evolution due to annealing, to avoid possible confusing with diffusion effects. The SEM images made for γ+γ′ and γ+β alloys annealed for 1 h and 26 h are presented in Fig. 40, next to the images of the alloys before annealing. Only the effects of grain growth are seen. Other differences in the microstructure can be due to various conditions of sample cooling. It is seen that no transformation took place during annealing of the alloys. For further experiments I decided to use unannealed samples to ensure conditions when the sizes of precipitates is comparable with area of diffusion zone.

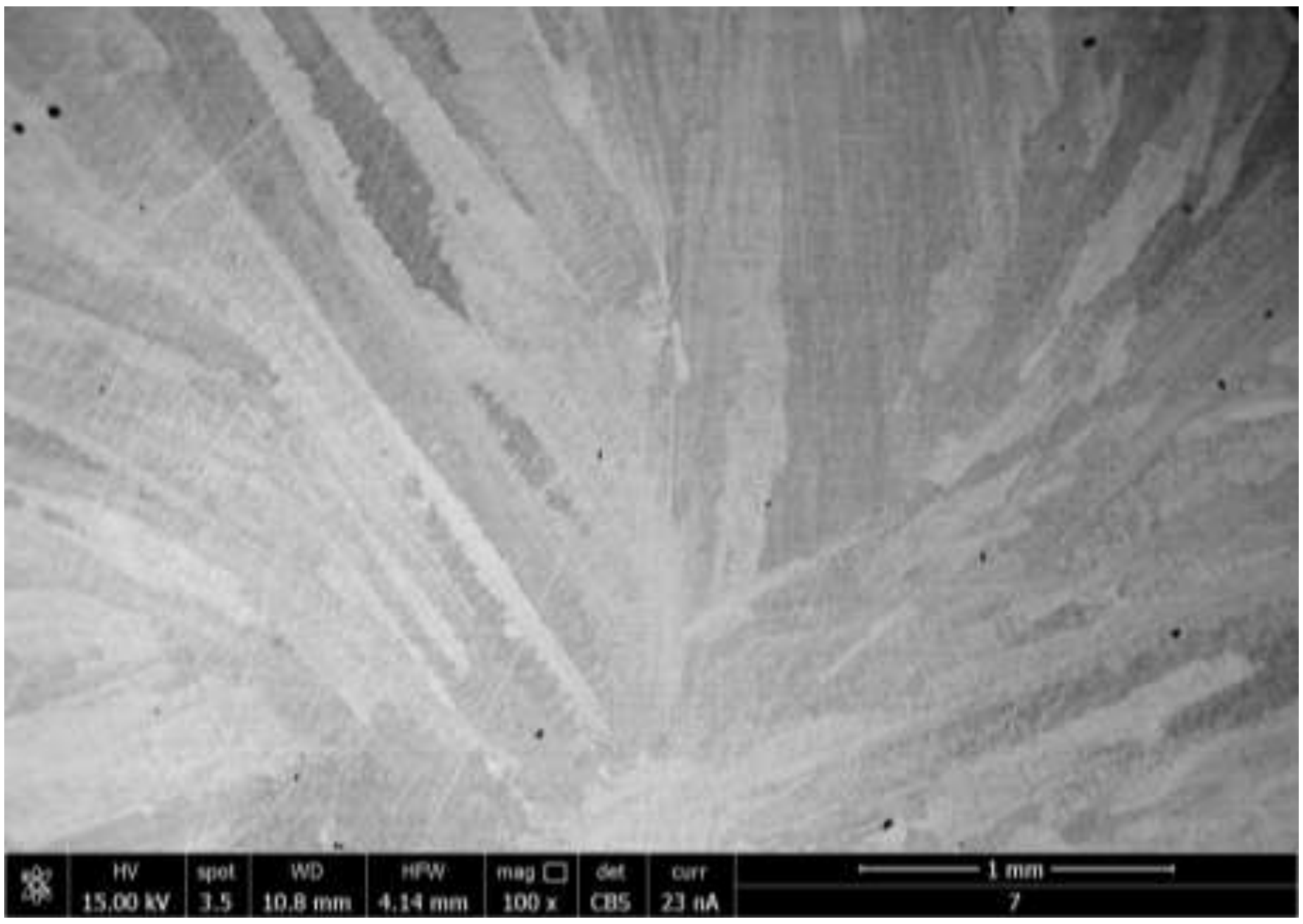

**Fig. 39. Microstructure of γ+γ′ alloy, low magnification.**

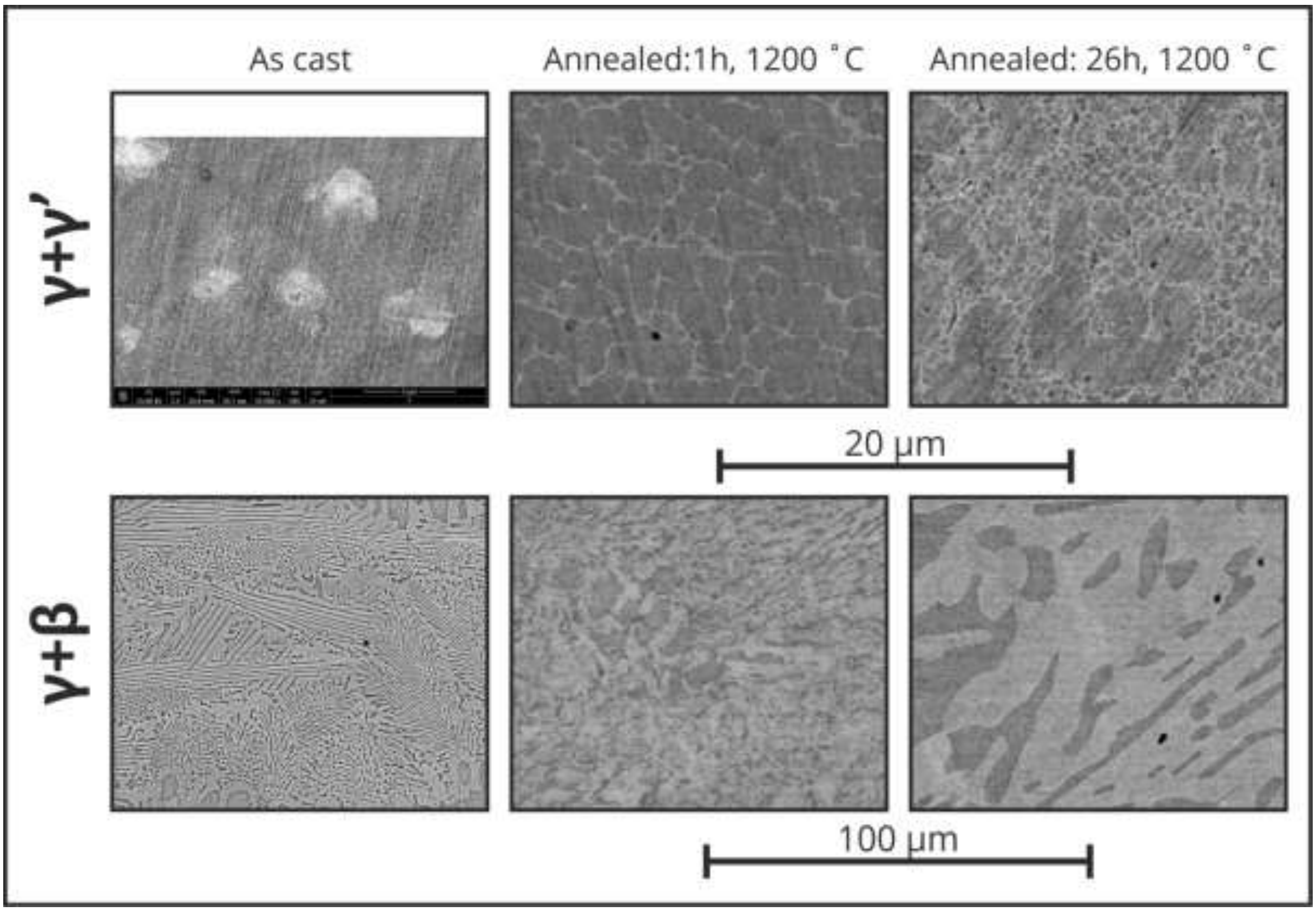

**Fig. 40. Microstructure of two-phase alloys: as delivered and after annealing for 1 and 26 h. Top – alloy γ+γ′, bottom – alloy γ+β.**

### 4.1.2 Preparation of the diffusion multiples

In the experiments dedicated to study interdiffusion in the ternary Ni-Cr-Al system I used diffusion multiples, prepared according to the method presented by Zhao in 2001 [87]. The method provides a lot of data from one experiment and can be used to investigate phase diagrams, precipitate-growth kinetics and interdiffusion [88-97]. The method, although time consuming in the initial stage of preparation, allows studying several diffusion couples in one experiment.

An advantage of the method is a possibility to obtain diffusion joining under low partial oxygen pressure, lower than when using protective atmosphere. It was especially beneficial in my experiments with the alloys containing aluminum, which oxidizes very easily even in helium protective atmosphere. Oxidation of aluminum could interfere diffusion.

A procedure which I used to prepare the Ni-Cr-Al multiples for diffusion studies is shown in Fig. 41. First, the alloys' rods were cut into bars 20 mm long and into the 1/4 or 1/2 sections of the circle. Using precise wire saw with SiC microabrasive I achieved dimensional accuracy within 50 μm. The resulting surfaces were further polished and washed with ethyl alcohol in ultrasonic washer. I put together the prepared bars so that they arrange full cylinder which I placed in the cylindrical container made of nickel alloy 200/201 with nickel content 99.0 %. The external diameter of the container was 16 mm, internal – 10 mm.

In total, I prepared three multiples, as presented in Fig. 42. The containers with the samples were vacuum welded with electron beam, (Tele & Radio Research Institute) under the pressure $2x10^{-5}$ mbar and then subjected to 1 hour hot isostatic pressing (HIP) at 1200°C, under the pressure 200 MPa, using Hot Isostatic Press EPSI furnace in the laboratory of Department of Ceramics and Refractories, WIMiC AGH. The final pressure was achieved in two stages. First, the pressure about 118 MPa was achieved at 600°C, with a use of high-pressure compressor and nitrogen as a medium bearing the pressure. Then the gas flow was stopped and the pressure was being increased only by raising the temperature with the rate 10 deg/min, up to 1200°C. After the process the samples were cooled with furnace. The pressed samples were cut into slices about 3 mm thick and then they were abraded to remove cut-offs. The cutting was made by use of electrical discharge machining (EDM), in the Institute of Advanced Manufacturing Technology in Kraków.

In total I obtained 12 multiples, i.e. three sets of different multiples, four slices for each type of the multiple. The schemes and compositions of the prepared multiples are presented in Fig. 42 and in Tab. 6.

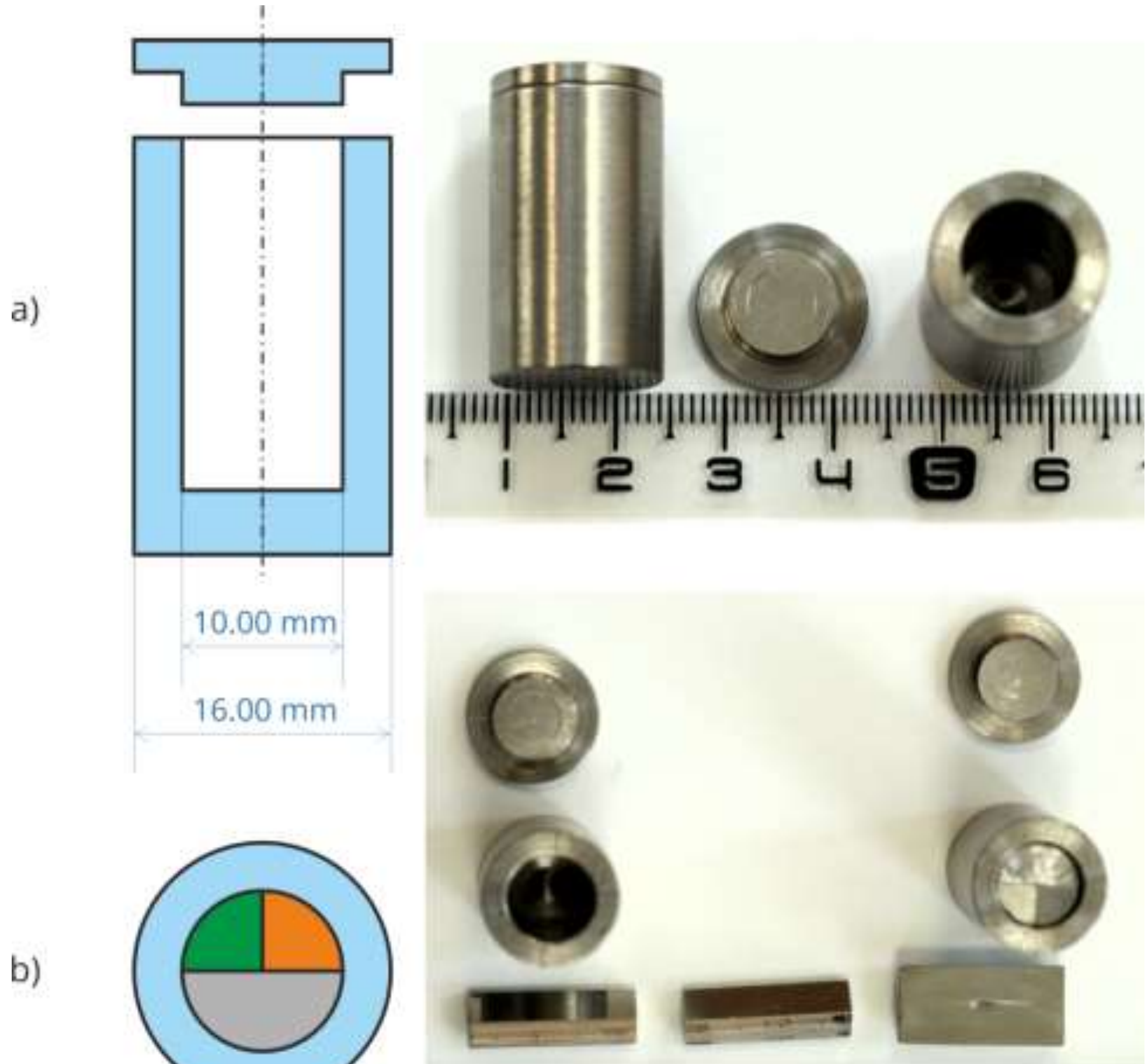

**Fig. 41. Preparation of the diffusion multiples: a) Nickel container, scheme and photo; b) Distribution of materials in the container, scheme and photo (empty container, container with a batch, exemplary bars.**

SEM observation of the multiples confirmed that the applied procedure allowed obtaining good connection between alloys. Deformation of alloys, seen in Fig. 43, is due to not perfect adaptation of the dimensions of the joined bars and the differences in their mechanical and thermal properties. The cracks occurred only in the centre of the multiple. Away from the centre the pores (~100 nm in diameter) were present however they do not lower quality of the diffusion joint. Which is important, I observed neither aluminum nor chromium oxides, which could disturb diffusion.

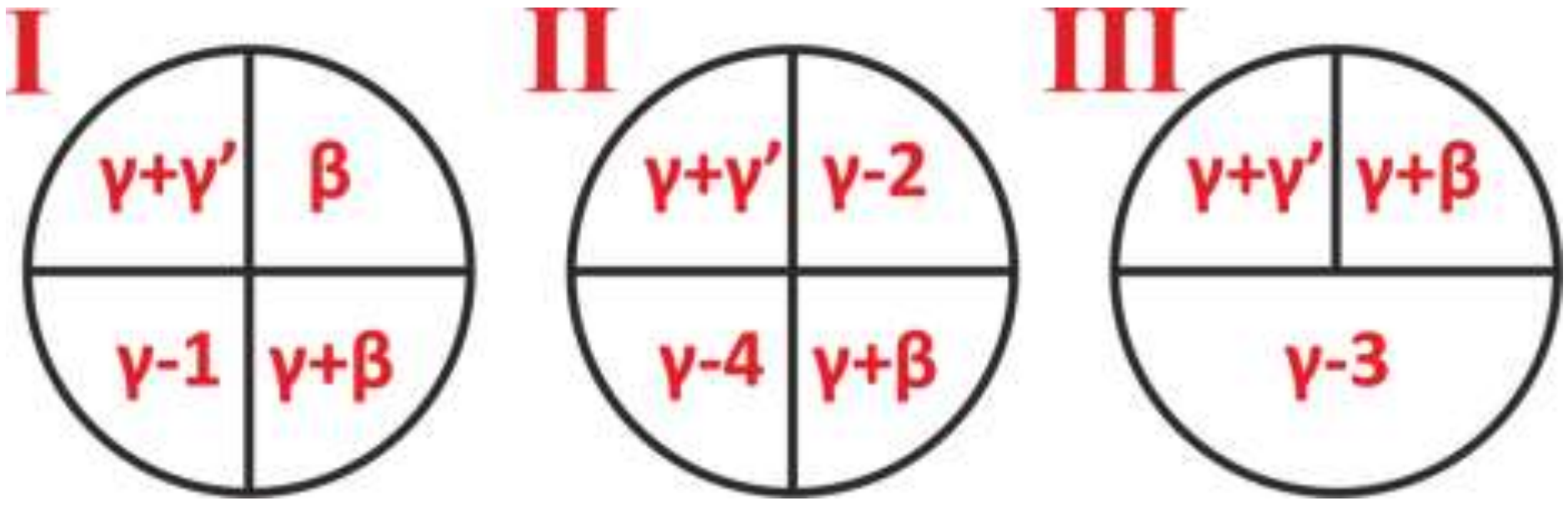

**Fig. 42. The scheme of the prepared multiples.**

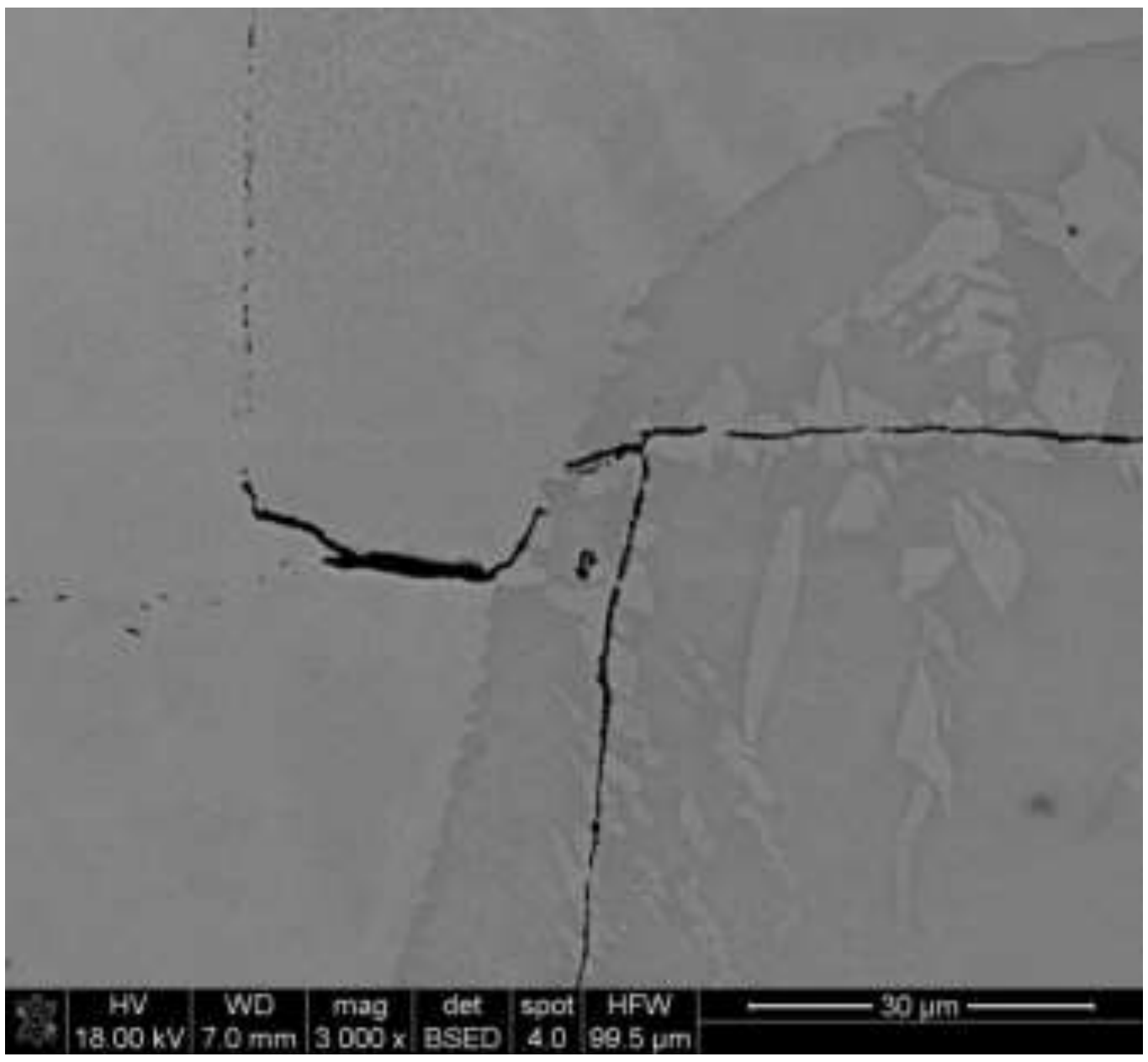

**Fig. 43. Central area of the multiple, subjected to HIP process at 1200°C for 1 hour, Multiple I in Fig. 42.**

To illustrate the effects of diffusion close to the centre of multiples, I made the respective 2D concentration maps. The maps were made from the central areas of the multiples, each of the dimensions 300 x 270 μm, covering central joins of all alloys, Fig. 42. The measurement was made with the resolution of 1024 x 921 point. First, the maps were averages for the squares 10 x 10 point. The noise was removed using Savitzky-Golay filter [98]. The results are presented in Fig. 44 as 2D concentration maps. It is seen that some of the isoconcentration lines overlap with the boundaries visible on SEM images and the range of 2D diffusion does not exceed 100 μm. Therefore the effects of 2D diffusion should not coincidence with the diffusion in one dimension. Irregular shape of isoconcentration lines and the presence of the loops are due to two-phase microstructure of the alloy, especially within γ+β region.

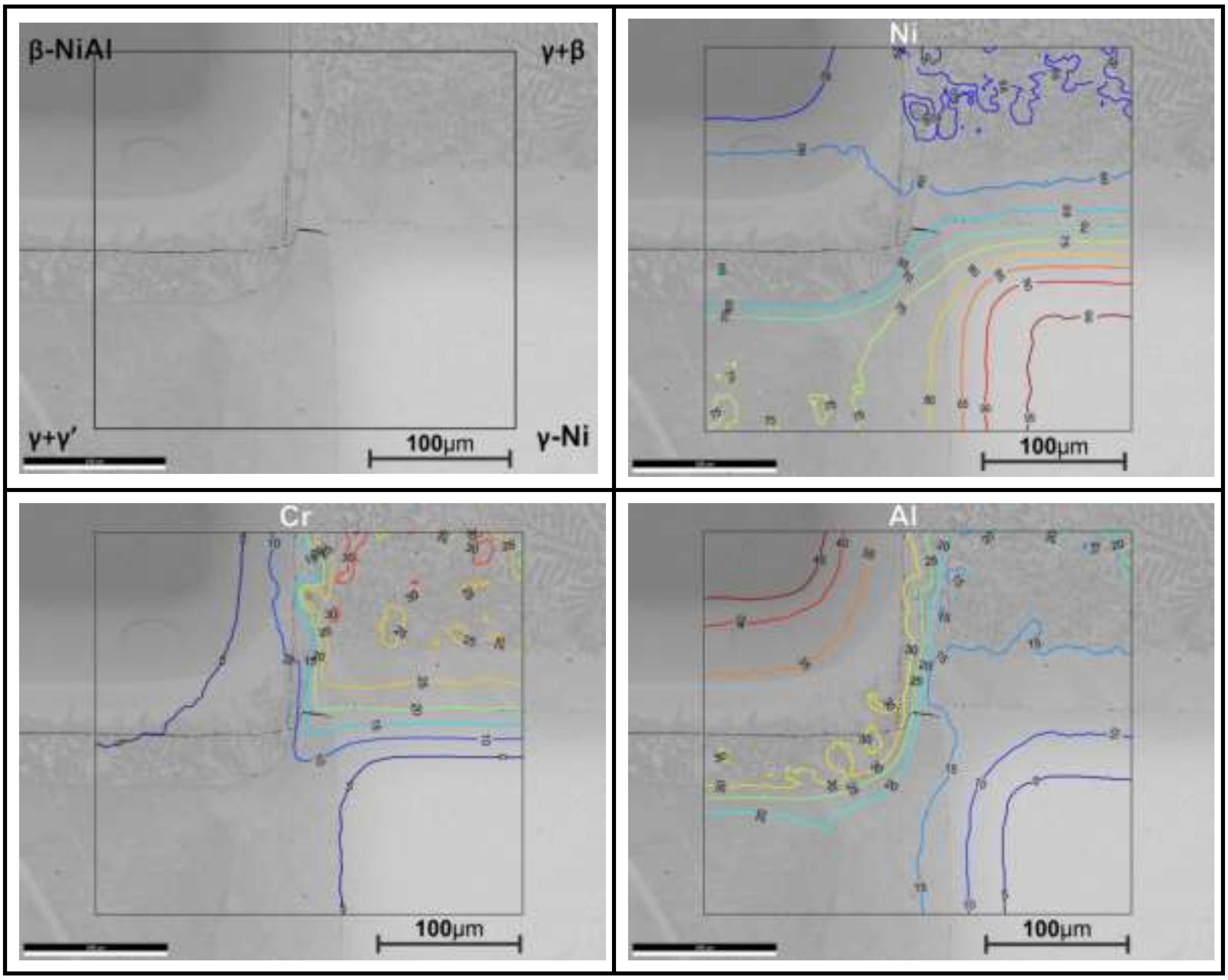

**Fig. 44. Isoconcentration 2D maps for the central area of the muiltple, annealed at 1200°C for 1 hour. Multiple I in Fig. 42.**

The samples dedicated to annealing by 26 h were vacuum closed in quartz ampoules and placed in chamber furnace, for 25 hours at 1200±5°C. After this time the samples were quenched in the water and then subjected to further examination.

### 4.1.3 Measurement method and the treatment of diffusion concentration profiles

Diffusion multiples were polished with polishing MD-materials, Struers. To eliminate an influence of corrosion or other surface phenomena on diffusion I removed, from each multiple, a layer about 1 mm thick using the sequence of grinding wheels MD-Piano 80, MD-Piano 220 and MD-Allegro. Next I polished the surface with MD-Mol, MD-Nap and MD-Chem cloths using the diamond suspensions of the granulation 3 and 1 μm, and as last step $SiO_2$ OP-S suspension of the grain size 0.25 μm was used. The samples were subjected to washing in ultrasonic washer using ethylene alcohol after polishing and directly before measurements.

SEM micrographs and EDS analysis of concentration profiles of the samples were made, using Scanning Electron Microscope FEI VERSA 3D equipped with field emission electron gun FEG and energy dispersive spectrometer EDS used in chemical composition analysis ( Academic Centre for Materials and Nanotechnology ACMiN in Kraków).

The key target of present studies was to determine concentration profiles in two-phase zone. The standard linear analysis, in which electron beam is moved along single straight line is not appropriate due to unavoidable scatter of the results. Within two-phase zone the electron beam, moving from one point to another, collects data from the two phases and the grain boundaries, Fig. 45. Obviously, one can eliminate the scattering by statistical analysis, but it would require multiple repetition of the measurements. A better method to ensure high quality measurements is offered by TEAM™ EDS Advanced Linescan, by EDAX Software. In the method, the electron beam moves perpendicularly to the base line and thanks to this, the data are collected from a given width band covering the area on both sides of base line (Fig 46 and 47). The collected data are averaged and each measure point represents the concentration averaged over the rectangle of the preset dimensions. Therefore the technique is called here the wide line EDS method (WL-EDS).

In the present measurements the single data point, represents the data collected from the rectangle 200 μm "wide" and approximately 1.5 μm thick (approximate spatial resolution of EDS). The wide line method does not allow total elimination of concentration scattering in two-phase zone and often increases the typical EDS-analysis noise in single-phase zone. Thus, I filtered the obtained data using Savitzky-Golay method [98].

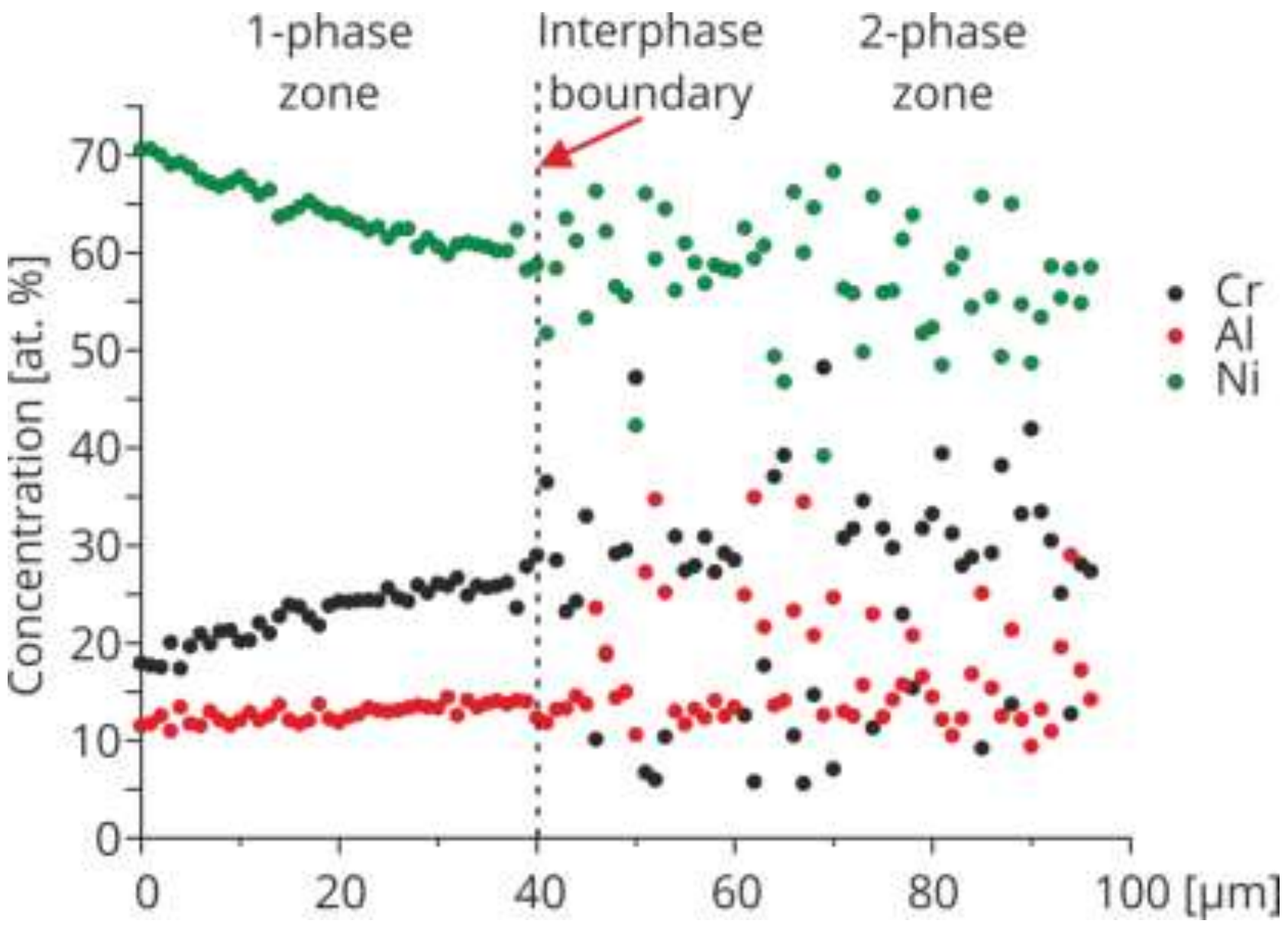

**Fig. 45. Concentration profiles measured using standard linear EDS analysis. Visible scattering of data within two-phase zone and noise in single-phase zone.**

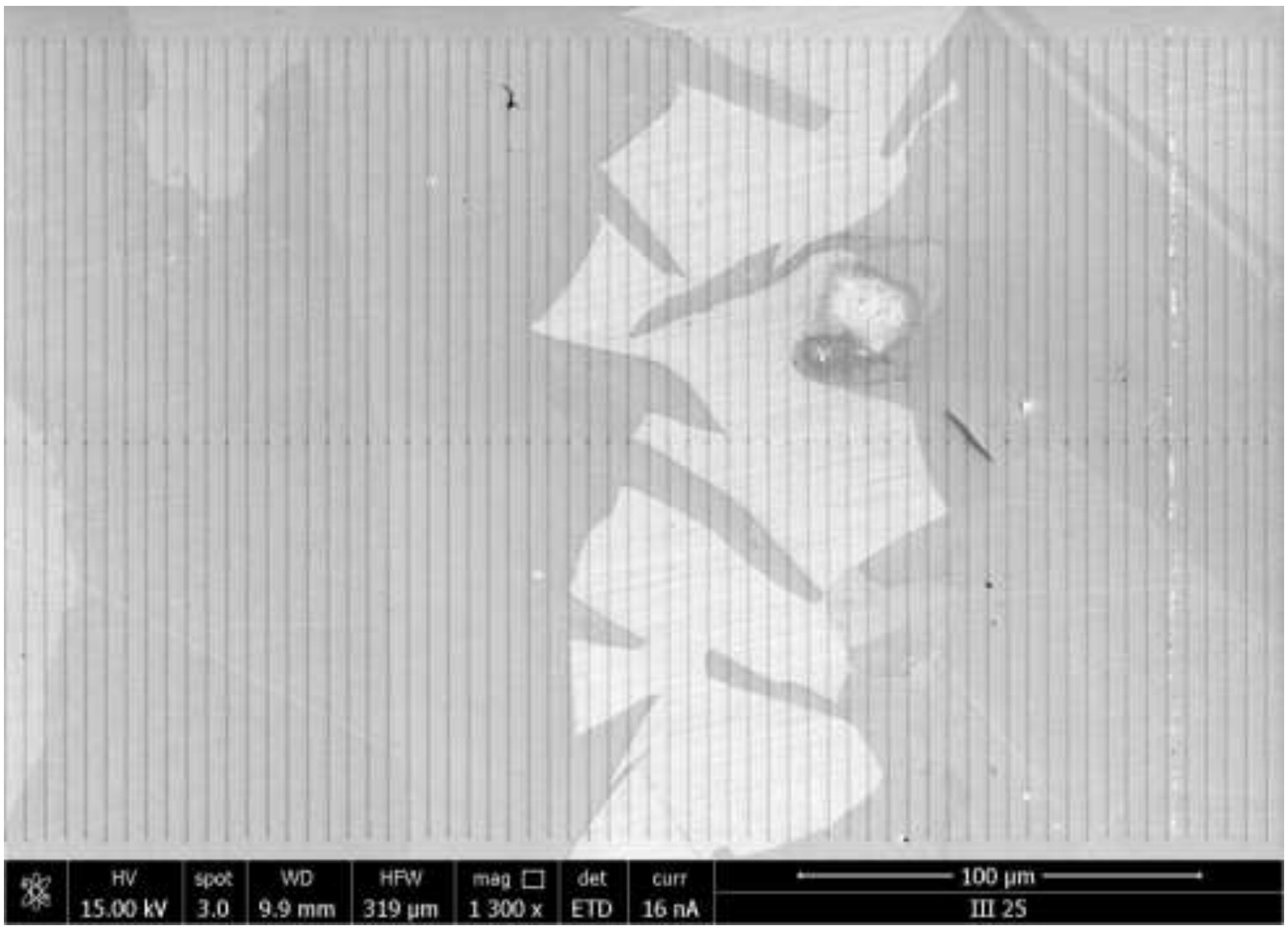

**Fig. 46. Trace of the electron beam in wide-line metod, recorded by the secondory electron mode. The width of the trace is ~200μm, measuring step 5μm. The image of (γ+γ')|(γ+β) couple, subjected to annealing at 1200°C, for 26 h. The concentration from more than 3·10$^4$ points are collected.**

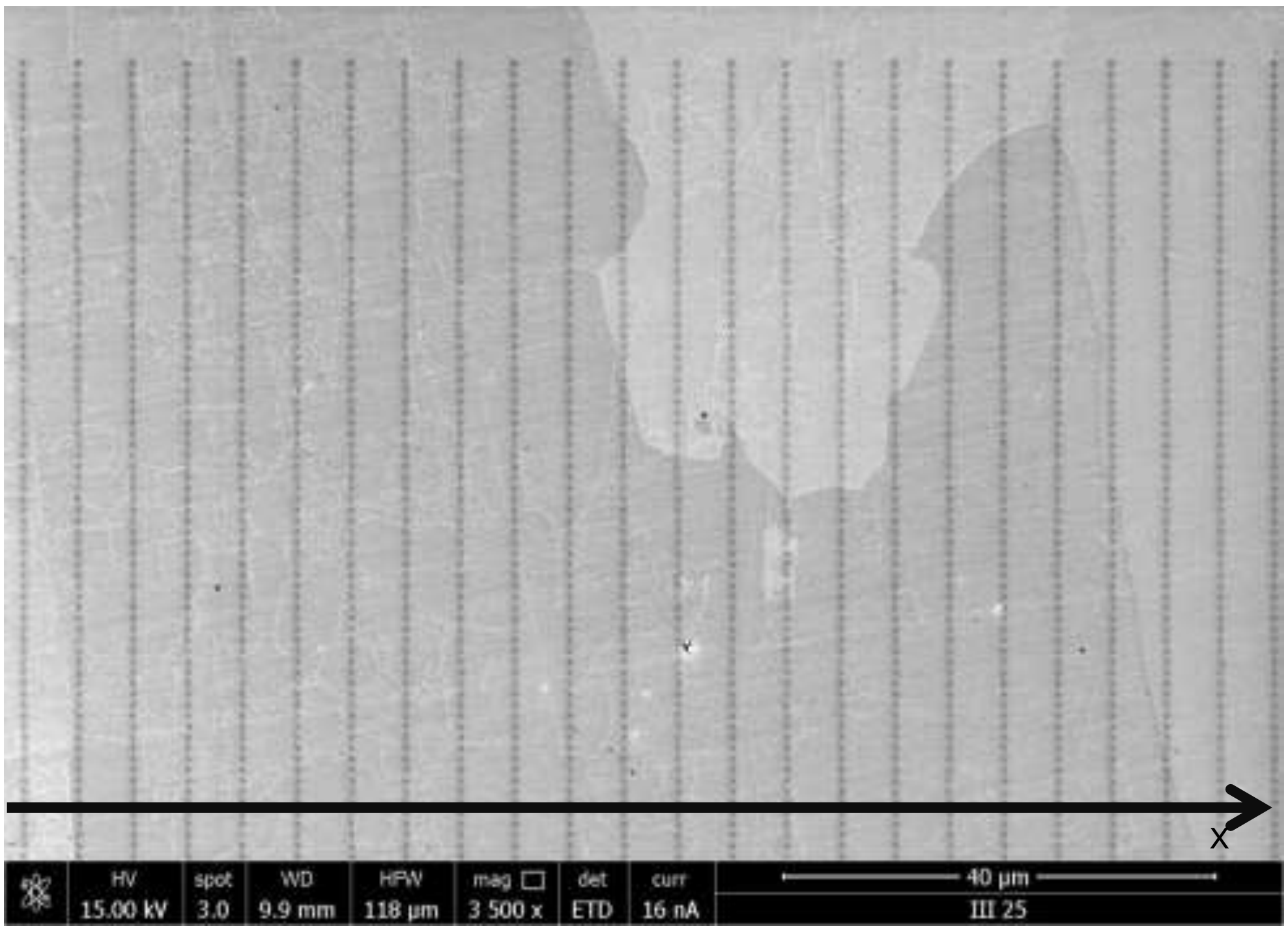

**Fig. 47. Trace of EDS analysis recorded by the secondory electron mode. Visible measurement points, $3\cdot10^4$ in total.**

The EDS measurements necessary to collect WL-EDS data are time-consuming. To make them more effective, without reducing quality of the measurement I compare the concentration profiles in (γ+γ')|(γ+β) couple obtained using various distances between electron-beam tracks (step size). The 4 sets of data, each presenting the analysis of different steps size: 0.1, 0.5, 1 and 2μm were obtained. Again, the Savitzky-Golay filter was used to reduce the remaining noise [98]. The filtering was made within the window 4 μm wide. The results are shown in Fig. 48. On the left the data before filtering are shown; on the right after filtering. The filtered data were further used to draw diffusion paths, Fig. 49.

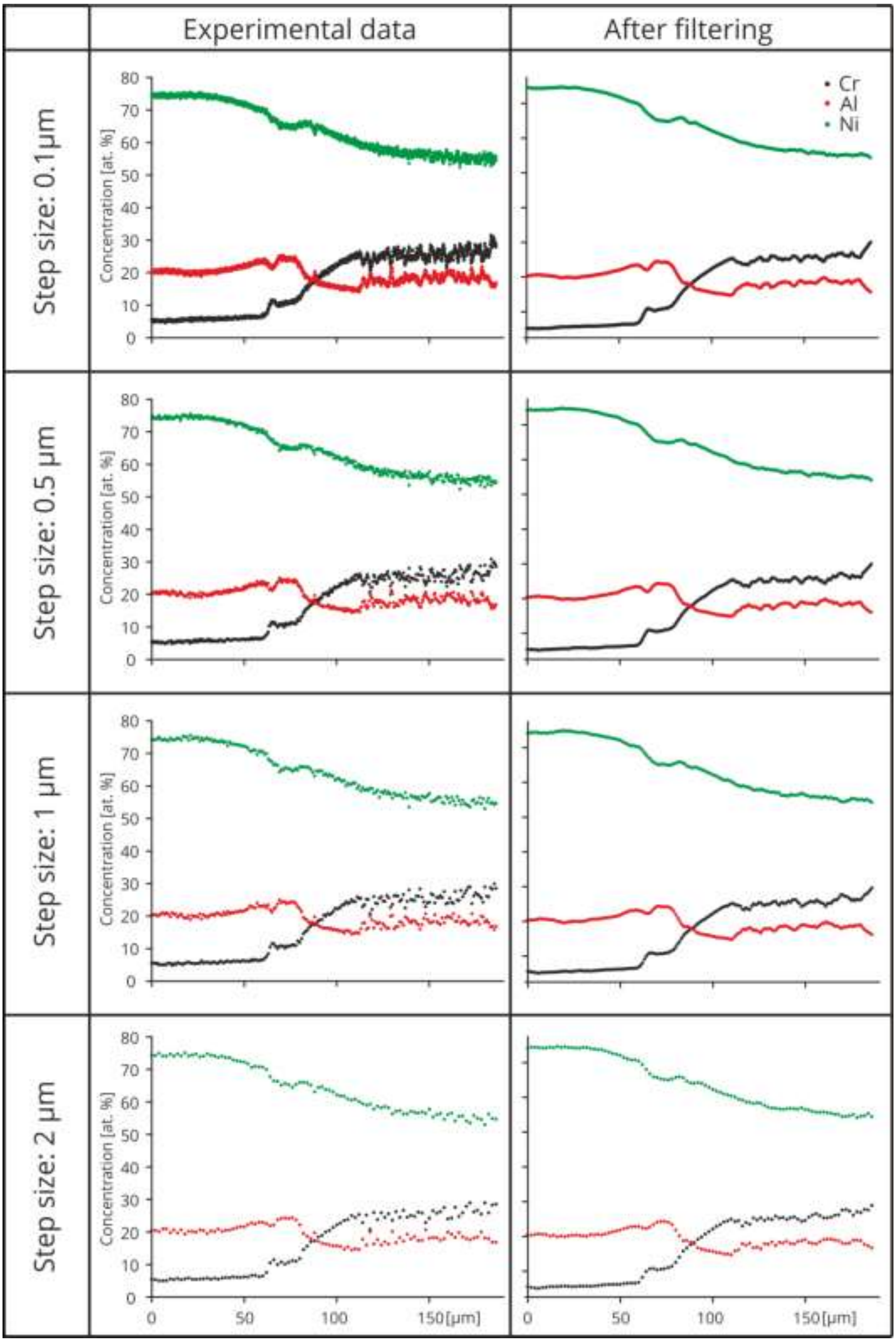

**Fig. 48. Comparison of concentration profiles for (γ+γ')|(γ+β) diffusion couple after annealing for 1 h determined by wide-line EDS analysis performed with different steps: 0.1, 0.5, 1 and 2 µm. Left column: the profiles not subjected to noise reducing; right column: the profiles subjected to filtering by Savicky-Golay filter with 4µm window [98].**

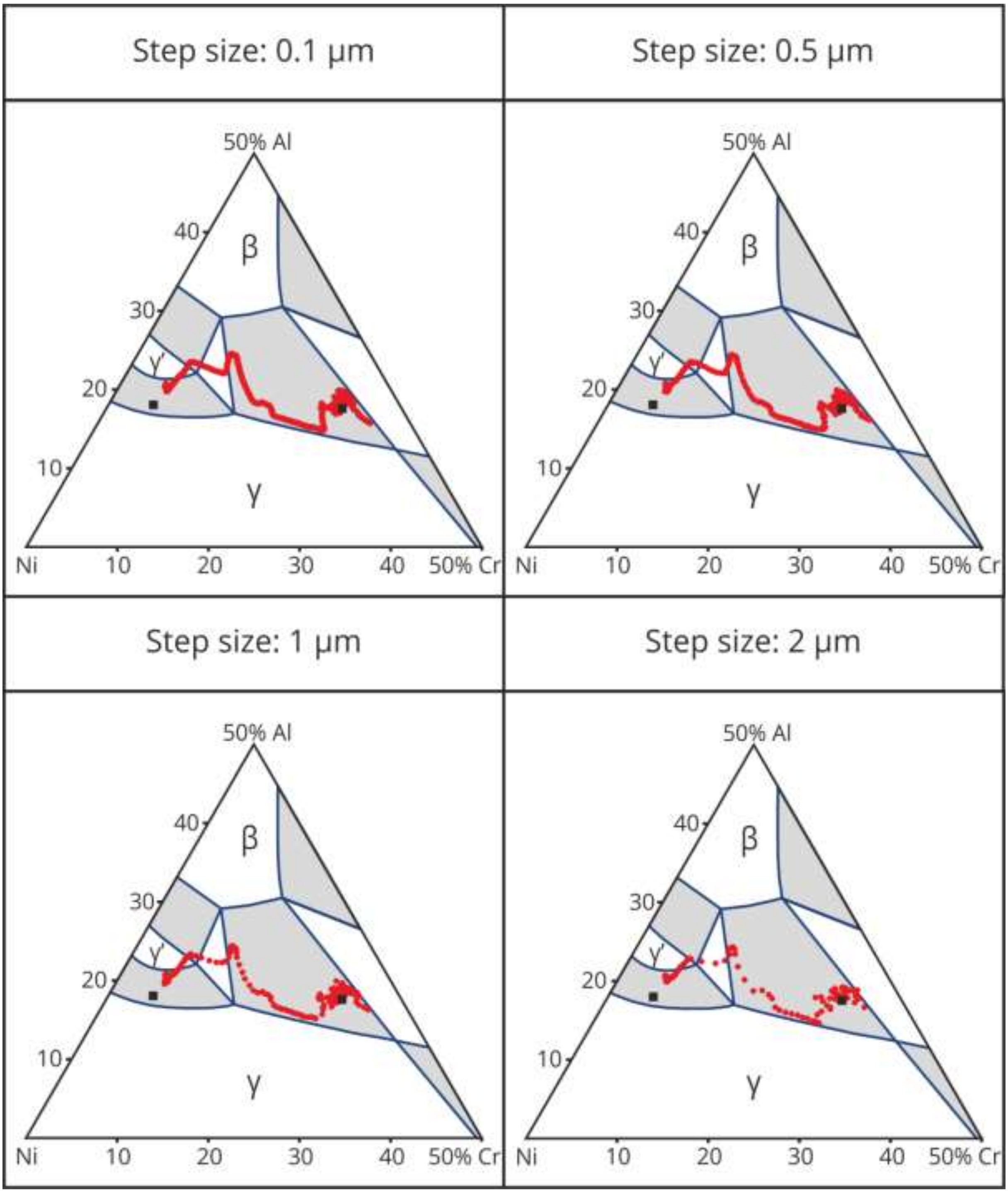

**Fig. 49. Comparison of diffusion paths in (γ+γ')|(γ+β) diffusion couple after annealing by 1 h determined from the concentration profiles measured by wide-line EDS analysis performed with different steps: 0.1, 0.5, 1 and 2 µm, respectively. Black squares represent terminal composition of the couples. The data scatter visible in two-phase fields is caused by grain microstructure of two-phase alloys.**

The presented that concentration profiles almost do not change when increasing the step size from 0.1 to 2 µm. The decrease of the data confirm step size below 1 µm ensures small increase of the quality of the data. The increase of the step up to 2 μm, causes only insignificant decrease of the measurement quality while the shortening of the measurement time is very small.

Finally, two step sizes and widths of the analyzed lines were chosen to examinee samples annealed for 1 and 26 h. They are respectively: 1 μm step size and 100 μm width for the samples annealed for 1 h; and 5 μm step size and 200 μm width for the samples annealed for 26 h.

## 4.2 Experimental results

The experimental results are presented in Figs. 50-63 as the sets of the results for the samples after 1 (on the left) and 26 h (on the right) are shown. The figures present from the top are: SEM image, the related concentration profiles and diffusion path. The scale of the concentration profile axis is the same as the scale of the relevant SEM image. Dashed and solid vertical lines represent Kirkendall [69] and Matano planes [99]. On the top of SEM image the positions of interphase boundaries are marked. Diffusion paths are indicated with red dots, and terminal compositions with black squares.

Comparison of the results confirms that the concentration profiles and diffusion paths for the samples annealed for 1 and 26 hour have similar shapes. Obviously, the diffusion zone is larger in the couples subjected to 26 h annealing.

In the further discussion of the results I will ignore following artefacts:

a) The scatter of the diffusion path in two-phase zone. It is due to coarse and inhomogeneous microstructure of the alloy in this zone, Figs. 50-55. It is especially pronounced in the case of the measurements made for the samples subjected to 1h annealing because in this case the WL-EDS analysis was in the range of 100 μm only.
b) Scatter of the results caused by the pores at Kirkendall plane and by other macrodefects, Figs. 50-55.
c) The presence of secondary precipitates formed during low cooling of the samples after HIP. In this case the decrease of the temperature from 1200 to 900°C occurred during 30 minutes. Such precipitates are not present in the case of the samples annealed by 26 h that quenched in water.

The WL-EDS technique eliminates concentration jumps at phase boundaries as well as non-continuous changes of concentration gradients.

Results for (γ+γ′)|(γ+β) diffusion couple are shown in Fig. 50. Three regions of different morphology have been formed during diffusion. Starting from the left terminal composition, γ+γ′, there are: γ′ (segment 1-2), γ+β (segment 3-4) and γ (segment 4-5). The γ′ zone is in contact on the left side with the γ+γ′ zone. This contact is formally defined as type 1a interphase boundary, (point 1).[8] On the right side the γ′ zone is in contact with the γ+β zone (points 2, 3), γ′|γ+β interphase is type 3. The diffusion path "jump" between points 2 and 3 is result of the passage of the diffusion path across three-phase region on the phase diagram.

The γ+β zone is formed on the left side of the Matano plane (segment 3-4 on the diffusion path). Big elongated precipitates of the β-phase can be seen in this zone.

The γ-phase zone is formed between two γ+β zones and is separated by two type 1 IBs. The diffusion path in γ-phase (segment 4-5) follows the γ-phase boundary line, $k^{\gamma}$.

[8] The jump of concentrations is not observed as a result of EDS resolution.

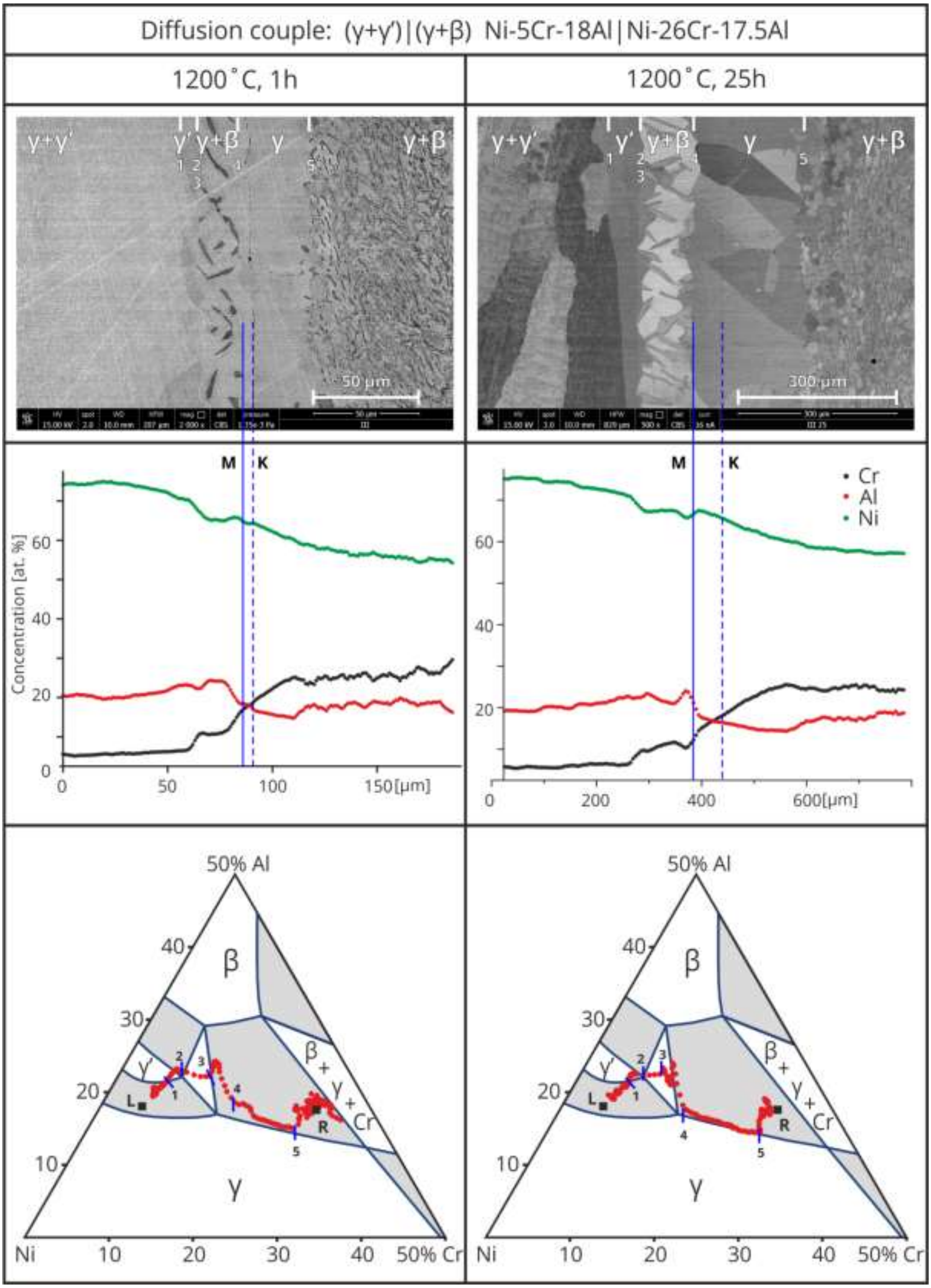

**Fig. 50. Interdiffusion in (γ+γ′)|(γ+β), Ni-5Cr-18Al|Ni-26Cr-17.5Al diffusion couple subjected to annealing for: 1 and 26 h, at 1200°C. The numbers are related to theinterphases on the diffusion path. Black squares represent terminal compositions: L-left, R-right. Grey areas on the phase diagrams represent two-**

**phase fields. Blue solid and dashed lines show the positions of the Matano and Kirkendall planes.**

In Fig. 51 the results for β|(γ+β), β-NiAl|Ni-26Cr-17.5Al diffusion couple are shown.

The diffusion path at 26 h have similar shape as the measured for 1 h. The differences are due to measurement errors for sample annealed for 26 h. The determined diffusion path enter three-phase region what it is forbidden due to thermodynamics. Such behavior of diffusion path is generated by EDS and WL-EDS method.

The microstructure analysis of the samples reveals a formation of single two-phase zone. Starting from the left end of the diffusion couple we can see the single-phase β-zone, (diffusion path segment between the stoichiometric NiAl and point 1). At the point 1, the diffusion path enters the two-phase γ+β region, then it goes along conode (segment 1-2). On the SEM image it is observed as type 1a interphase boundary (point 1, 2). Then diffusion path deviates from the conode (segment 2-3) due to formation of the two-phase β+α-Cr zone, and traverses γ+β+α-Cr three-phase region and finally enters the γ+β two-phase region. The traverse through three-phase region is measured as a jump of the concentration at the type 2 IB.

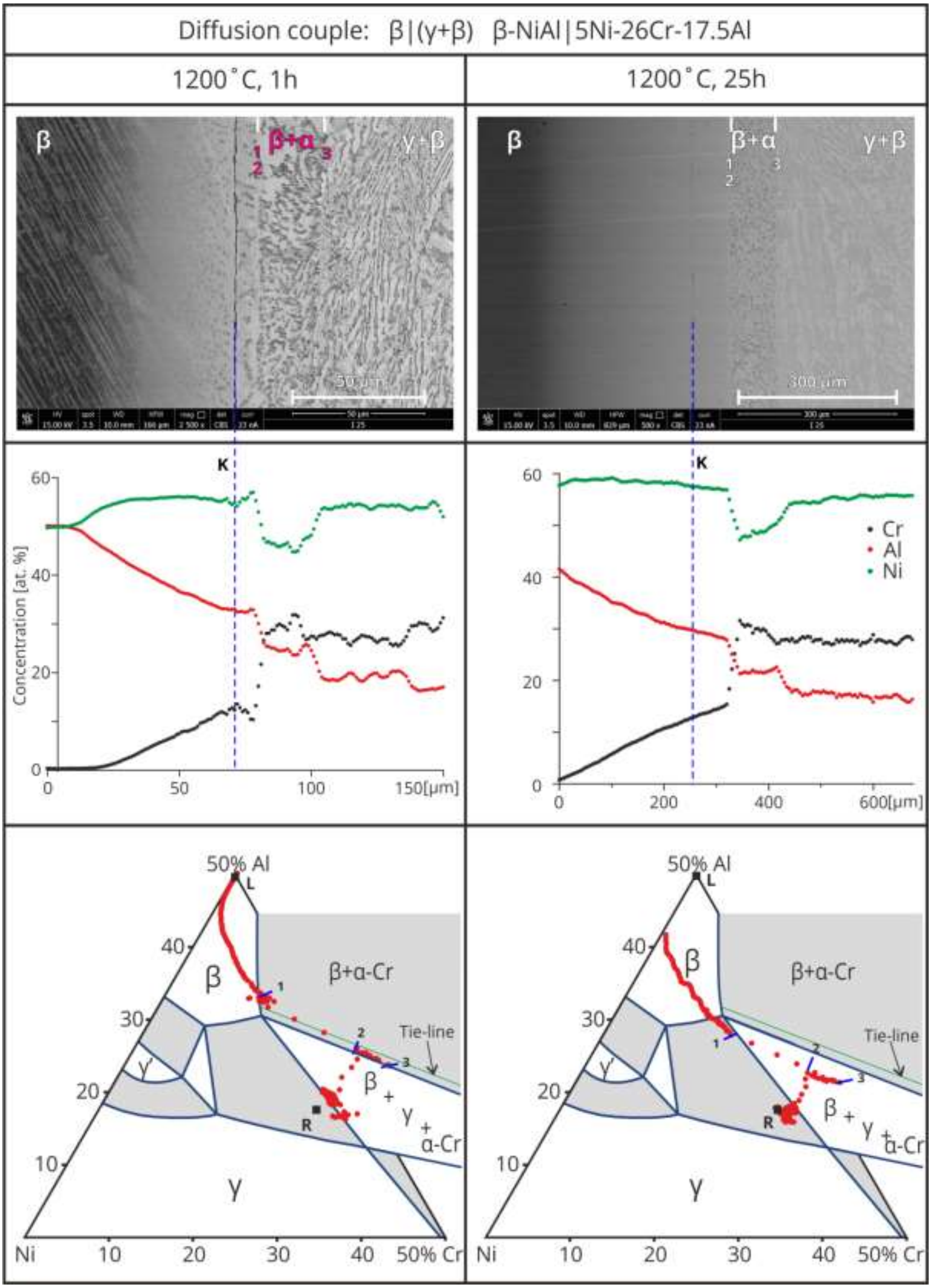

**Fig. 51. Interdiffusion in β|(γ+β), β-NiAl|Ni-26Cr-17.5Al diffusion couple subjected to annealing for 1 and 26 h, at 1200°C. The sequence of the figures like in Fig. 50.**

Figures 52 to 55, present the results measured for the γ+β two-phase alloys (right composition) coupled with four different γ alloys (γ-1 to γ-4): Ni, Ni-10Cr, Ni-25Cr and Ni-40Cr.

The results for all couples are similar. On the SEM image a characteristic recession of β-phase in γ+β zone is seen. It is measured as a shift of the type 1a interphase boundary, γ|(γ+β), from the initial position (Matano plane). A diffusion path within the γ single-phase region assumes typical S-shape, while within γ+β two-phase region has zigzag form. The small differences between the results for different diffusion couples concern behavior of the diffusion path within two-phase region are be caused by the secondary precipitates formation during cooling, Fig. 56.

In the case of the γ-1|(γ+β) and γ-2|(γ+β) diffusion couples annealed for 1 hour (Figs. 52 and 53) dispersion of the data at the phase boundary and within two-phase zone is caused by nonplanar boundary and grain microstructure of the alloy.

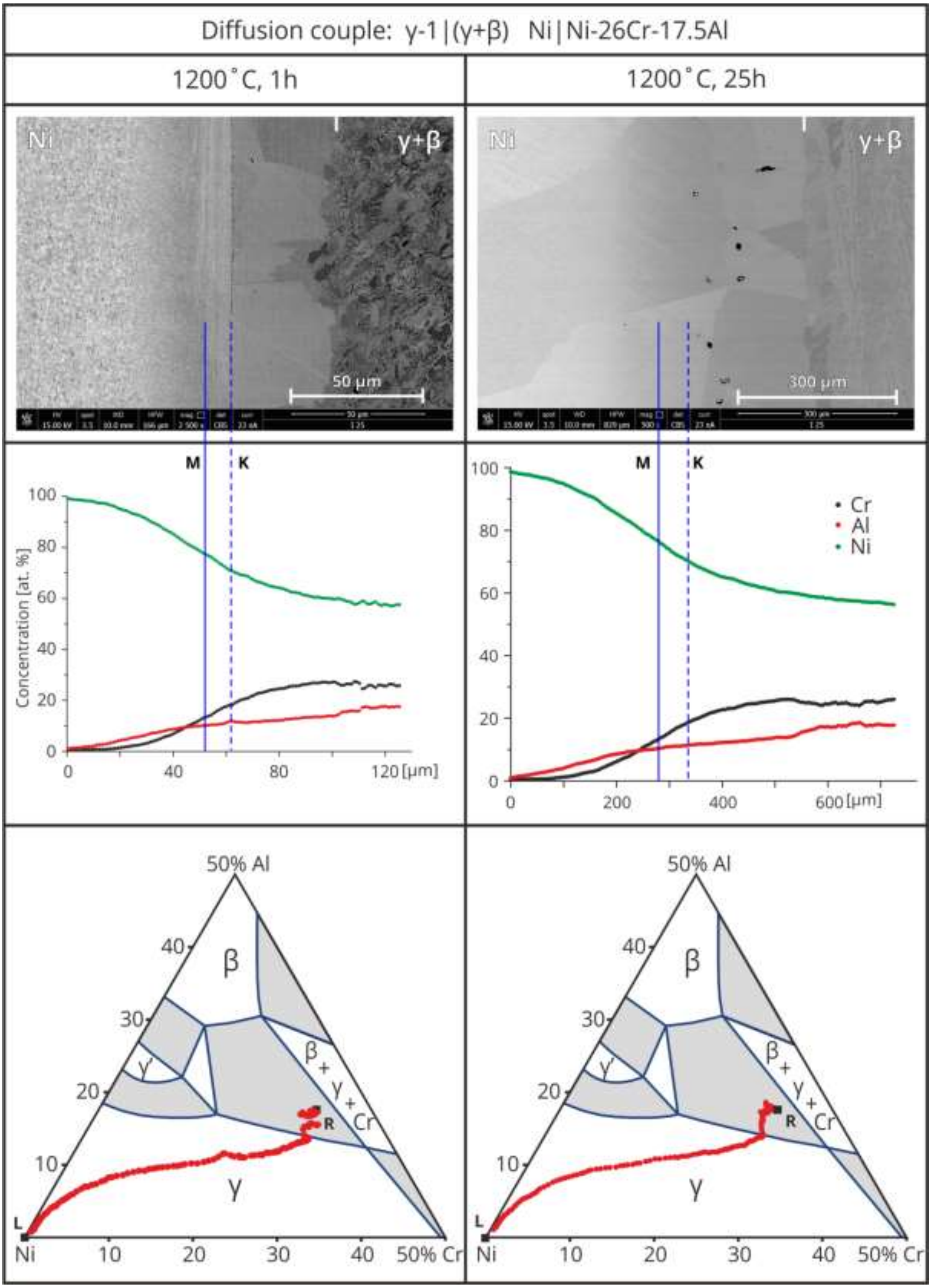

**Fig. 52. Interdiffusion in γ-1|(γ+β), Ni|Ni-26Cr-17.5Al diffusion couple subjected to to annealing for 1 and 26 h, at 1200°C. The sequence of the figures like in Fig. 50.**

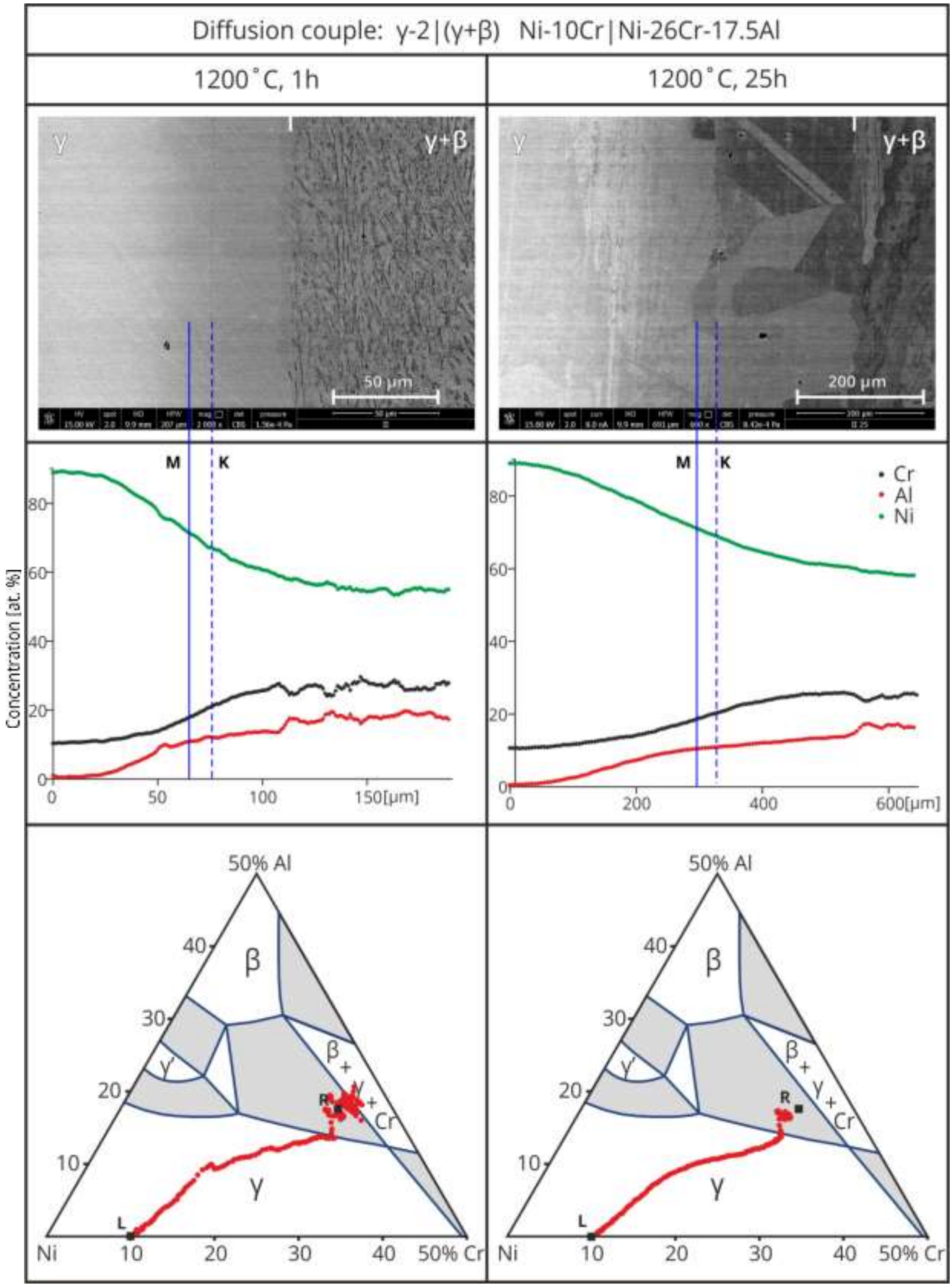

**Fig. 53. Interdiffusion in γ-2|(γ+β), Ni-10Cr|Ni-26Cr-17.5Al diffusion couple subjected to annealing for 1 and 26 h, at 1200°C. The sequence of the figures like in Fig. 50.**

The results for γ-3|(γ+β) diffusion couple annealed by 1 h, Fig. 54, show the presence of γ+β+α three-phase zone. This zone was formed due to slow cooling of the sample, γ+β zone formed as a result of diffusion. Such precipitates (γ+β+α zone) are not observed in the sample annealed by 26 h, which was quenched in water.

Similar results were obtained for the γ-4|(γ+β) couple, Fig. 55, with clear concentration jump at the interphase. Also in this case the formation of precipitates of the β-phase (in γ-phase zone) was caused by slow cooling of the sample, Fig. 56.

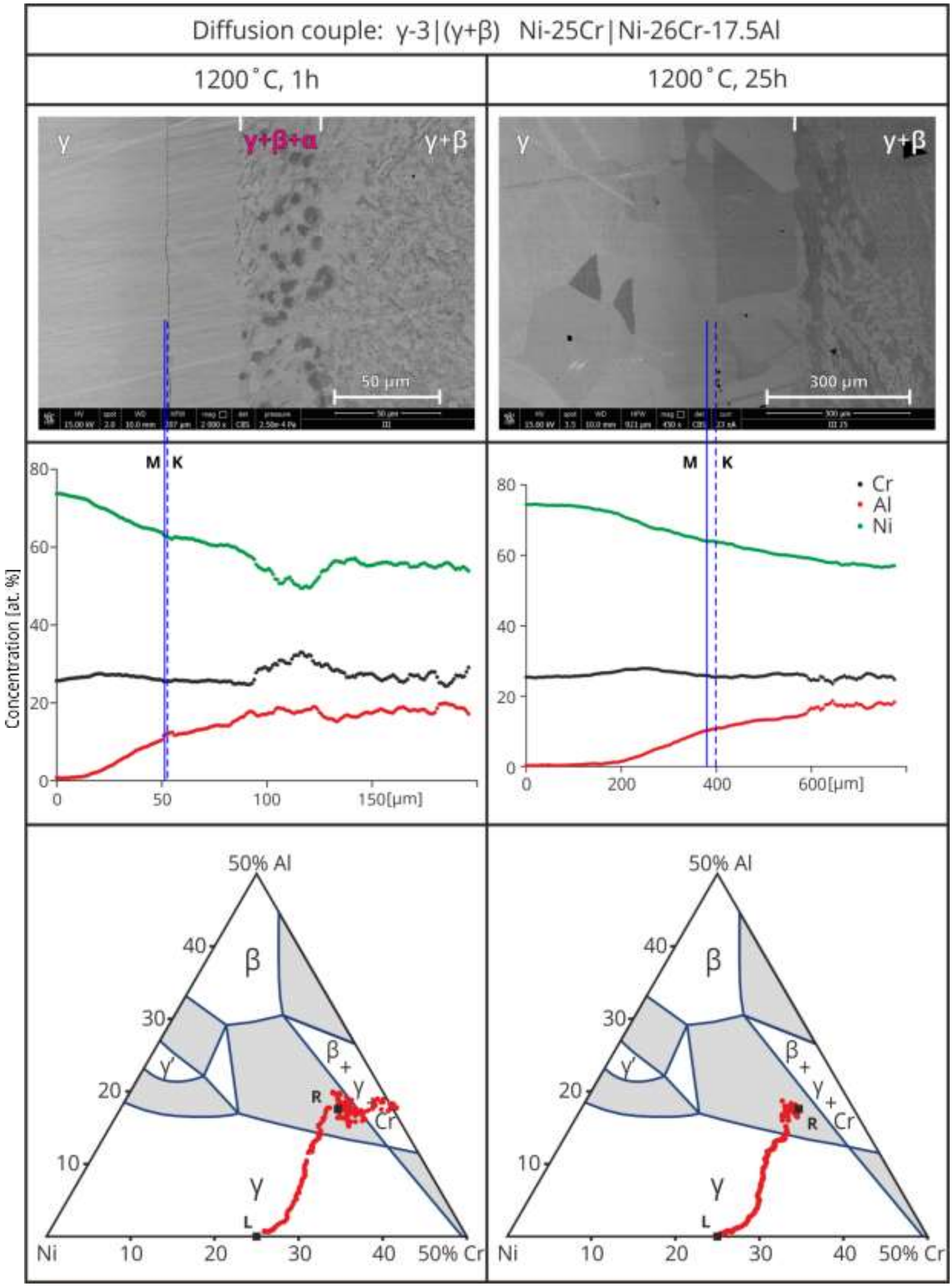

**Fig. 54. Interdiffusion in γ-3|(γ+β), Ni-25Cr|Ni-26Cr-17.5Al diffusion couple subjected to annealing for 1 and 26 h, at 1200°C. The sequence of the figures like in Fig. 50.**

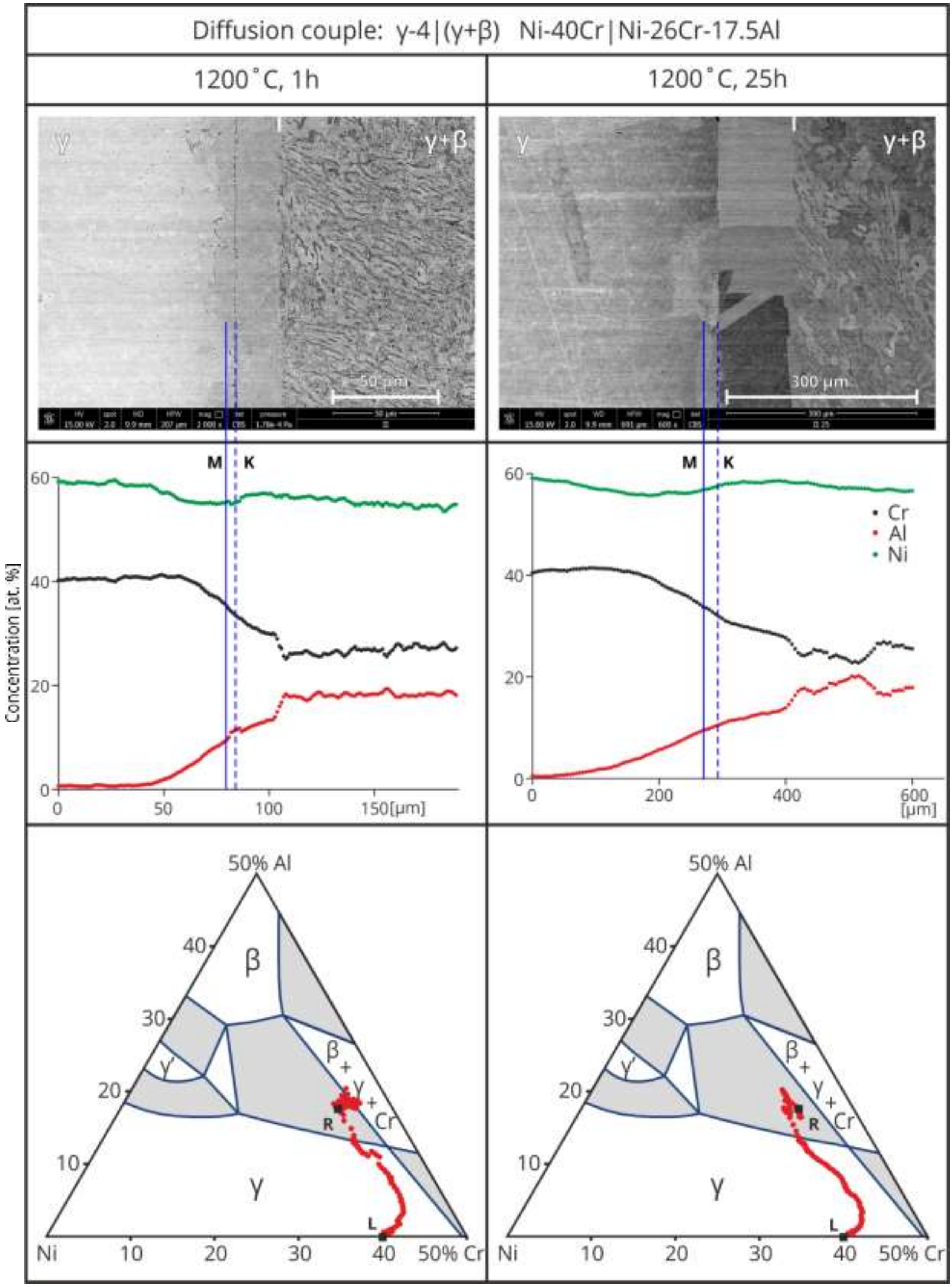

**Fig. 55. Interdiffusion in γ-4|(γ+β), N-40Cr|Ni-26Cr-17.5Al diffusion couple subjected to annealing for 1 and 26 h, at 1200°C. The sequence of the figures like in Fig. 50.**

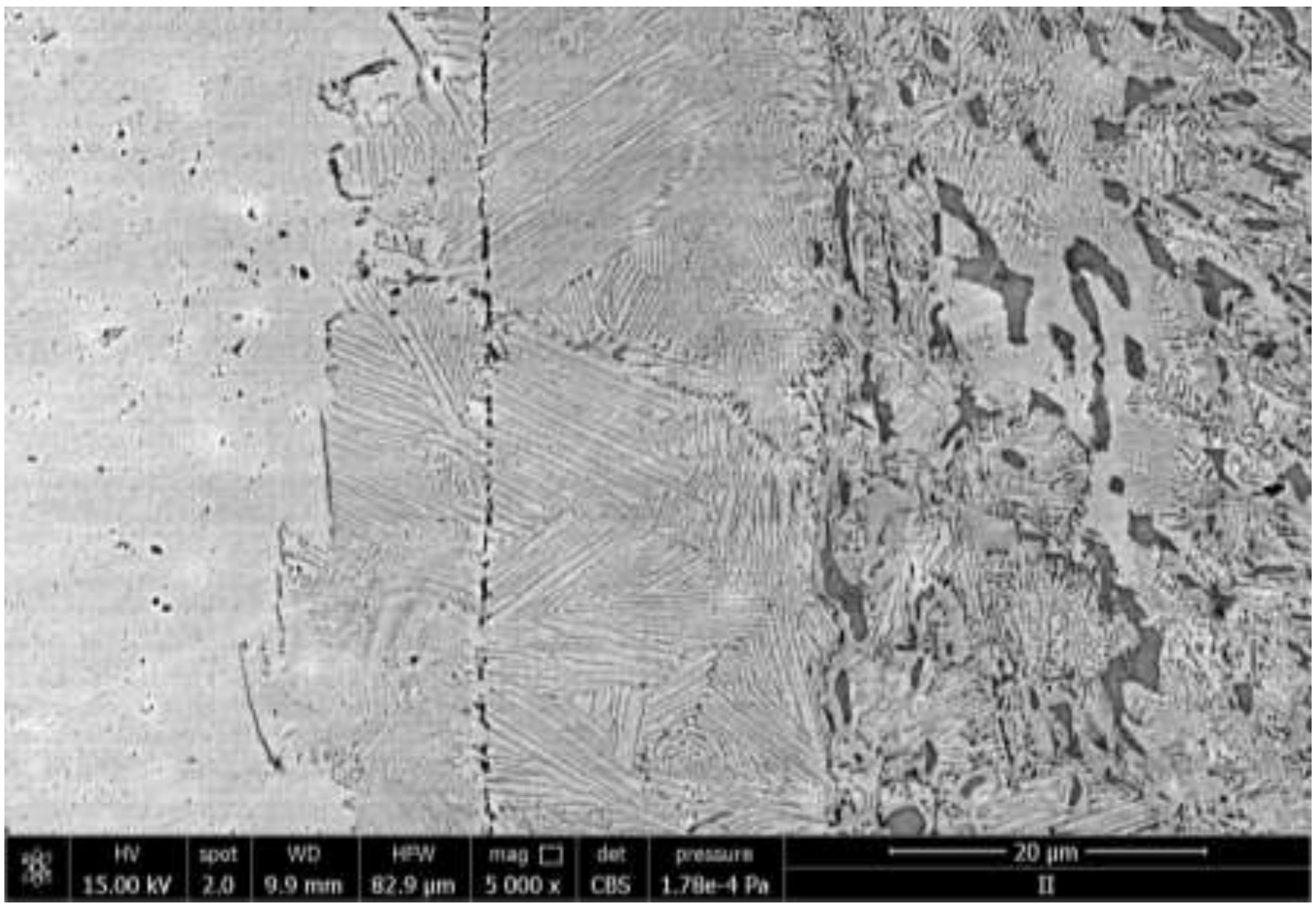

**Fig. 56. Secondary precipitates of β-phase in γ-3|(γ+β) diffusion couple, caused by a slow cooling after annealing by 1h.**

In Fig. 57 the results for the (γ+γ′)|β couple are shown. In this couple one single-phase γ′ zone has been formed during diffusion. Its thickness is about 15 µm in the sample annealed for 1 and for 26 h, Fig. 58. The γ′ zone from the left side is bordered with the γ+γ′ zone (the γ+γ′|γ′, type 1 IB, point 1). On the right site it is bordered with β zone (the γ′|β, type 2 IB (points 2 and 3)). The jump of concentration along conode across γ′+β two-phase region is clearly visible.

The microstructure of β-phase with characteristic big grains, about 2 mm in the diameter, changes close to the Kirkendall plane. It transforms to the microstructure with twin boundaries, Fig. 59 inside the centre of the diffusion couple.

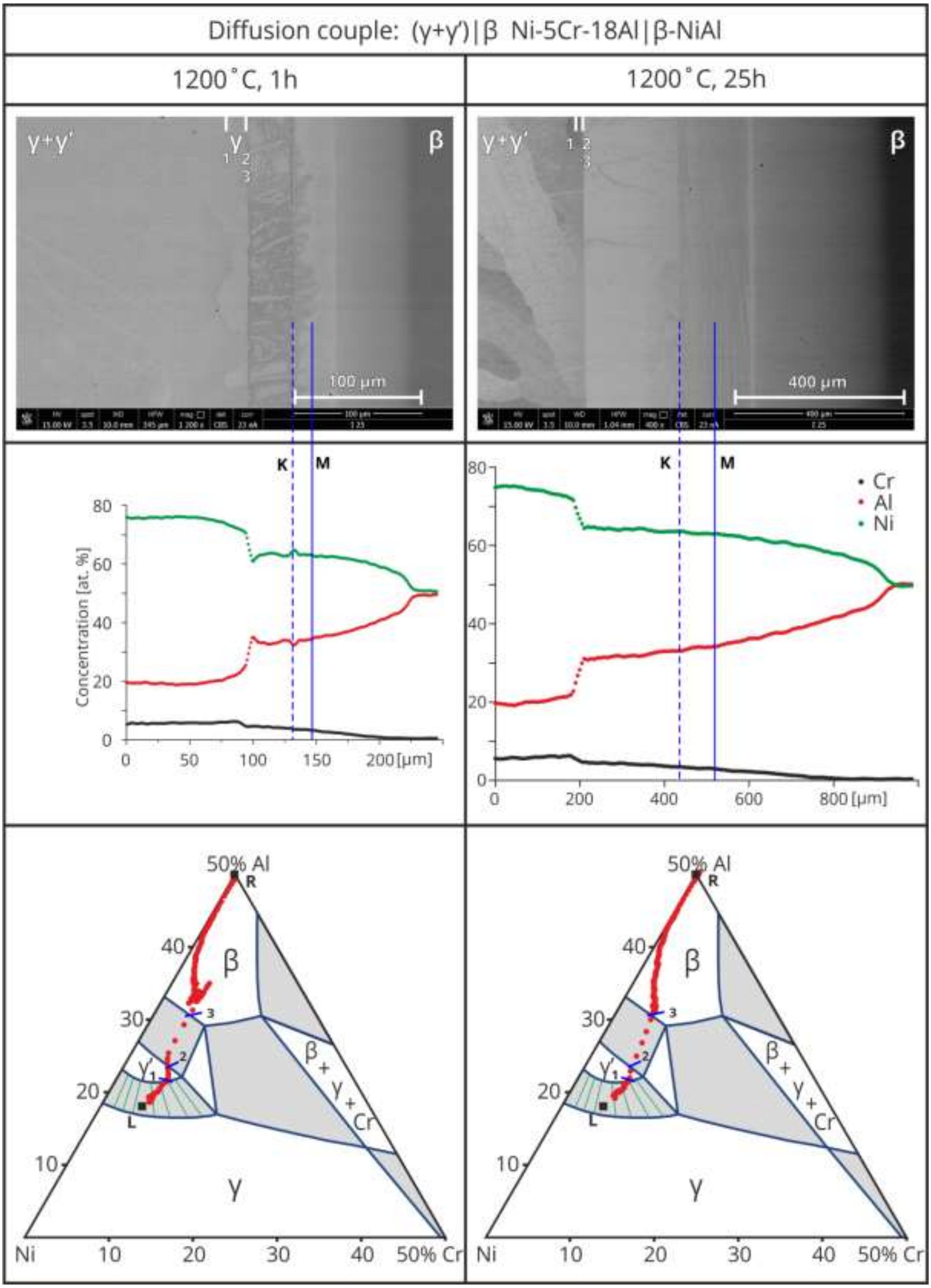

**Fig. 57. Interdiffusion in (γ+γ′)|β, Ni-5Cr-18Al|β-NiAl diffusion couple subjected to annealing for 1 and 26 h, at 1200°C. The sequence of the figures like in Fig. 50.**

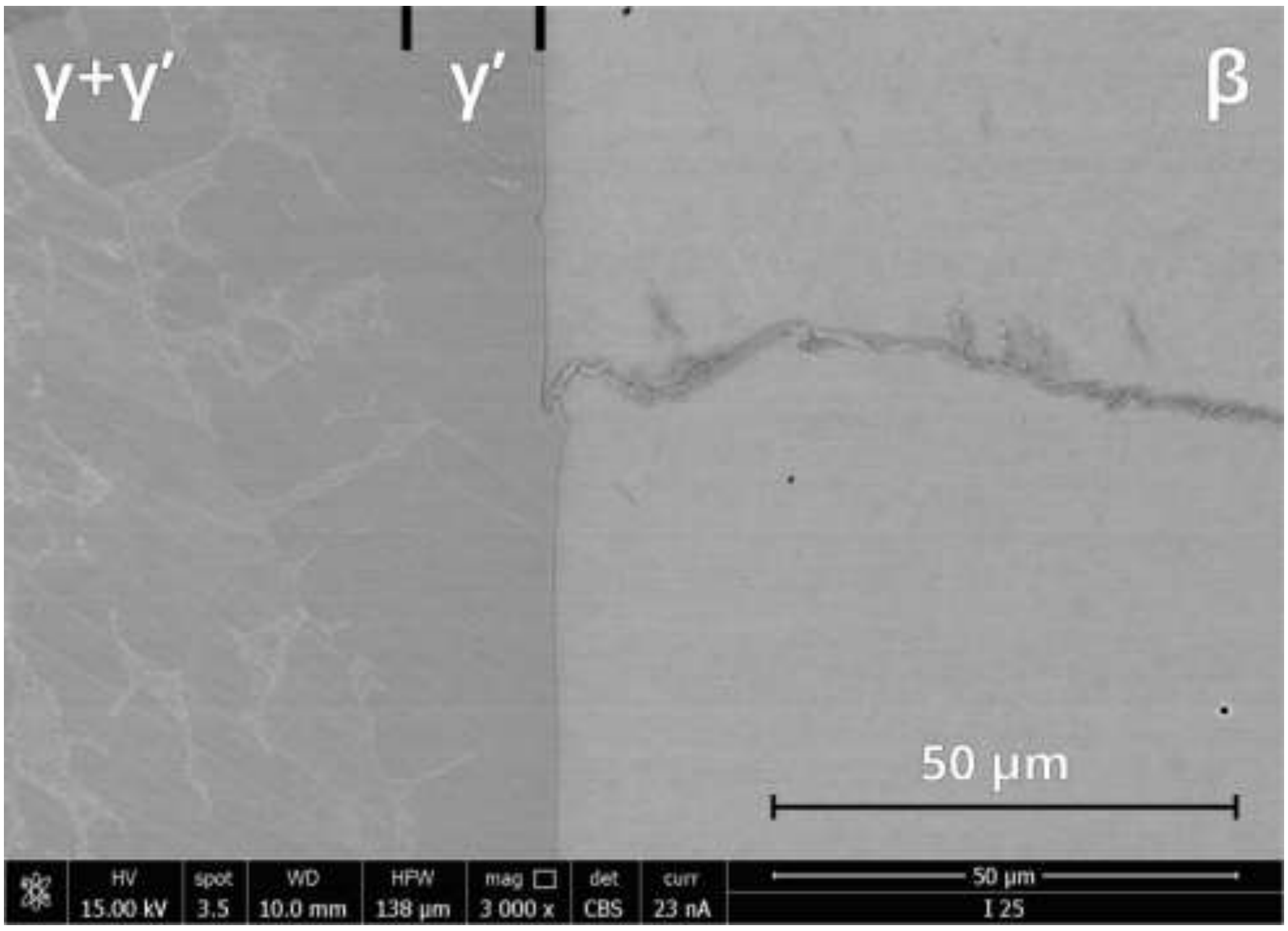

**Fig. 58. SEM image of diffusion zone in the (γ+γ′)|β diffusion couple after annealing by 26 h. A narrow γ′-phase layer is seen.**

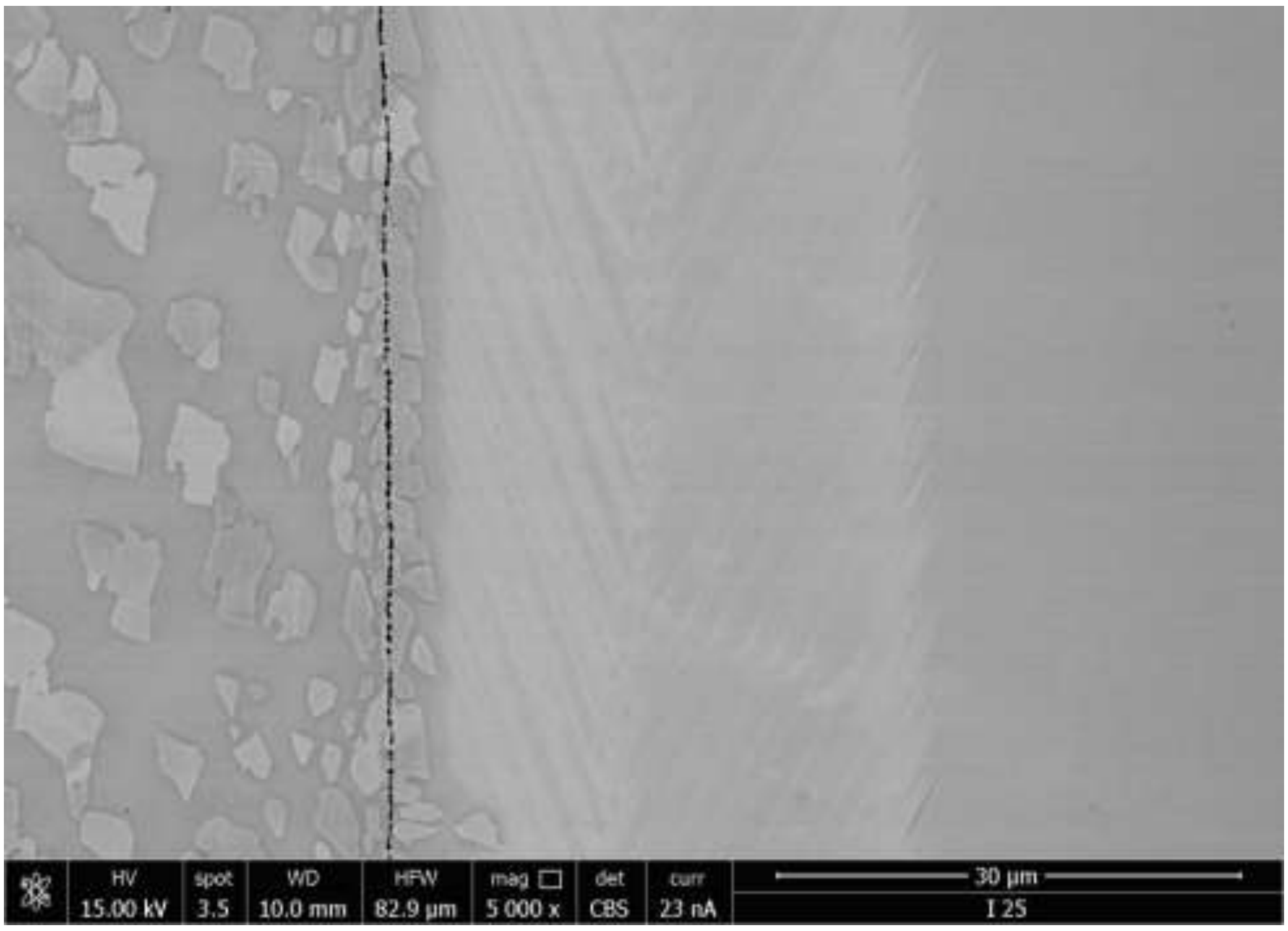

**Fig. 59. SEM image of the (γ+γ′)|β couple subjected to annealing by 1 h. The secondary precipitates of γ′-phase are seen on the left side of the Kirkendall plane where pores are visible. Twin grain boundaries are seen on the right side.**

Figs. 60-63, present the results measured for the γ+γ' alloy (right composition) coupled with four different γ alloys (γ-1 to γ-4): Ni, Ni-10Cr, Ni-25Cr and Ni-40Cr. The behaviour of all samples is similar. In all cases the recession of γ'-phase is observed and the diffusion path within γ single-phase field has typical S-shape. Visible differences of the microstructure of the samples annealed by 1 and 26 h are due to different cooling conditions. The samples annealed for 1 h were cooled slowly and their microstructure exhibits a presence of secondary γ'-phase precipitates, Fig. 64. Such precipitates do not occur in the quenched samples.

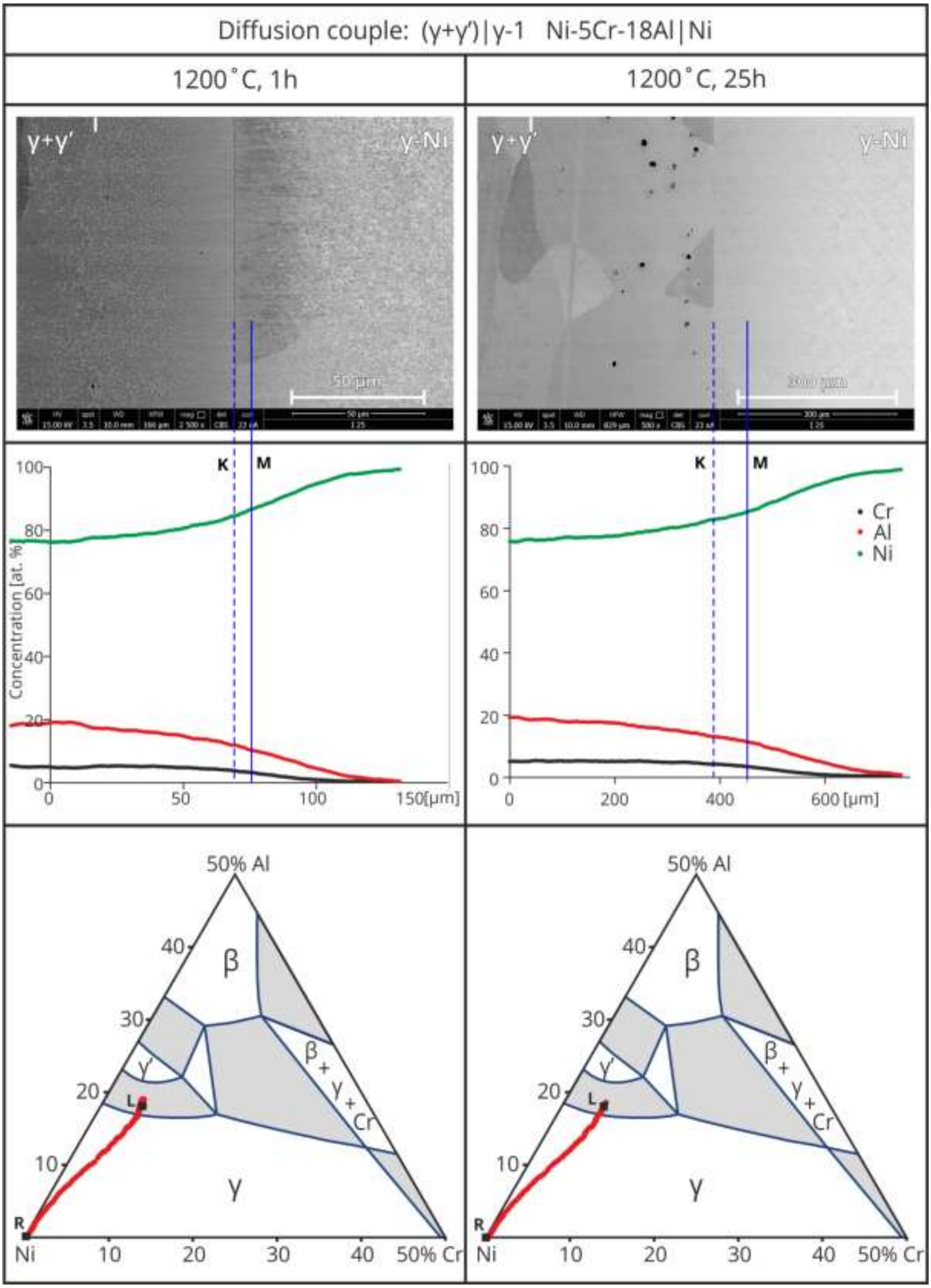

**Fig. 60. Interdiffusion in (γ+γ′)|γ-1, Ni-5Cr-18Al|Ni diffusion couple subjected to annealing for 1 and 26 h, at 1200°C. The sequence of the figures like in Fig. 50.**

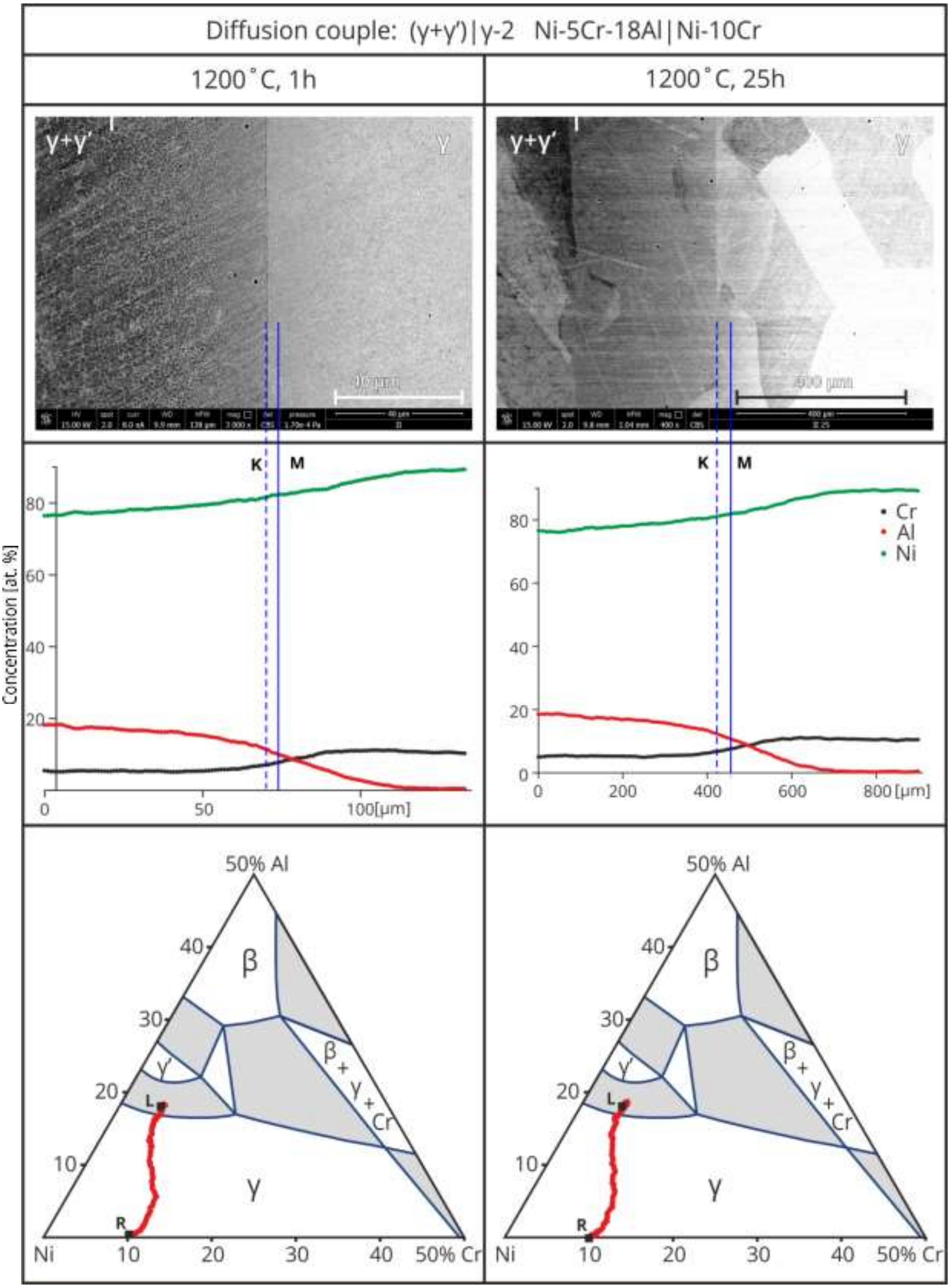

**Fig. 61. Interdiffusion in (γ+γ')|γ-2, Ni-5Cr-18Al|Ni-10Cr diffusion couple subjected to annealing for 1 and 26 h, at 1200°C. The sequence of the figures like in Fig. 50.**

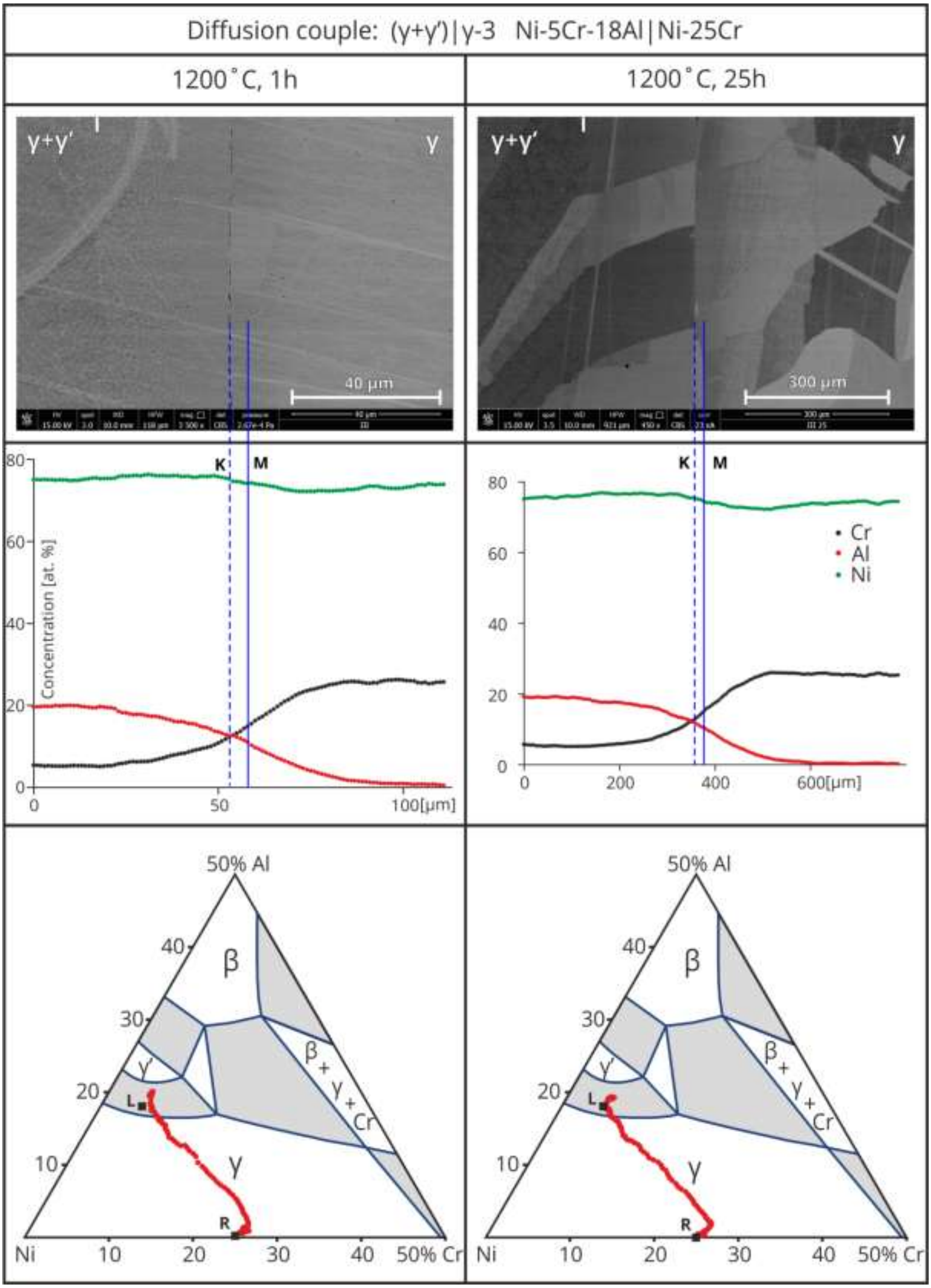

**Fig. 62. Interdiffusion in (γ+γ′)|γ-3, Ni-5Cr-18Al|Ni- 25Cr diffusion couple subjected to annealing for 1 and 26 h, at 1200°C. The sequence of the figures like in Fig. 50.**

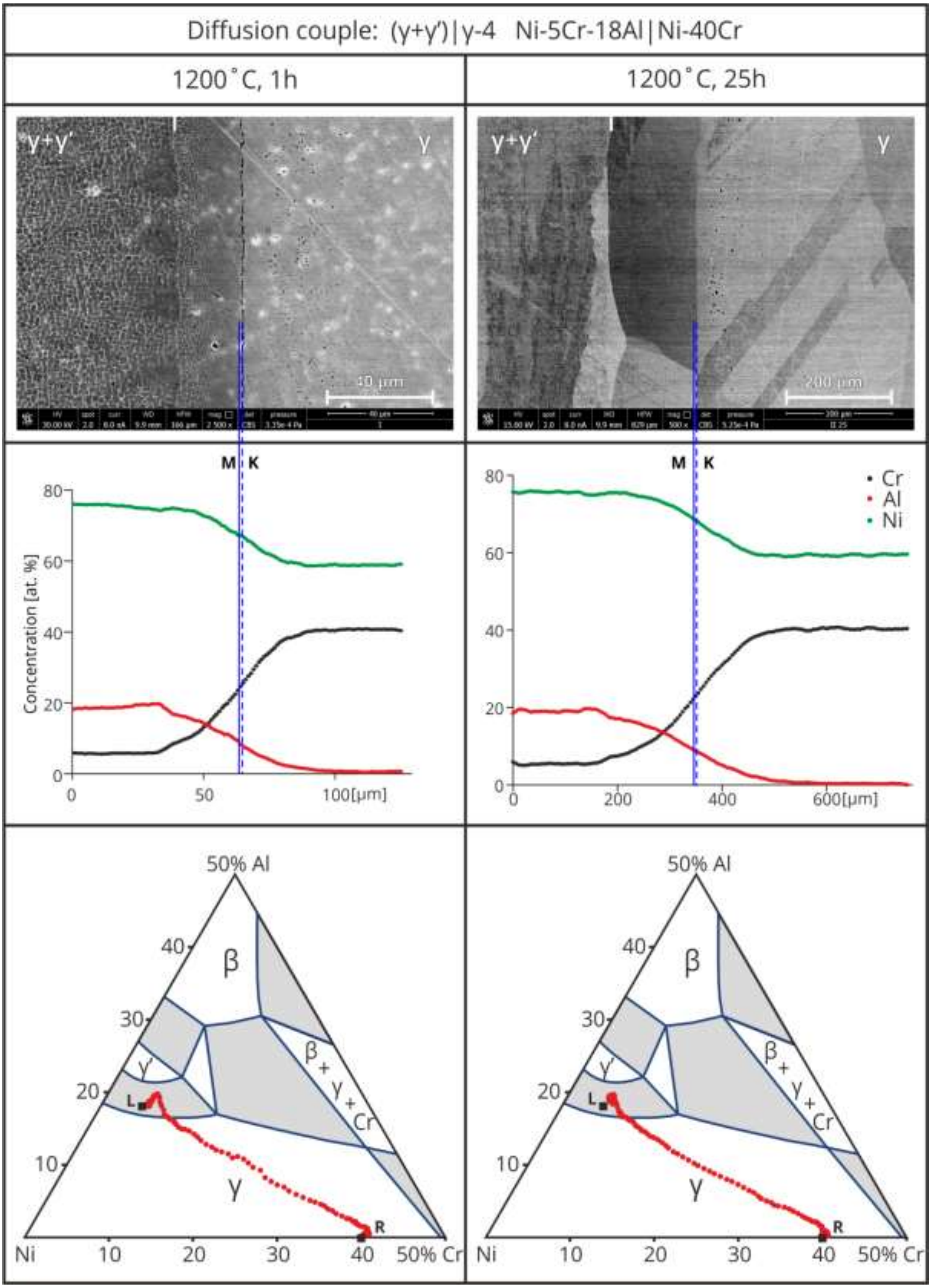

**Fig. 63. Interdiffusion in (γ+γ')|γ-4, Ni-5Cr-18Al|Ni -40Cr diffusion couple subjected to annealing for 1 and 26 h, at 1200°C. The sequence of the figures like in Fig. 50.**

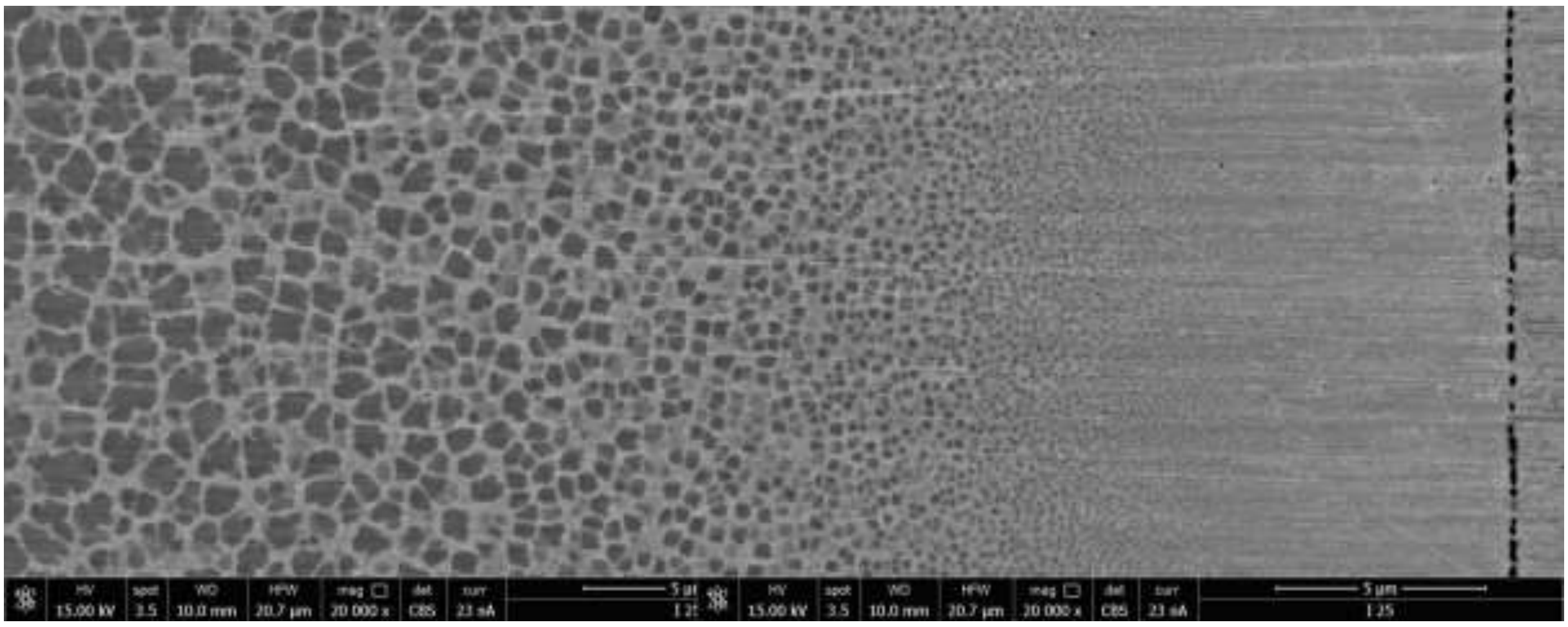

**Fig. 64. Secondary precipitates of γ'-phase in the sample (γ+γ')|γ-1, subjected to annealing by 1 h.**

The experimental Kirkendall and computed Matano plane positions have been used to calculate shift of the Kirkendall plane. The results are presented in Tab. 7. The error in determination of Matano plane position is high due to scatter of the measured concentrations in two-phase zone and other measurement errors. Therefore the obtained results are only qualitative. These results will be discussed in section 5.

**Tab. 7. Shift of Kirkendall plane after annealing by 1 and 26 h.**

| Diffusion couple | | Kirkendall plane shift [μm] | |
|---|---|---|---|
| | | 1 h | 26 h |
| Ni | Ni-26Cr -17.5Al | 9.8 | 55.8 |
| Ni-10Cr | | 10.9 | 31.2 |
| Ni-25Cr | | 1.3 | 18.0 |
| Ni-40Cr | | 4.5 | 22.6 |
| NiAl | | --- | --- |
| Ni | Ni-5Cr-18Al | -6.4 | -64.0 |
| Ni-10Cr | | -3.9 | -32.8 |
| Ni-25Cr | | -4.9 | -19.4 |
| Ni-40Cr | | 1.2 | 5.4 |
| NiAl | | -15.4 | -83.2 |
| Ni-26Cr-17.5Al | Ni-5Cr-18Al | 4.8 | 54.2 |

## 4.3 Summary

a) The multi-couple method applied in the studies of interdiffusion in multi-phase alloys allows obtaining high quality, planar, repetitive and oxide-free diffusion couples. Some imperfections in sample preparation, such as mismatch of the dimensions and shapes or the presence of cavities are reduced by applying HIP procedure. Unfortunately the method is time consuming and does not allow execution of sample quenching to freeze its microstructure achieved by diffusion during HIP process.

b) The precise measurement of the concentration profiles in the multi-phase alloys is possible by application of wide-line EDS analysis. It is a first reported application of WL-EDC method to measure concentration profiles in the multi-phase samples formed by interdiffusion.

c) The diffusion paths in single-phase region, shown in Figs. 52-55 and 60-63 present the shapes which are consistent with results presented by Nesbitt and Heckel [63-65].

d) The results for the γ+β alloy coupled with γ alloys (Figs. 52-55) confirm that the diffusion path within two-phase region has zig-zag shape as predicted by the theory. (Seen as a jump of concentration, Figs. 52 to 55). Similar behaviour is observed for the (γ+γ')|γ couples, Figs. 60-63. The diffusion path for (γ+γ')|β couple, goes within γ'+β field along the conode which is observed as a concentration jump at the γ'|β interphase.

e) The Kirkendall plane positions were found by using "nanoporosity" (~100 nm in diameter) as marker of initial contact interface. However the final precision of the Kirkendall plane shift is limited due to the errors in determination of the Matano plane position.

# Chapter 5. Simulations of diffusion in the two-phase region of the Ni-Cr-Al system in R1

The model presented in section 3.1 was applied to simulate diffusion in Ni-Cr-Al system. Simulations were made for the following Ni-Cr-Al couples: γ+γ′ and γ+β alloys coupled with four different γ single-phase alloys: Ni, Ni-10Cr, Ni-25Cr and Ni-40Cr, and limited to the two-phase regions. The initial concentrations were shown in Tab. 6.

The simulations were made for the diffusion times 1 and 26 h, according to time of the annealing. To accelerate simulation speed evolutionary time step was used. The chosen widths of the diffusion couple were 300 (1 h) and 1600μm (26 h), with respectively: 200 and 400 grid points. The polynomials describing phase boundaries and positions of common tie-lines points were calculated basing on phase diagram by Hashimoto et al. [23] and are shown in Tab. 8.

The diffusion coefficients of the components in γ-phase were numerically extracted from DICTRA [31, 35], using the relationship between intrinsic diffusion coefficient and the matrix of interdiffusion coefficients presented in [72]. The computed values of diffusion coefficient which were used in simulations are shown in Figs. 65 to 72. The intrinsic diffusion coefficients in γ′ and β phases are specified in Tab. 9 [100]. In the same table the partial molar volumes for all phases are presented. The partial molar volumes of the components in γ′ and β phases are assumed the same for each component. For γ-phase they differ [101].

**Tab. 8. Phase boundaries relations and common tie-lines points, $\mathbf{p}^{tie}$, used to calculate local equilibrium in two-phase zone [23].**

| Phase boundary | | Relation | Common tie-lines point: $\mathbf{p}^{tie}$ |
|---|---|---|---|
| β | β+γ | $k^{\beta}$: $N_2=0.2216N_1 + 0.2752$ | (-0.0077, 0.4195) |
| γ | β+γ | $k^{\gamma}$: $N_2=-0.2264N_1 + 0.2003$ | |
| γ′ | γ+γ′ | $k^{\gamma'}$: $N_2=8.2533(N_1)^2 - 0.7857N_1 + 0.2323$ | (0, 0.2793) |
| γ | γ+γ′ | $k^{\gamma}$: $N_2=2.0422(N_1)^2 - 0.4041N_1 + 0.1845$ | |

**Tab. 9. Diffusion coefficients in β and γ′ phases at 1200°C [100] and partial molar volumes in γ, β and γ′ [101].**

| | Phase | Ni | Cr | Al |
|---|---|---|---|---|
| Diffusion coefficient $[cm^2/s]$ | β | $1.2\cdot10^{-9}$ | $1.2\cdot10^{-9}$ | $1.4\cdot10^{-9}$ |
| | γ′ | $3.2\cdot10^{-11}$ | $1.2\cdot10^{-10}$ | $1.0\cdot10^{-10}$ |
| Partial molar volume $[cm^3/mol]$ | γ | 7.2 | 10.0 | 6.6 |
| | β | 8.3 | 8.3 | 8.3 |
| | γ′ | 7.4 | 7.4 | 7.4 |

The comparison between experiments and simulations is presented in Figs. 65-72. For transparency the presented number of measured concentration points was reduced by half.

## 5.1 Results of simulation and comparison with experimental data for couples with β+γ alloy

The comparison of simulation and experimental results for the diffusion couples formed by the two-phase γ+β alloy coupled with the four different γ alloys (Ni, Ni-10Cr, Ni-25Cr and Ni-40Cr) is shown in Figs. 65 to 68. For all couples very good agreement of the position of γ|γ+β type 1a interphase boundary was obtained. The calculated diffusion path and concentration profiles within the γ-phase zone presented in Figs. 65 to 67 are in good agreement with the measured ones. The agreement for the couple presented in Fig. 68, is satisfactory only for the results after 26 h of annealing. Due to artefacts present in short annealed samples, (described in section 4), the diffusion paths differ.

The agreement of the diffusion paths in the β+γ region, Figs. 65 to 68, is satisfactory. The simulations show bigger concentration jump than the measured one. There are three main sources of this “smoothing” effect:

- the WL-EDS and filtering procedure result in smoothing and reduce the extrema;
- the simple approximation of the constant diffusion coefficients;
- the use of the Fickian flux formula in non-ideal system (gradient of chemical potential is the true driving forces).

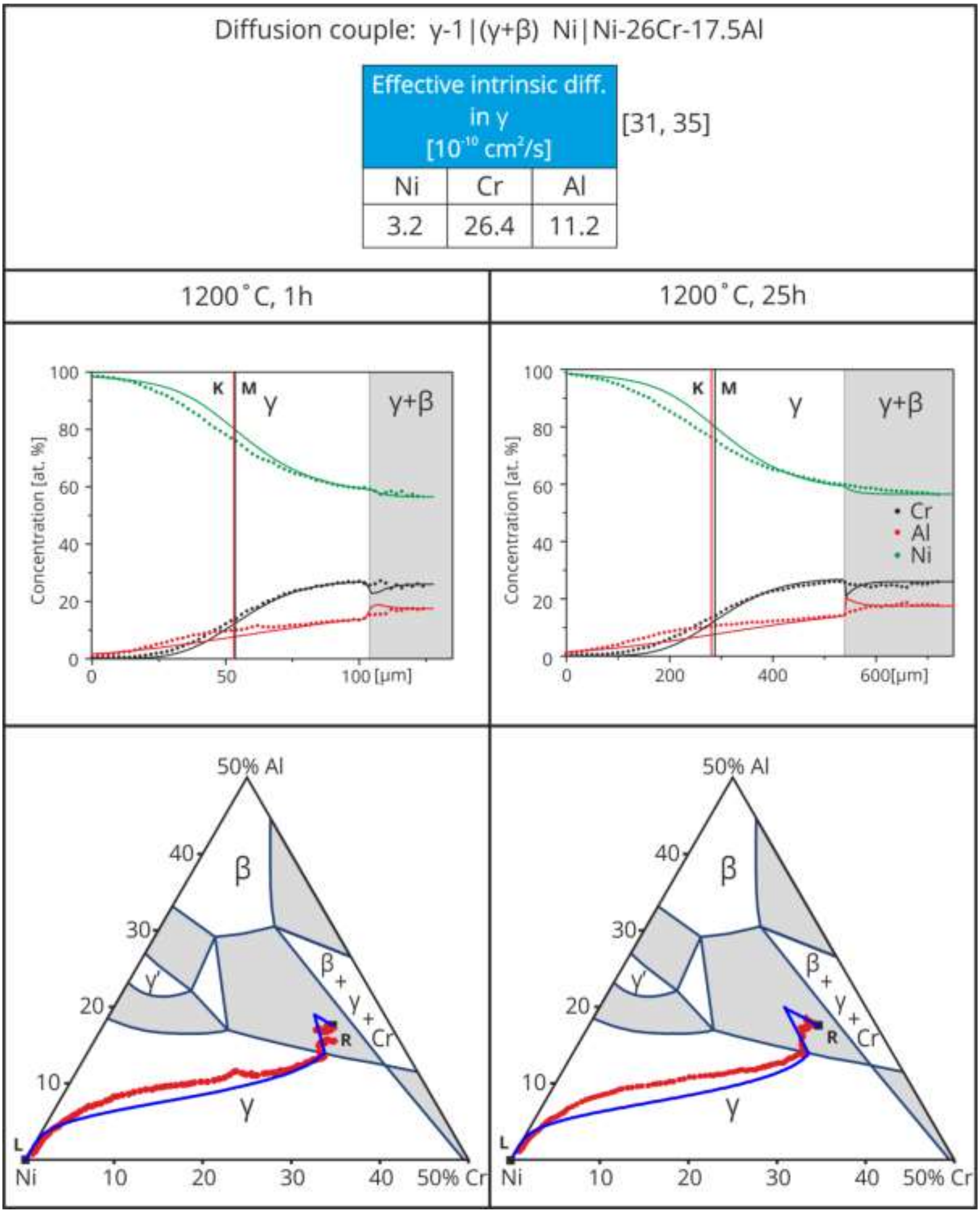

**Fig. 65. Experimental (dots) and simulated (lines) concentration profiles, and diffusion paths for γ-1|(γ+β), Ni|Ni-26Cr-17.5Al diffusion couple (shown in Fig. 52, section 4.2).**

Diffusion couple: γ-2|(γ+β) Ni-10Cr|Ni-26Cr-17.5Al

| Effective intrinsic diff. in γ [$10^{-10}$ $cm^2/s$] | | |
|---|---|---|
| Ni | Cr | Al |
| 11.2 | 26.2 | 3.2 |

[31, 35]

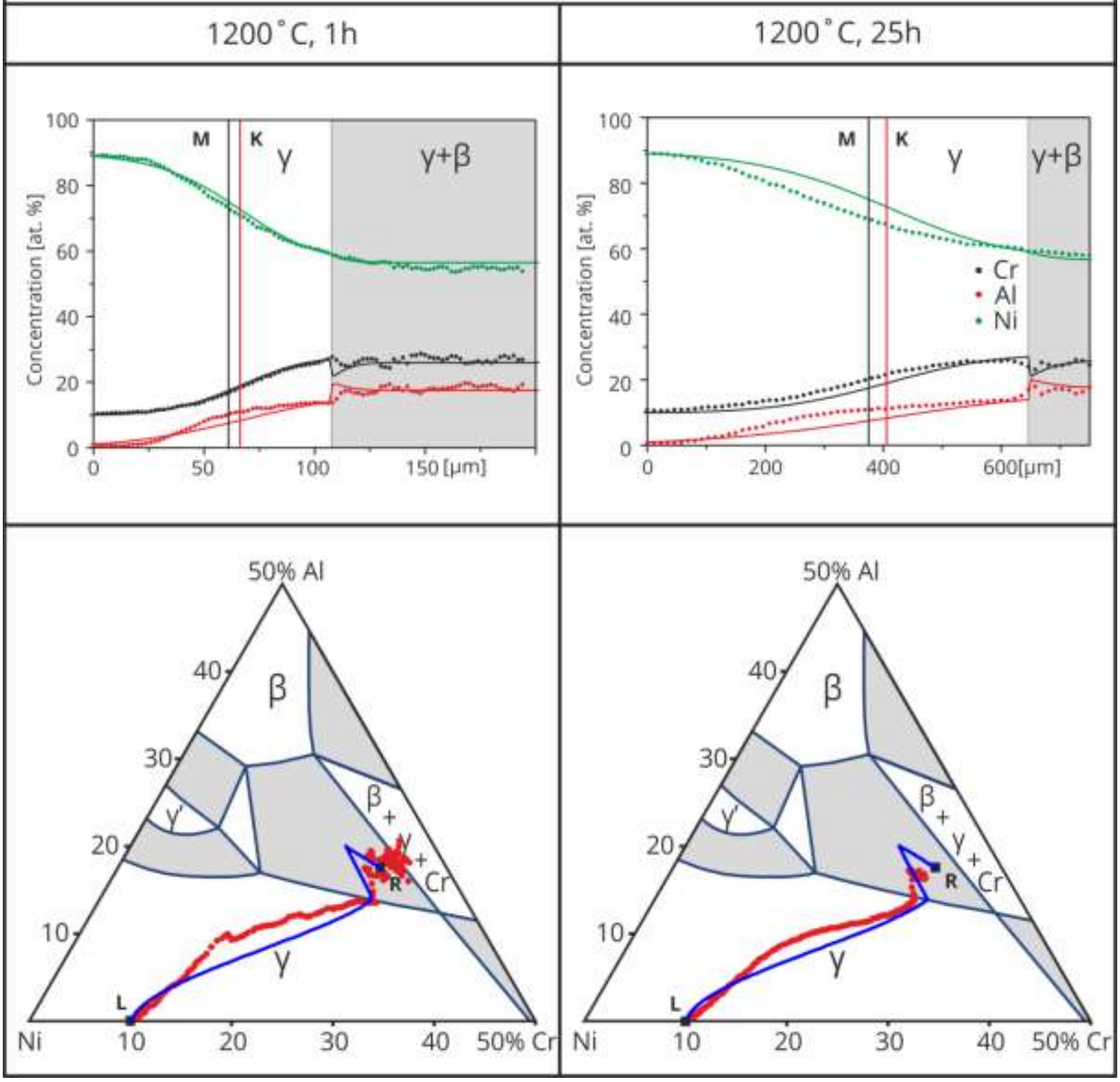

Fig. 66. Experimental (dots) and simulated (lines) concentration profiles, and diffusion paths for γ-2|(γ+β), Ni-10Cr|Ni- 26Cr-17.5Al diffusion couple (shown in Fig. 53, section 4.2).

Diffusion couple: γ-3|(γ+β) Ni-25Cr|Ni-26Cr-17.5Al

| Effective intrinsic diff. in γ [$10^{-10}$ cm²/s] | | | |
|---|---|---|---|
| Ni | Cr | Al | [31, 35] |
| 22.4 | 6.4 | 52.8 | |

| 1200˚C, 1h | 1200˚C, 25h |
|---|---|

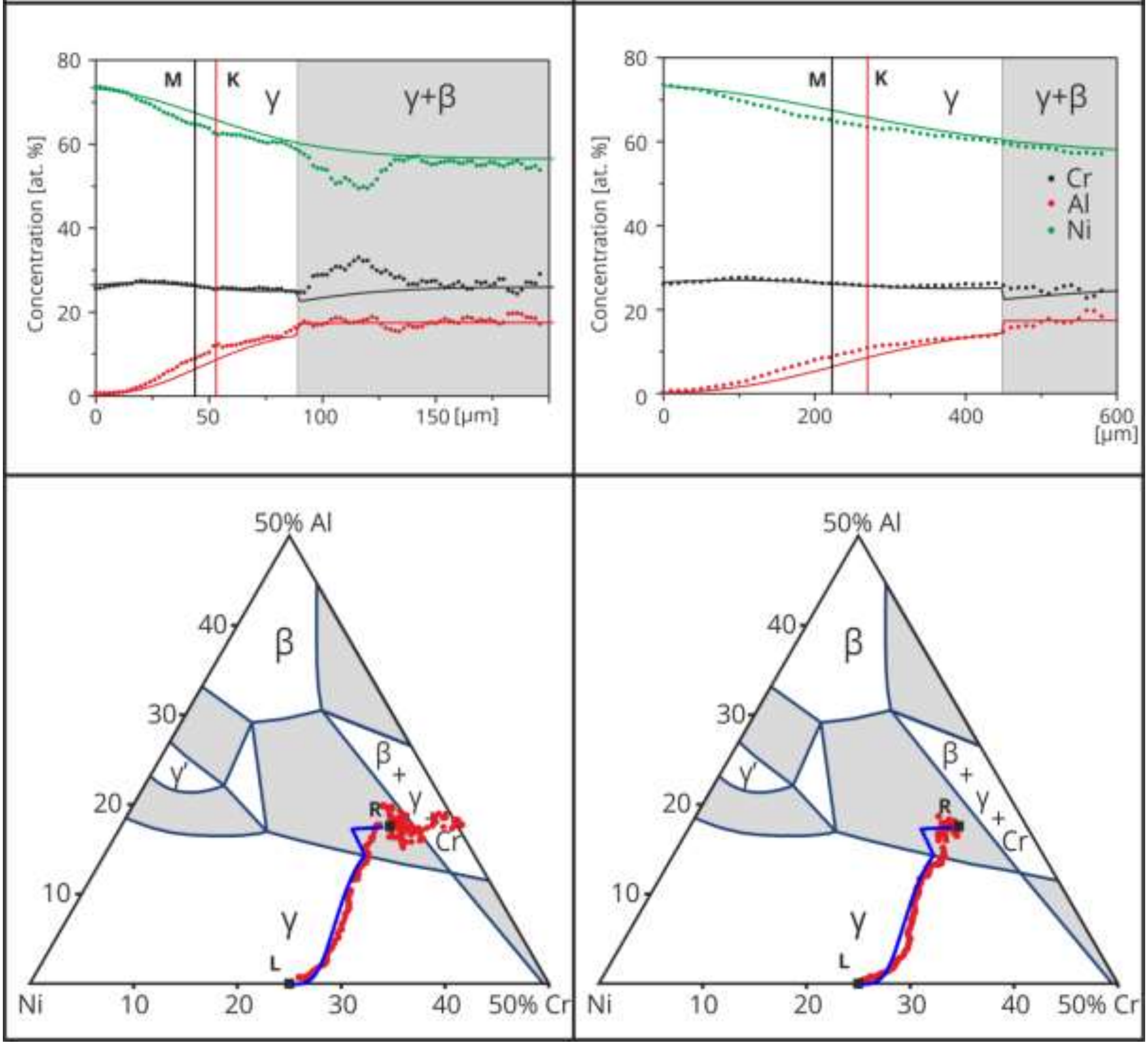

**Fig. 67. Experimental (dots) and simulated (lines) concentration profiles, and diffusion paths for γ-3|(γ+β), Ni-25Cr|Ni-26Cr-17.5Al diffusion couple (shown in Fig. 54, section 4.2).**

Diffusion couple: γ-4|(γ+β) Ni-40Cr|Ni-26Cr-17.5Al

| Effective intrinsic diff. in γ [$10^{-10}$ $cm^2/s$] | | |
|---|---|---|
| Ni | Cr | Al |
| 3.2 | 11.2 | 26.4 |

[31, 35]

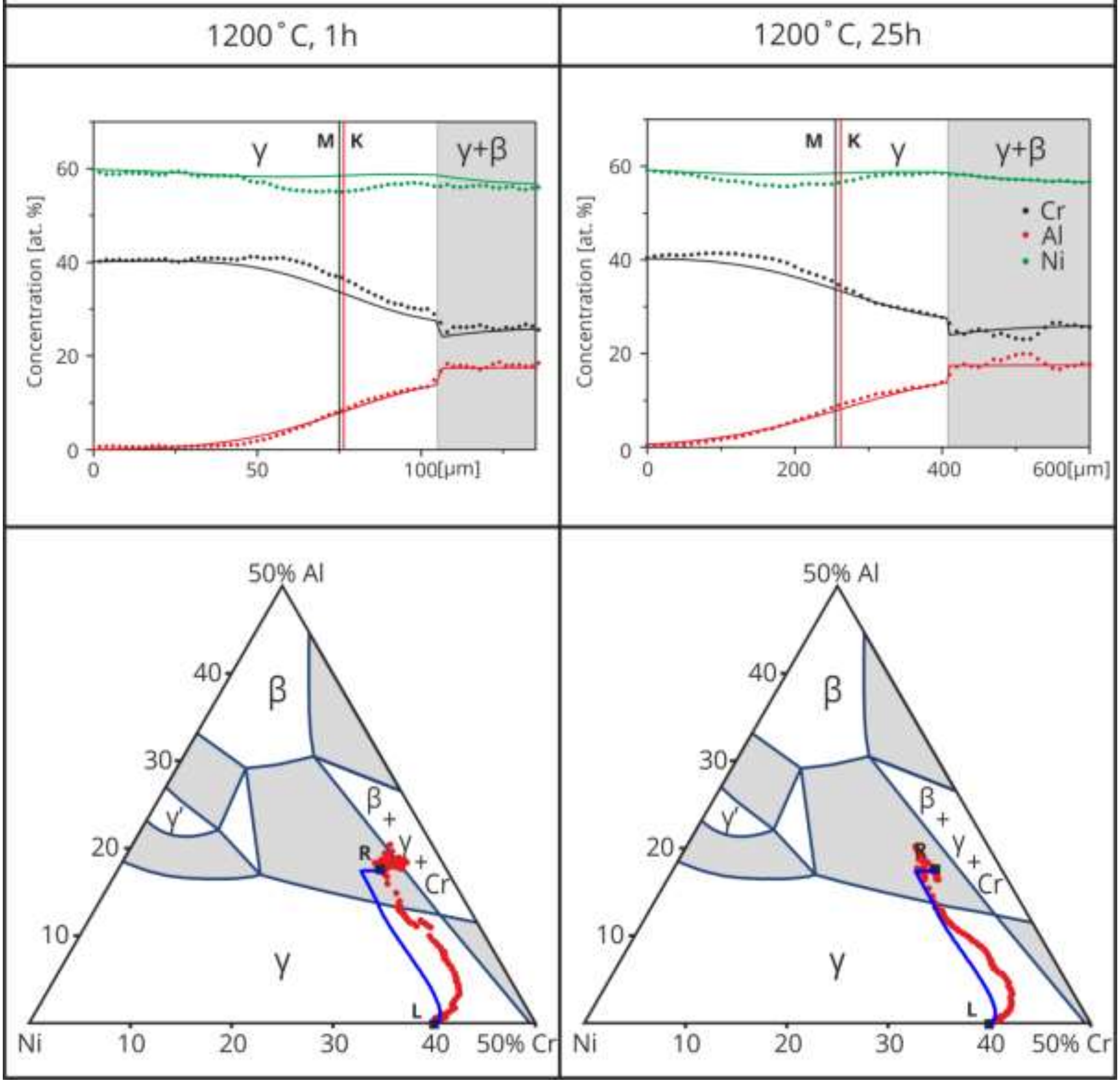

**Fig. 68. Experimental (dots) and simulated (lines) concentration profiles, and diffusion paths for γ-4|(γ+β), Ni40Cr|Ni-26Cr-17.5Al diffusion couple (shown in Fig. 55, section 4.2).**

## 5.2 Results of simulation and comparison with experimental data for couples with γ+γ′ alloy

The comparison of simulation and experimental results for the diffusion couples formed by the two-phase γ+γ′ alloy coupled with the four different γ alloys (Ni, Ni-10Cr, Ni-25Cr and Ni-40Cr) is shown in Figs. 69 to 72. For all couples very good agreement of the position of γ|γ+γ′ type 1a interphase boundary was obtained.

The experimental results for the couples presented in Fig. 71 show deviation of measured terminal compositions from the initial values shown in Tab. 6.

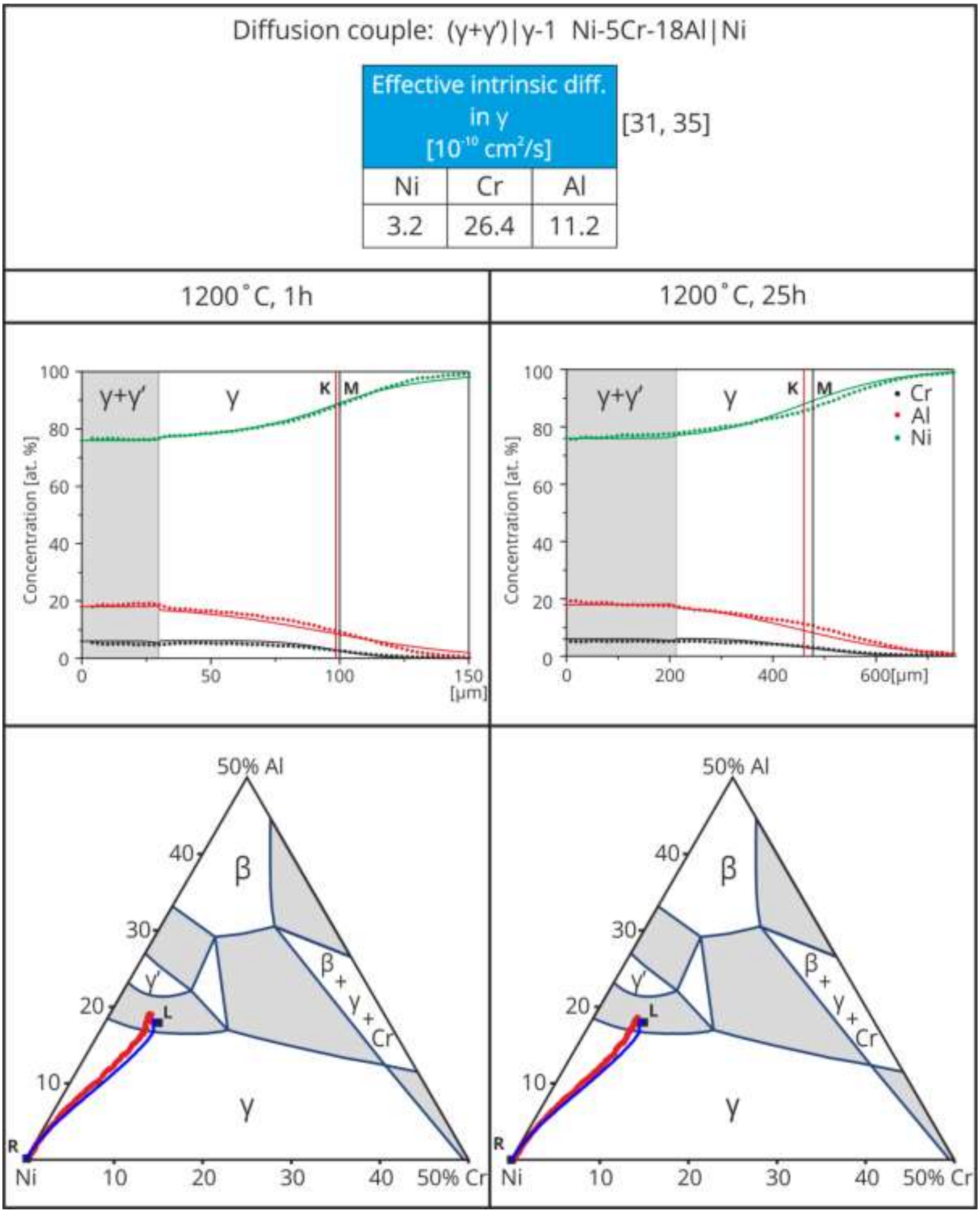

**Fig. 69. Experimental (dots) and simulated (lines) concentration profiles, and diffusion paths path for (γ+γ′)|γ-1, Ni-5Cr-18Al|Ni diffusion couple (shown in Fig. 60, section 4.2).**

Diffusion couple: (γ+γ′)|γ-2 Ni-5Cr-18Al|Ni-10Cr

| Effective intrinsic diff. in γ [$10^{-10}$ $cm^2/s$] | | |
|---|---|---|
| Ni | Cr | Al |
| 1.6 | 13.2 | 5.6 |

[31, 35]

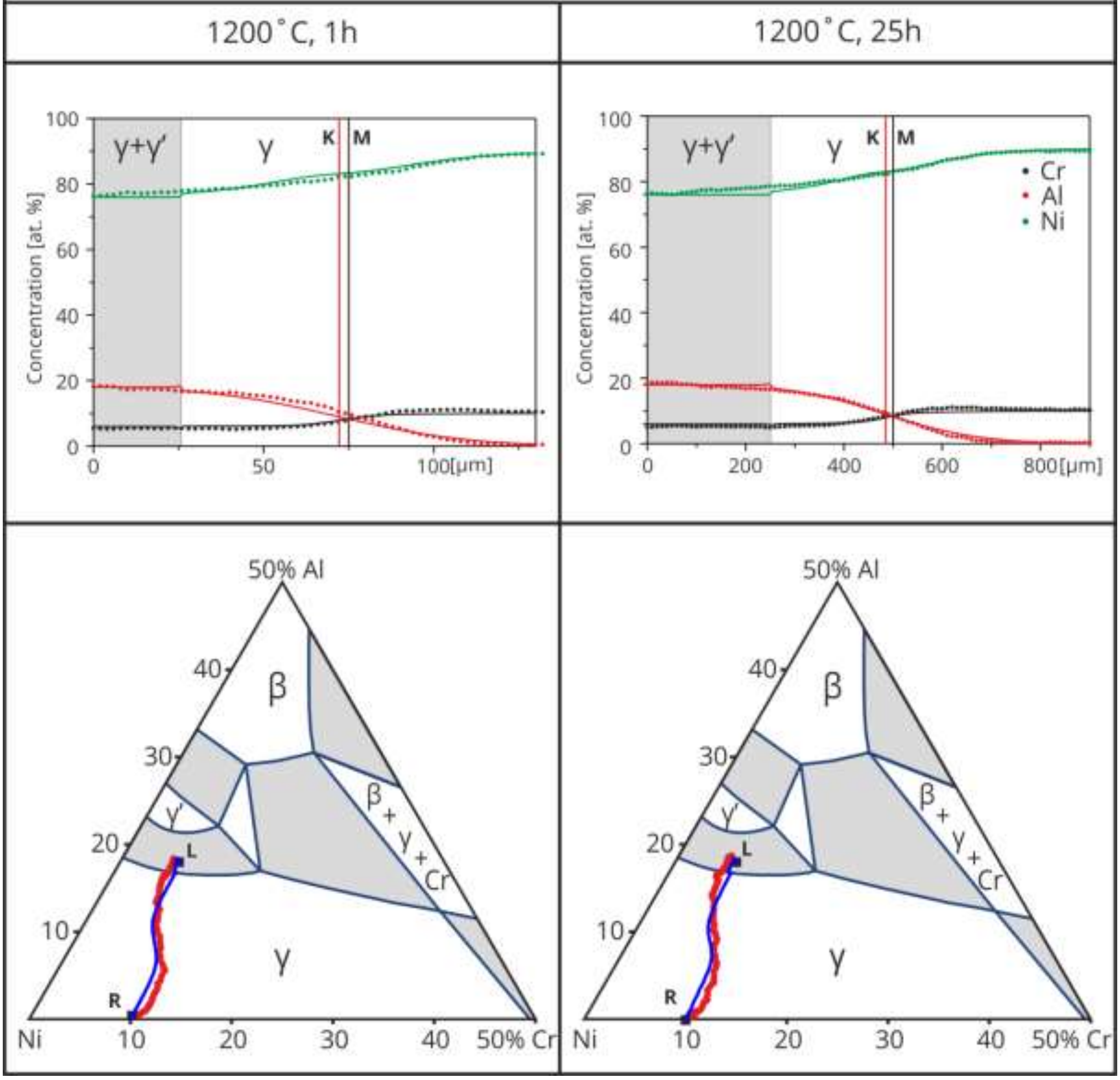

**Fig. 70. Experimental (dots) and simulated (lines) concentration profiles, and diffusion paths for (γ+γ′)|γ-2, Ni-5Cr-18Al|Ni-10Cr diffusion couple (shown in Fig. 61, section 4.2).**

Diffusion couple: (γ+γ')|γ-3 Ni-5Cr-18Al|Ni-25Cr

| Effective intrinsic diff. in γ [$10^{-10}$ $cm^2/s$] | | |
|---|---|---|
| Ni | Cr | Al |
| 2.24 | 5.28 | 0.64 |

[31, 35]

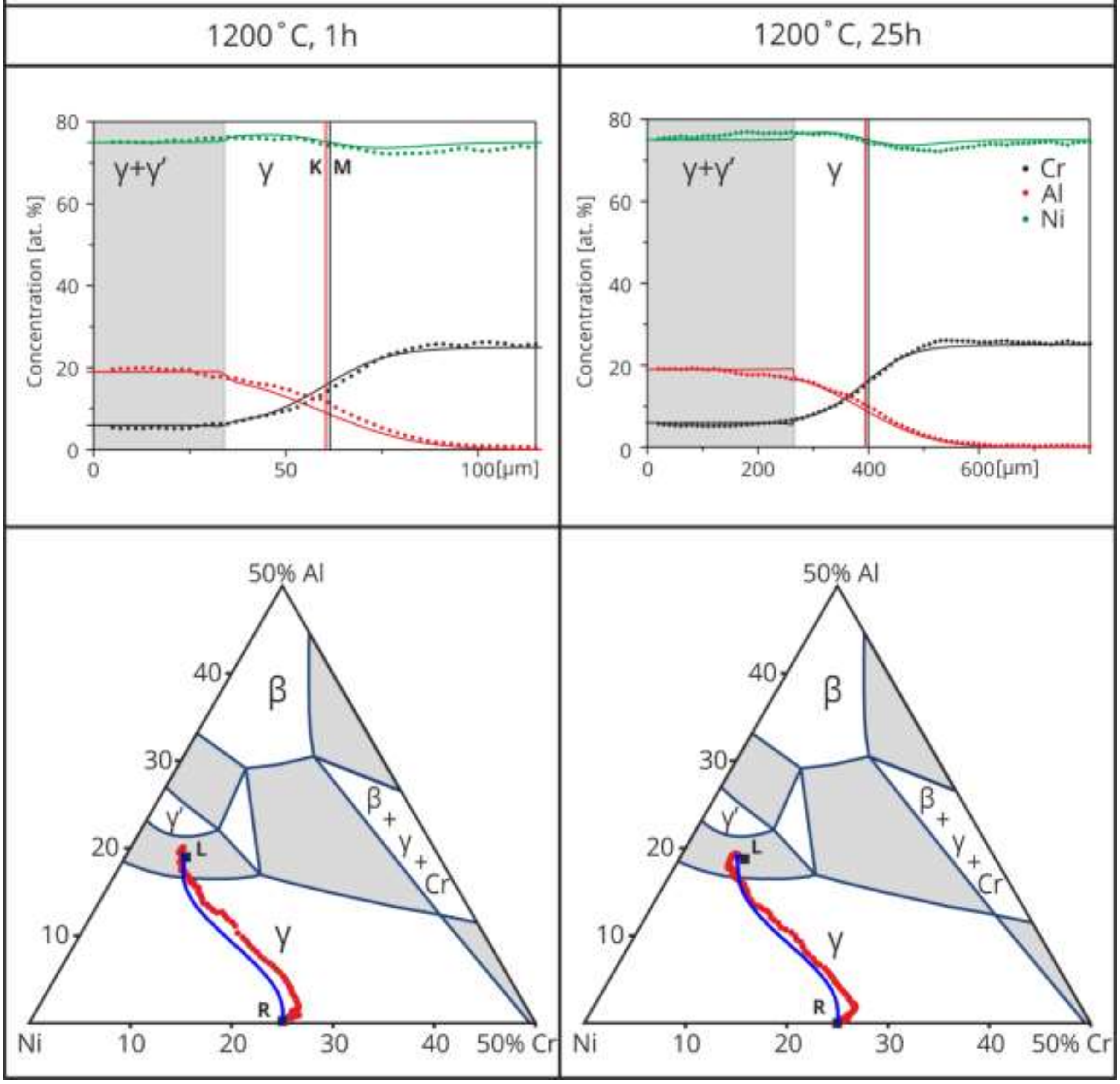

**Fig. 71. Experimental (dots) and simulated (lines) concentration profiles, and diffusion paths for (γ+γ')|γ-3, Ni-5Cr-18Al|Ni-25Cr diffusion couple (shown in Fig. 62, section 4.2).**

Diffusion couple: (γ+γ′)|γ-4 Ni-5Cr-18Al|Ni-40Cr

| Effective intrinsic diff. in γ [$10^{-10}$ $cm^2/s$] | | |
|---|---|---|
| Ni | Cr | Al |
| 2.4 | 2.4 | 2.4 |

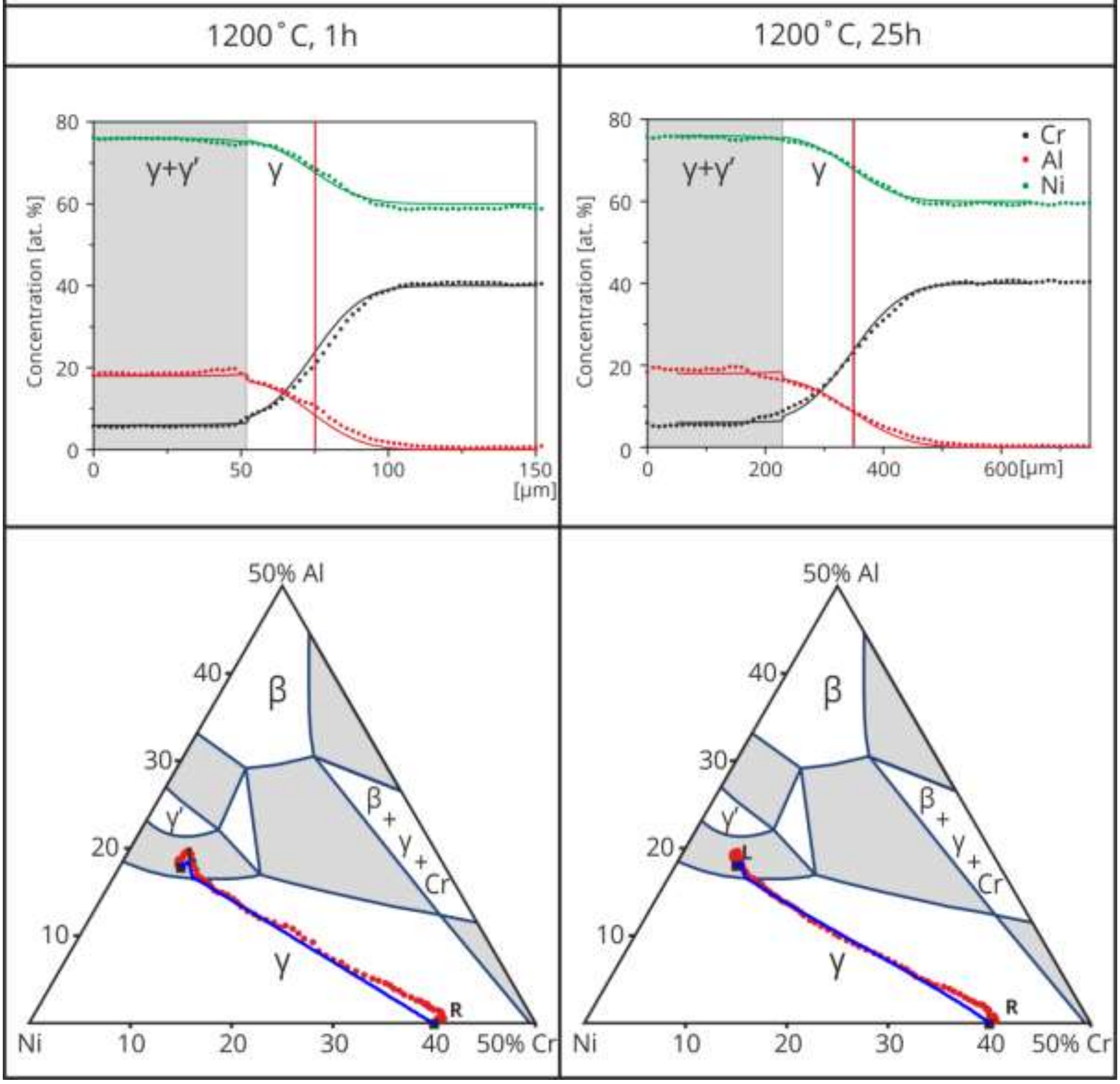

**Fig. 72. Experimental (dots) and simulated (lines) concentration profiles, and diffusion paths for (γ+γ′)|γ-1, Ni-5Cr-18Al|Ni-40Cr diffusion couple (shown in Fig. 63, section 4.2).**

## 5.3 Summary

Using the multi-multi model a good agreement between experimental and simulated results has been obtained. Good agreement between concentration profiles, diffusion paths and position of interphase boundaries can be seen.

For the first time, the shift of Kirkendall plane is calculated in multiphase diffusion couple and compared with experimental data, Tab. 10. The qualitative agreement has been achieved for majority of samples. No agreement between measured and computed shifts in samples containing pure nickel as one of terminal alloys is caused by unsatisfactory determination of Matano plane position. The comparison of calculated Kirkendall plane with SEM image for γ-3|(γ+β) diffusion couple is presented in Fig. 73. The satisfactory agreement between position of Kirkendall plane and interphase boundary is observed.

**Tab. 10. Comparision of measured and calculated shifts of Kirkendall plane.**

| | | Kirkendall plane shift [μm] | | | |
|---|---|---|---|---|---|
| Diffusion couple | | Experimental | | Simulated | |
| | | 1 h | 26 h | 1 h | 26h |
| Ni | Ni-26Cr -17.5Al | 9.8 | 55.8 | -3.4 | -7.3 |
| Ni-10Cr | | 10.9 | 31.2 | 5.2 | 26.8 |
| Ni-25Cr | | 1.3 | 18 | 9.2 | 47.1 |
| Ni-40Cr | | 4.5 | 22.6 | 1.4 | 7.3 |
| Ni | Ni-5Cr -18Al | -6.4 | -64 | 1.56 | 17.8 |
| Ni-10Cr | | -3.9 | -32.8 | -2.8 | -14.9 |
| Ni-25Cr | | -4.9 | -19.4 | -1.1 | -5.3 |
| Ni-40Cr | | 1.2 | 5.4 | 0 | 0 |

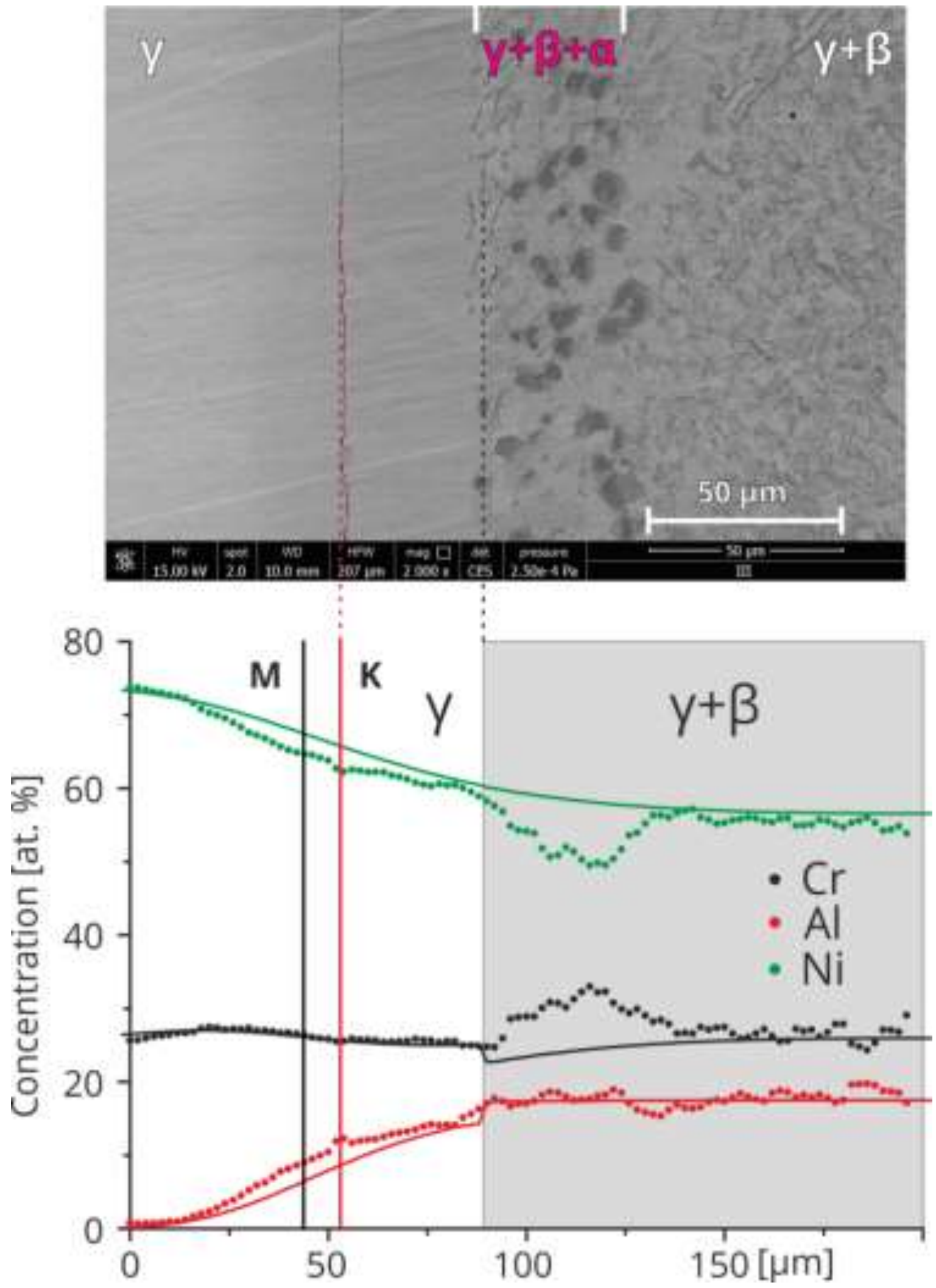

**Fig. 73. The comparison of experimental and simulated results for γ-3|(γ+β) diffusion couple (shown in Fig. 54, section 4.2).**

# Chapter 6. Summary and conclusions

The diffusion in ternary, multiphase systems is studied theoretically and experimentally. Two phenomenological models of such processes are presented:

- Generalized multi-phase, multi-component model based on the bi-velocity method in R1. The model is an extension of the Hopfe, Morral model through including the diffusion in both phases present in the system, the different partial molar volumes of components and the drift.
- Dual-scale two-phase model, that includes interdiffusion in matrix (global scale in R1) and the diffusion between matrix and precipitates (local scale in R3).

The models were used in numerical experiments in which the concentration profiles, the diffusion paths, the local entropy production rate and the volume fractions of the phases present in the system were calculated for various initial conditions. In this way a diversity of diffusion paths in ternary systems was confirmed.

For the first time the Kirkendall plane shift in two-phase system was computed and compared with experimental findings.

It has been shown that the dual-scale two-phase model of interdiffusion in two-phase zone is consistent with generalized multi-phase multi-component model. Both allow simulating evolution of the model microstructure during diffusion.

The multi-phase, multi-component model was applied to simulate diffusion in Ni-Cr-Al ternary, two-phase diffusion couples and the results were compared with experiment. Experimental and numerical results are in agreement. Calculated and measured concentration profiles and position of the IBs agree well.

Eleven diffusion couples between 7 terminal alloys were prepared, annealed at 1200°C for 1 and 26 hours and analyzed. The samples were prepared by the multiple method. It has been shown that the method allows obtaining good quality, planar, reproducible and oxide-free diffusion couples. Some imperfections in sample preparation, such as mismatch of the dimensions and shapes or the presence of cavities has been reduced by applying HIP procedure.

The concentration profiles were measured by the wide-line EDS analysis. The method allows measuring the overall concentrations in two-phase zones and the obtained results present high accuracy. According to my knowledge this is the first use of wide-line EDS method to study interdiffusion in multi-phase systems. The subsequent original conclusions follow from the performed studies:

1) The type 0 interphase boundary shifts during diffusion, contrary to the Morral A6 theorem [3].
2) In the case when the terminal compositions of the two-phase alloys (e.g. α+β|α+β) are close to the phase boundary (e.g. α+β|β) then the single β-phase zone can grow between two α+β zones and diffusion path enters the β region.
3) The diffusion path strongly depends on the shape of the phase boundary line and the accuracy of thermodynamical data is a key factor in modeling.
4) The occurrence of the local extrema in distribution of the entropy production at the interphase boundaries, as reported by Wierzba et al. [78], is a result of numerical errors and is related to the discontinuity of concentration at IB. Such errors can be eliminated as was shown in dual-scale two-phase model.

The performed studies open new challenges for theory end experiment, especially in the following areas:

1) theory
- implementation of the equilibrium parameters for the three-phase zone (tie-triangle);
- introducing the open boundary conditions, for applications like: carburizing, nitriding, ultra-high gravity treatment and others;
- considering the concentration dependent diffusivities;
- integration of model with the thermodynamic database to consider the chemical potentials as a true driving force;
- generalization of Matano method for multi-phase systems with volume change.

2) experiment

- extended studies of diffusion couples formed by the two two-phase alloys to reveal the horns near the type 0 interphase boundary, and microstructure evolution in diffusion couple. Especially, investigation aimed at the movement of the type 0 IB are necessary.
- long time diffusion experiments to collect data allowing to analyze stochastisation of diffusion along conode;
- sandwich type diffusion couple, i.e., $\beta|\gamma+\beta|\beta$, to investigate Kirkendall plane shift without limitation due to errors in Matano plane determination;
- modification of the HIP process to implement sample quenching in order to freeze the microstructure achieved by diffusion.

# Appendix 1. Numerical schema of the multi-multi model

To solve differential problem the following assumption were made:

$$w(x,t)=w\big(c_1(x,t),c_2(x,t),c_3(x,t)\big) \text{ for } t\geq 0,\ x\in[-\Lambda,\Lambda]. \tag{98}$$

where "$w$" can denotes: $c_i^j$, $\upsilon$, $\upsilon^j$, $\varphi^j$...,

The one dimensional space were discretized by $N$ spatial nodes, $\left(N\in\mathbb{N}\setminus\{0\}\right)$, defined at t=0 as: $x_k=-\Lambda+(k-1)h$ $\quad (k=1,2,...,N)$ and $x_{k+1/2}=\left(x_{k+1}+x_k\right)/2$ $\left(k=1,2,...,N-1\right)$.

The following denotations were introduced $w_k^t:=w(t,x_k)$ and $w_{k+1/2}^t=\left(w_k^t+w_{k+1}^t\right)/2$, where *t* denotes time and *k* is numbers spatial nodes.

The partial derivatives were approximated as follow $\dfrac{\partial w_{k+1/2}^t}{\partial x}=\left(w_{k+1}^t-w_k^t\right)/\left(x_{k+1}-x_k\right)$.

In this model the following differential equations, (49), (50) were considered:

$$c_{i,k}^{t+\Delta t}=c_{i,k}^t-\left(\tilde{J}_{i,k+1/2}^t-\tilde{J}_{i,k-1/2}^t\right)\Big/\left(\frac{x_{k+1}+x_{k-1}}{2}\right)\Delta t \text{ for } i=1,2,3, \tag{99}$$

$$x_k^{t+\Delta t}=x_k^t+\upsilon_k^t\Delta t, \tag{100}$$

$$x_{Kirk}^{t+\Delta t}=x_{Kirk}^t+\upsilon_{k+1/2}^{Drift,t}\Delta t, \text{ where } k \text{ is the nodes on left side of } x_{Kirk}^t, \tag{101}$$

where:

Eq. (48): $$\tilde{J}_{i,k+1/2}^t=\varphi_{k+1/2}^t\cdot\tilde{J}_{i,k+1/2}^{\alpha,t}+\left(1-\varphi_{k+1/2}^t\right)\tilde{J}_{i,k+1/2}^{\beta,t}+c_{i,k+1/2}^t\upsilon_{k+1/2}^t,\ i=1,2,3, \tag{102}$$

Eq. (46): $$\tilde{J}_{i,k+1/2}^{j,t}=J_{i,k+1/2}^{i,t}+c_{i,k+1/2}^{j,t}\upsilon_{k+1/2}^{j,t},\ j=\alpha,\beta,\ i=1,2,3, \tag{103}$$

Eq. (44): $$J_{i,k+1/2}^{j,t}=-D_i^j\left(c_{i,k+1/2}^{j,t}\right)',\ j=\alpha,\beta,\ i=1,2,3, \tag{104}$$

Eq. (45): $$\upsilon_{k+1/2}^{j,t}=-\sum_i\Omega_i^jJ_{i,k+1/2}^{j,t},\ j=\alpha,\beta, \tag{105}$$

Eq. (47): $$\upsilon_{k+1}^t=\frac{\upsilon_k^tc_k^{\alpha,t}\dfrac{\Omega_k^{\alpha,t}-\Omega_k^{\beta,t}}{\Omega_k^t}\dfrac{\varphi_{k+1}^tc_{k+1}^{\alpha,t}-\varphi_{k+1}^{t-\Delta t}c_{k+1}^{\alpha,t-\Delta t}}{\Delta t}\left(x_{k+1}^t-x_k^t\right)}{c_{k+1}^{\alpha,t}}, \text{ where } \upsilon_1^t=0. \tag{106}$$

The parameters of $\Omega_k^{\alpha,t}$, $\Omega_k^{\beta,t}$ (or $\Omega^\alpha$, $\Omega^\beta$) and $\Omega_k^t$ follows from Eqs. (29) and (34), the other parameters can be calculated from phase diagrams according to section 3.1.3.

## Bibliography

[1] A. M. Gusak, *Diffusion-controled Solid State Reactions* (Wiley-VCH, 2010)

[2] W. D. Hopfe and J. E. Morral, *ZIGZAG DIFFUSION PATHS IN MULTIPHASE DIFFUSION COUPLES,* Acta metall, mater. Vol. 42 (1994) p. 3887-3894

[3] J. E. Morral, *Diffusion Path Theorems for Ternary Diffusion Couples,* Metallurgical and Materials Transactions A Vol. 43A (2012) p. 3462-3470

[4] M. A. Dayananda and C. W. Kim, *An analytical representation of ternary diffusion paths,* Scripta Metallurgica Vol. 16 (1982) p. 815-818

[5] D. E. Coates and J. S. Kirkaldy, *Morphological stability of α-β phase interfaces in the Cu−Zn−Ni system at 775°C,* Metallurgical and Materials Transactions B Vol. 2 (1971) p. 3467-3477

[6] R. C. Reed, *The superalloys : fundamentals and applications* (Cambridge University Press, Cambrige, 2006)

[7] B. Mikółowski, *Stopy Żaroodporne i Żarowytrzymałe -Nadstopy-* (Wydawnictwo AGH, Kraków, 1997)

[8] A. Taylor and R. W. Floyd, *The constitution of nickel-rich alloys of the nickel-chromium-aluminum system,* J. Inst. Met. Vol. 81 (1952) p.

[9] I. I. Kornilov and R. S. Mints, Izv. Sekt Fiz -Khim Anal Vol. 22 (1953) p. 111-116

[10] A. Yu, Bagaryatskiy Russ J Inorg Chem Vol. 3 (1958) p. 247-252

[11] D. C. Tu and L. L. Seigle, *Kinetics of formation and microstructure of aluminide coatings on NiCr alloys,* Thin Solid Films Vol. 95 (1982) p. 47-56

[12] S. Ochiai, Y. Oya and T. Suzuki, *SOLUBILITY DATA IN Ni//3Al WITH TERNARY ADDITIONS,* Bulletin of Research Laboratory of Precision Machinery and Electronics Vol. (1983) p. 1-17

[13] S. M. Merchant and M. R. Notis, *A review: Constitution of the AlCrNi system,* Materials Science and Engineering Vol. 66 (1984) p. 47-60

[14] Y. M. Hong, H. Nakajima, Y. Mishima and T. Suzuki, *γ solvus surface in Ni-Al-X (X:Cr, Mo, and W) ternary systems,* ISIJ International Vol. 29 (1989) p. 78-84

[15] P. Rogl, *Ternary alloysa: comprehensive compendium of evaluated constitutional data and phase diagrams* (Weinheim, New York, 1991) 400-415

[16] C. C. Jia, K. Ishida and T. Nishizawa, *Partition of alloying elements between γ (A1), γ' (L12), and β (B2) phases in Ni-Al base systems,* Metallurgical and Materials Transactions A Vol. 25 (1994) p. 473-485

[17] E. Rosell-Laclau, M. Durand-Charre and M. Audier, *Liquid-solid equilibria in the aluminium-rich corner of the Al-Cr-Ni system,* Journal of Alloys and Compounds Vol. 233 (1996) p. 246-263

[18] D. Tu, Ph. D. Dissertation. state University of New York at Stony Brok. 1982

[19] W. Boettinger and U. Kattner, *Report to Investment Casting Cooperative Arrangement*, NIST, Year

[20] I. Ansara, N. Dupin, H. L. Lukas and B. Sundman, *Thermodynamic assessment of the Al-Ni system,* Journal of Alloys and Compounds Vol. 247 (1997) p. 20-30

[21] W. Huang and Y. A. Chang, *A thermodynamic analysis of the Ni-Al system,* Intermetallics Vol. 6 (1998) p. 487-498

[22] W. Huang and Y. A. Chang, *Thermodynamic properties of the Ni–Al–Cr system,* Intermetallics Vol. 7 (1999) p.

[23] K. Hashimoto, T. Abe and Y. Sawada, National Institute for Materials Science, http://www.nims.go.jp/cmsc/pst/database/Multicomponent/alcrni/alcrni.htm

[24] J. S. Kirkaldy and L. C. Brown, *Diffusion Behaviour in Ternary, Multiphase Systems,* Canadian Metallurgical Quarterly Vol. 2 (1963) p. 89-115

[25] J. E. Morral, C. Jin, A. Engstrom and J. Agren, *Three Typesof Planar Boundary in Multiphase Diffusion Couples,* Scripta Materialia Vol. 34 (1996) p. 1661-1666

[26] C. Jin and J. E. Morral, *Microstructures From Ni-Cr-Al Diffusion Couples Illustrating "THree Types of Boundaries",* Scripta Materialia Vol. 37 (1997) p. 621-626

[27] G. W. Roper and D. P. Whitle, *Interdiffusion in two-phase thernary solid systems,* Material Science Vol. 148 (1981) p. 148-153

[28] G. W. Roper and D. P. Whitle, *Theoretical diffusion profiles in single-phase ternary systems,* Metal Science Vol. 14 (1980) p. 541-549

[29] J. Crank, *The Mathematics of Diffusion 2 ed.* (Oxford Univ. Press, Qxford, 1979)

[30] A. M. Gusak and Y. A. Lyashenko, *Unique features encountered in the solution of equations for the diffusion transfer of mass in the two-phase zone of a triple system,* Journal of Engineering Physics Vol. 59 (1991) p. 1044-1049

[31] H. Chen and J. E. Morral, *VARIATION OF THE EFFECTIVE DIFFUSIVITY IN TWO-PHASE REGIONS,* Acta Materialia Vol. 47 (1999) p. 1175-1180

[32] J. O. Andersson, L. Höglund, B. Jönsson and J. Ågren, *Computer simulation of multicomponent diffusional transformations in steel* (Pergamon, Oxford, 1990)

[33] A. Engstrom, J. E. Morral and J. Agren, *Computer Simulations of Ni-Cr-Al Multiphase Diffusion Couples,* Acta Materialia Vol. 45 (1997) p. 1189-1199

[34] J. E. Morral and H. Chen, *On the Composition Difference at Boundaries Betwen a and a+b Regions in Multicomponent Diffusion Couples,* Scripta Materialia Vol. 43 (2000) p. 699-703

[35] H. Yang, J. E. Morral and Y. Wang, *On diffusion paths with "horns" and the formation of single phase layers in multiphase diffusion couples,* Acta Materialia Vol. 53 (2005) p. 3775-3781

[36] K. Wu, J. E. Morral and Y. Wang, *Horns on diffusion paths in multiphase diffusion couples,* Acta Materialia Vol. 54 (2006) p.

[37] J. E. Morral, X. Pan, N. Zhou, H. Larsson and Y. Wang, *Singularities in multiphase diffusion couples,* Scripta Materialia Vol. 58 (2008) p.

[38] K. Wu, J. E. Morral and Y. Wang, *A phase field study of microstructural changes due to the Kirkendall effect in two-phase diffusion couples,* Acta Materialia Vol. 49 (2001) p. 3401-3408

[39] L. Q. Chen, *Phase-field models for microstructure evolution,* Annual Review of Materials Science Vol. 32 (2002) p. 113-140

[40] Y. H. Sohn and M. A. Dayanda, *Diffusion studies in the β (B2 ), β' (bcc), and γ (fcc) Fe-Ni-Al alloys at 1000°C,* Metallurgical and Materials Transactions A Vol. 33 (2002) p. 3375-3392

[41] K. Wu, Y. A. Chang and Y. Wang, *Simulating interdiffusion microstructures in Ni-Al-Cr diffusion couples: A phase field approach coupled with CALPHAD database*, Scripta Materialia Vol. 50 (2004) p. 1145-1150

[42] K. Wu, J. E. Morral and Y. Wang, *Movement of Kirkendall markers, second phase particles and the Type 0 boundary in two-phase diffusion couple simulations,* Acta Materialia Vol. 52 (2004) p. 1917-1925

[43] J. E. Morral, *Predicting interdiffusion in high temperature coatings,* Tsinghua Science and Technology Vol. 10 (2005) p. 704-708

[44] B. Böttger, V. Witusiewicz and S. Rex, *Phase-field method coupled to calphab: Quantitative comparison between simulation and experiments in ternary eutectic In-Bi-Sn*, Modeling of Casting, Welding and Advanced Solidification Processes - XI, Year

[45] R. R. Mohanty and Y. Sohn, *Phase-field investigation of multicomponent diffusion in single-phase and two-phase diffusion couples,* Journal of Phase Equilibria and Diffusion Vol. 27 (2006) p. 676-683

[46] A. Ekhlakov, S. Dimitrov, T. A. Langhoff and E. Schnack, *Phase-field model for deposition of pyrolytic carbon,* Communications in Numerical Methods in Engineering Vol. 24 (2008) p. 2139-2154

[47] R. R. Mohanty, A. Leon and Y. H. Sohn, *Phase-field simulation of interdiffusion microstructure containing fcc-γ and L12-γ' phases in Ni-Al diffusion couples,* Computational Materials Science Vol. 43 (2008) p. 301-308

[48] R. R. Mohanty and Y. H. Sohn, *Microstructural stability of fcc-γ+B2-β coatings on γ substrate in Ni-Cr-Al system - A phase field model study,* Surface and Coatings Technology Vol. 203 (2008) p. 407-412

[49] K. Wu, N. Zhou, X. Pan, J. E. Morral and Y. Wang, *Multiphase Ni–Cr–Al diffusion couples: A comparison of phase field simulations with experimental data,* Acta Materialia Vol. 56 (2008) p. 3854-3761

[50] U. R. Kattner and C. E. Campbell, *Modelling of thermodynamics and diffusion in multicomponent systems,* Materials Science and Technology Vol. 25 (2009) p. 443-459

[51] Y. H. Wen, J. V. Lill, S. L. Chen and J. P. Simmons, *A ternary phase-field model incorporating commercial CALPHAD software and its application to precipitation in superalloys,* Acta Materialia Vol. 58 (2010) p. 875-885

[52] J. Heulens, B. Blanpain and N. Moelans, *Phase-field analysis of a ternary two-phase diffusion couple with multiple analytical solutions,* Acta Materialia Vol. 59 (2011) p. 3946-3954

[53] Y. S. Li, X. L. Cheng, F. Xu and Y. L. Du, *Interdiffusion Flux and Interface Movement in Metallic Multilayers,* Chinese Physics Letters Vol. 28 (2011) p.

[54] S. Tang, J. Wang, G. Yang and Y. Zhou, *Phase field modeling the growth of Ni3Al layer in the β/γ diffusion couple of Ni-Al binary system,* Intermetallics Vol. 19 (2011) p. 229-233

[55] L. Zhang, I. Steinbach and Y. Du, *Phase-field simulation of diffusion couples in the Ni-Al system,* International Journal of Materials Research Vol. 102 (2011) p. 371-380

[56] Y. Liu, J. Wang, Y. Du, G. Sheng, Z. Long and L. Zhang, *Phase boundary migration, Kirkendall marker shift and atomic mobilities in fcc Au-Pt alloys,* Calphad: Computer Coupling of Phase Diagrams and Thermochemistry Vol. 36 (2012) p. 94-99

[57] X. Q. Ke, J. E. Morral and Y. Wang, *Type n Boundaries in n-Component Diffusion Couples,* Acta Materialia Vol. 61 (2013) p. 2339-2347

[58] Y. Li, L. Zhang, H. Zhu and Y. Pang, *Effects of applied strain on interface microstructure and interdiffusion in the diffusion couples,* Metallurgical and Materials Transactions A: Physical Metallurgy and Materials Science Vol. 44 (2013) p. 3060-3068

[59] X. Q. Ke, J. E. Morral and Y. Wang, *Fundamentals of interdiffusion microstructure maps for dual-alloy systems,* Acta Materialia Vol. 76 (2014) p. 463-471

[60] R. S. Qin and H. K. Bhadeshia, *Phase field method,* Materials Science and Technology Vol. 26 (2010) p. 803-811

[61] S. R. Le Vine, *Reaction diffusion in the NiCrAl and CoCrAl systems,* Metallurgical and Materials Transactions A Vol. 9 (1978) p.

[62] L. A. Carol, A Study of Interdiffusion in β+y/y+y' Ni-Cr-Al Alloys a t 1200°C. Michigan Technological University. 1985

[63] J. A. Nesbitt and H. R.W., *Interdiffusion in Ni-Rich, Ni-Cr-Al Alloys at 1100 and 1200 C: Part I. Diffusion Paths and Microstructures,* Metallurgical Transactions A Vol. 18A (1987) p. 2061-2073

[64] J. A. Nesbitt and H. R.W., *Interdiffusion in Ni-Rich, Ni-Cr-Al Alloys at 1100 and 1200 C: Part II. Diffusion Coefficients and Predicted Concentration Profiles,* Metallurgical Transactions A Vol. 18A (1987) p. 2075-2086

[65] J. A. Nesbitt and R. W. Hckel, *Predicting Diffusion Paths and Interface Motion in γ/γ +β, Ni-Cr-Ai Diffusion Couples,* METALLURGICAL TRANSACTIONS A Vol. 18A (1987) p. 2087-2094

[66] S. M. Merchant, M. R. Notis and J. I. Goldstein, *Interface Stability in the Ni-Cr-Al System: Part I. Morphological Stability of β-γ Diffusion Couple Interfaces at 1150°C* Metallurgical Transactions A Vol. 21 (1990) p. 1901-1910

[67] S. M. Merchant, M. R. Notis and J. I. Goldstein, *Interface Stability in the Ni-Cr-Al System: Part II. Morphological Stability of β-Ni50Al vs γ-Ni40Cr Diffusion Couple Interfaces at 1150°C,* Metallurgical Transactions A Vol. 21 (1990) p. 1911-1919

[68] X. Qiao, Universsity of Connecticut. 1998

[69] A. C. Smigelkas and E. O. Kirkendall, *Zinc Diffusion in Alpha Brass,* Trans. AIME Vol. 171 (1947) p. 130

[70] L. S. Darken, *A Commentary on "Diffusion, Mobility and Their Interrelation through Free Energy in Binary Metallic Systems",* Trans. AIME Vol. 175 (1948) p.

[71] K. Holly and M. Danielewski, *Interdiffusion and free-boundary problem for r-component (r2) one-dimensional mixtures showing constant concentration,* Physical Review B Vol. 50 (1994) p. 13336-13346

[72] B. Bożek, M. Danielewski, K. Tkacz-Śmiech and M. Zajusz, *Interdiffusion: compatibility of Darken and Onsager formalisms,* Materials Science and Technology Vol. (2015) p.

[73] M. Danielewski, B. Wierzba, K. Tkacz-Śmiech, A. Nowotnik, B. Bożek and J. Sieniawski, *Bi-velocity model of mass transport in two-phase zone of ternary system,* Philosophical Magazine Vol. 93 (2013) p. 2044-2056

[74] M. Danielewski, B. Wierzba, K. Tkacz-Śmiech and A. Nowotnik, *Bi-velocity phase field method; reactive diffusion in Ni–Cr–Al,* Computational Materials Science Vol. 69 (2013) p. 1-6

[75] B. Wierzba, K. Tkacz-Śmiech and A. Nowotnik, *Reactive Mass Transport during the Aluminization of Rene-80,* Chemical Vapor Deposition Vol. (2013) p.

[76] B. Wierzba and K. Tkacz-Śmiech, *Diffusion zone formation in aluminized MAR-247 from Bi-velocity Phase Field Method,* Physica A: Statistical Mechanics and its Applications Vol. 392 (2013) p. 1100-1110

[77] B. Wierzba, M. Danielewski, A. Nowotnik and J. Sieniawski, *Bi-velocity Phase Field Method,* Defect and Diffusion Forum Vol. 333 (2013) p. 83-89

[78] B. Wierzba, K. Tkacz-Śmiech, A. Nowotnik and K. Dychtoń, *Aluminizing of nickel alloys by CVD. the effect of HCl flow,* Chemical Vapor Deposition Vol. 20 (2014) p. 80-90

[79] B. Wierzba, J. Romanowska, M. Zagula-Yavorska, J. Markowski and J. Sieniawski, *The Ni-Al-Zr Multiphase Diffusion Simulations,* High Temperature Materials and Processes Vol. 34 (2015) p. 495-502

[80] M. Danielewski, A. M. Gusak, B. Bożek and M. Zajusz, *Model of Diffusive Interaction between Two-phase Alloys with Explicit Fine-tuning of the Morphology Evolution,* (Under rewiew) p.

[81] M. Planck, *Ueber die Potentialdifferenz zwischen zwei verdünnten Lösungen binärer Electrolyte,* Annalen der Physik Vol. 276 (1890) p. 561-576

[82] B. Wierzba, *The Kirkendall effect in binary diffusion couples,* Physica A: Statistical Mechanics and its Applications Vol. 392 (2013) p. 2860-2867

[83] I. Prigogine and D. Kondepudi, *Modern Thermodynamics; From Heat Engine to Dissipative Structures* (John Willey & Sons, New York, 1998)

[84] X. Pan, N. Zhou, J. E. Morral and Y. Wang, *Microstructural stability in multi-alloy systems: Nanostructured two-phase, dual alloy multilayers,* Acta Materialia Vol. 58 (2010) p. 4149-4159

[85] L. Onsager, *Reciprocal Relations in Irreversible Processes. I,* Physical Review Vol. 37 (1931) p. 405-426

[86] F. S. Ham, *THeory of diffusion-limited precipitation,* J. Phys. Chem. Solids Vol. 6 (1958) p. 335-351

[87] J.-C. Zhao, *A combinatorial approach for efficient mapping of phase diagrams and properties,* Journal of Materials Research Vol. 16 (2001) p. 1565-1578

[88] J.-C. Zhao, *Document A combinatorial approach for structural materials,* Advance Engineering Materials Vol. 3 (2001) p. 143

[89] J.-C. Zhao, *A combinatorial approach for efficient mapping of phase diagrams and properties,* Journal of Material Research Vol. 16 (2001) p. 1565

[90] J.-C. Zhao, M. R. Jacson, L. A. Peluso and L. Brewear, *A diffusion multiple approach for the accelerated design of structural materials,* MRS Bulletine Vol. 27 (2002) p. 324

[91] J.-C. Zhao, M. R. Jacson, L. A. Peluso and L. Brewear, *A diffusion-multiple approach for mapping phase diagrams, hardness, and elastic modulus,* JOM Vol. 57 (2002) p. 42

[92] J.-C. Zhao, M. R. Jackson and L. A. Peluso, *Determination of the Nb-Cr-Si phase diagram using diffusion multiples,* Acta Materialia Vol. 51 (2003) p. 6395-6405

[93] J.-C. Zhao, *Reliability of the diffusion-multiple approach for phase diagram mapping,* Journal of Materials Science Vol. 39 (2004) p. 3913-3925

[94] J.-C. Zhao, X. Zheng and D. G. Cahill, *High-throughput diffusion multiples,* Materials Today Vol. (2005) p. 28-37

[95] C. E. Campbell, J.-C. Zhao and M. F. Henry, *Examination of Ni-base superalloy diffusion couples containing multiphase regions,* Materials Science and Engieniering A Vol. 407 (2005) p. 135-146

[96] J.-C. Zhao, *Combinatorial approaches as effective tools in the study of phase diagrams and composition-structure-property relationships,* Progress in Material Science Vol. 174 (2006) p. 557-631

[97] S. Kobayashi and S. Zaefferer, *Determination of phase equilibria in the Fe3Al-Cr-Mo-C semi-quaternary system using a new diffusion-multiple technique,* Journal of Alloys and Compounds Vol. 452 (2008) p. 67-72

[98] A. Savitzky and M. J. E. Golay, *Smoothing and differentiation of data by simplified least squares procedures,* Analytical Chemistry Vol. 36 (1964) p. 1627-1639

[99] C. Matano, *On the Relation betwen the Diffusion-Coefficient and Concentration of Solid Materials,* Japanes Journal of Physics Vol. 9 (1933) p. 109-113

[100] C. E. Campbell, *Assessment of the diffusion mobilites in the γ' and B2 phases in the Ni-Al-Cr system,* Acta Materialia Vol. 56 (2008) p. 4277-4290

[101] R. L. David, *CRC Handbook of Chemistry and Physics* (CRC Press, Boca Raton, FLs, 2005)

## List of figures

## List of tables